\documentclass[aps,prd,twocolumn,superscriptaddress,showpacs,amsmath,amssymb,nofootinbib,eqsecnum, preprintnumbers]{revtex4-1}
\usepackage{graphicx}

\usepackage{epstopdf}
\usepackage{color}
\usepackage{enumerate}

\newcommand{\be}{\begin{equation}}
\newcommand{\ee}{\end{equation}}
\newcommand{\ba}{\begin{eqnarray}}
\newcommand{\ea}{\end{eqnarray}}

\newcommand{\beq}{\begin{equation}}
\newcommand{\eeq}{\end{equation}}
\newcommand{\beqa}{\begin{eqnarray}}
\newcommand{\eeqa}{\end{eqnarray}}

\begin{document}
\title{Thermodynamics of rotating black holes and black rings: phase transitions and thermodynamic volume}

\author{Natacha Altamirano}
\email{naltamirano@famaf.unc.edu.ar}
\affiliation{Perimeter Institute, 31 Caroline St. N., Waterloo,
Ontario, N2L 2Y5, Canada}
\affiliation{Facultad de Matem\'atica, Astronom\'ia y F\'isica, FaMAF, Universidad Nacional de C\'ordoba,
Instituto de F\'isica Enrique Gaviola, IFEG, CONICET,
Ciudad Universitaria (5000) C\'ordoba, Argentina}
\author{David Kubiz\v n\'ak}
\email{dkubiznak@perimeterinstitute.ca}
\affiliation{Perimeter Institute, 31 Caroline St. N., Waterloo,
Ontario, N2L 2Y5, Canada}
\affiliation{Department of Physics and Astronomy, University of Waterloo,
Waterloo, Ontario, Canada, N2L 3G1}
\author{Robert B. Mann}
\email{rbmann@sciborg.uwaterloo.ca}
\affiliation{Department of Physics and Astronomy, University of Waterloo,
Waterloo, Ontario, Canada, N2L 3G1}
\author{Zeinab Sherkatghanad}
\email{zsherkat@uwaterloo.ca}
\affiliation{Perimeter Institute, 31 Caroline St. N., Waterloo,
Ontario, N2L 2Y5, Canada}
\affiliation{Department of Physics and Astronomy, University of Waterloo,
Waterloo, Ontario, Canada, N2L 3G1}
\affiliation{Department of Physics, Isfahan University of Technology, Isfahan, 84156-83111, Iran}

\date{January 11, 2014}  

\begin{abstract}
 In this review we summarize, expand, and set in context recent developments on the thermodynamics of black holes in extended phase space,  
where the cosmological constant is interpreted as thermodynamic pressure and treated as a thermodynamic variable in its own right. We specifically consider
 the thermodynamics of higher-dimensional rotating asymptotically flat and AdS black holes and black rings in a canonical (fixed angular momentum) ensemble.  
We plot the associated thermodynamic potential--the Gibbs free energy--and study its behaviour to uncover possible thermodynamic phase transitions in these black hole spacetimes. We show that the multiply-rotating Kerr-AdS black holes exhibit a rich set of interesting thermodynamic phenomena analogous to the `every day thermodynamics' of simple substances, such as reentrant phase transitions of multicomponent liquids, multiple first-order solid/liquid/gas phase transitions, and liquid/gas phase transitions of the Van der Waals type. Furthermore, the reentrant phase transitions also occur  for   multiply-spinning asymptotically flat Myers--Perry black holes. The thermodynamic volume, a quantity conjugate to the thermodynamic pressure, is studied for AdS black rings and demonstrated to satisfy the reverse isoperimetric inequality; this provides a first example 
of calculation confirming the validity of isoperimetric inequality conjecture for a black hole with non-spherical horizon topology.
The equation of state $P=P(V,T)$ is studied for various black holes both numerically and analytically---in the ultraspinning 
and slow rotation regimes.
\end{abstract}


\maketitle

\section{Introduction}\label{sec:Intro}
The subject of black hole thermodynamics continues to be one of import in gravitational physics.  Despite  long-established notions that black holes have a temperature proportional to their surface gravity, an entropy proportional to their horizon area, and obey a version of the first law of thermodynamics, the subject is still not fully understood.  Investigations in asymptotically anti de Sitter  (AdS) spacetimes have been carried out in considerable detail for three decades since it was pointed out that radiation/large AdS black hole phase transitions can take place \cite{HawkingPage:1983}.
Thermodynamic equilibrium can be defined, and  many additional corroborative results have been obtained that provide strong evidence that such objects indeed behave as thermodynamic systems.  

The proposal that the  mass of an AdS black hole should be interpreted as the {\em enthalpy} of the spacetime represents an interesting new development in this subject.  It emerged from geometric derivations of the Smarr formula for AdS black holes \cite{KastorEtal:2009} that suggested the cosmological constant should be considered as a thermodynamic variable \cite{CreightonMann:1995} analogous to pressure in the first law \cite{CaldarelliEtal:2000}.  This has led to a number of investigations of black hole thermodynamics in this extended phase space 
 \cite{ Dolan:2010, Dolan:2011a, Dolan:2011b, Dolan:2012, Dolan:2013, CveticEtal:2010, KastorEtal:2010, LarranagaCardenas:2012, LarranagaMojica:2012, Gibbons:2012ac, KubiznakMann:2012,GunasekaranEtal:2012,BelhajEtal:2012,LuEtal:2012,Smailagic:2012cu,Spallucci:2013osa, HendiVahinidia:2012, 
Chen:2013ce, Zhao:2013oza, Belhaj:2013ioa, AltamiranoEtal:2013a, AltamiranoEtal:2013b,Cai:2013qga, Belhaj:2013cva, Spallucci:2013jja, Mo:2013sxa, Xu:2013zea, MoLiu:2013, Zou:2013owa, Ma:2013aqa, Ulhoa:2013ffa, Castro:2013pqa, El-Menoufi:2013pza, Lu:2013ura, Lu:2012xu, MoLiu:2014}.  
Extensions of this idea to asymptotically de Sitter space times have also been recently explored, with the resultant thermodynamic notions for black holes remaining intact \cite{DolanEtal:2013} (see also \cite{GibbonsHawking:1977, Hayward:1997jp, Padmanabhan:2002sha, Cai:2002, Cai:2002b, Sekiwa:2006, UranoEtal:2009, Gibbons:2005vp, Bhattacharya:2013tq}).

The purpose of this paper is to provide a detailed overview and investigation of 
 the thermodynamics of  rotating asymptotically flat and AdS black holes and black rings in a canonical (fixed angular momentum) ensemble.  We find a variety of novel features emerge.  In particular, we shall show that the thermodynamics of multiply-rotating Kerr-AdS black holes  are, under certain circumstances, analogous to the `every day thermodynamics' of simple substances, such as reentrant phase transitions of multicomponent liquids, multiple first-order solid/liquid/gas phase transitions, and liquid/gas phase transitions of the Van der Waals type.  We also find that some of these phenomena, such as  reentrant phase transitions, also occur  for   multiply-spinning asymptotically flat Myers--Perry black holes. We compute  the thermodynamic volume (conjugate to the pressure), and 
  demonstrate  that in all  examples we consider it satisfies the reverse isoperimetric inequality.  In particular, we find that this inequality remains true even for the  thin asymptotically AdS black rings---providing a first confirmation of the reverse isoperimetric inequality conjecture for a black hole with non-spherical horizon topology.

\subsection{Canonical ensemble and phase transitions}

Concentrating entirely on asymptotically flat and AdS black holes,  we consider the proposal that {\em thermodynamic pressure} is given by
\be\label{PLambda}
P = - \frac{1}{8 \pi} \Lambda=\frac{(d-1)(d-2)}{16 \pi l^2}\,,
\ee
where $d$ stands for the number of spacetime dimensions, $\Lambda$ is the cosmological constant, and $l$ denotes the AdS radius.  
The asymptotically flat case has $P=0$.

We shall study the thermodynamics in a {\em canonical ensemble}, that is for fixed black hole angular momenta $J_i$, and/or, in the case of charged black holes for fixed charge $Q$.  The stability
of supersymmetric AdS vacua \cite{Weinberg:1982id} ensures that 
these kinds of thermodynamic ensembles can in principle be prepared by coupling an asymptotically AdS space-time to a bath that allows
transfer of energy but not angular momentum or charge by coupling only to operators with $J_i = 0$, $Q = 0$. 
Hence within this context one can prepare a canonical ensemble with any fixed $J_i$ and $Q$.
 
The equilibrium thermodynamics is governed by the {\em Gibbs free energy}, $G=G(T,P,J_i, Q)$,
whose {\em global minimum} yields the state of a system
for a fixed $(T,P, J_i, Q)$.\footnote{Note that there are states of matter, for example   supercooled water, which do not correspond to the global minimum of $G$ but  rather are `metastable'.}
Since mass of the black hole $M$ is interpreted as the enthalpy, we have the following thermodynamic relation:
\be\label{G}
G=M-TS=G(P,T,J_1,\dots,J_N, Q)\,. 
\ee
Here $T$ and $S$ stand for the horizon temperature and black hole entropy.

Understanding the behaviour of $G$ is essential for uncovering
possible {\em thermodynamic phase transitions}.  We shall see that it is often impossible to write the 
relation $G=G(T,P,J_i,Q)$ explicitly. For this reason we fix the pressure and angular momenta (and charge)--- usually expressed parametrically through the horizon radius $r_+$---to particular values and plot $G$ numerically. In this way we obtain $G-T$ diagrams which incorporate the information about possible phase transitions.\footnote{An interesting alternative proposal which attracted attention in past few years is to consider an ``effective thermodynamic geometry'' and study its curvature singularities \cite{Weinhold:1975a, Weinhold:1975b, Ruppeiner:1979, Ruppeiner:1995, Quevedo:2006xk,Liu:2010sz}. These then provide an information about presence of critical points and hence possible phase transitions in the given spacetime. However, in this paper we do not follow this strategy and concentrate entirely on analyzing the Gibbs free energy.}  

The {\em local thermodynamic stability} of a canonical ensemble is characterized by positivity
of the specific heat at constant pressure
\be\label{CP}
C_P\equiv C_{P,J_1,\dots,J_N,Q}=T\left(\frac{\partial S}{\partial T}\right)_{P,J_1,\dots,J_N,Q}\,.
\ee
We take negativity of $C_P$ as a sign of local thermodynamic instability.
Note that  in this relation we automatically assume fixed $J_i$ (and/or $Q$ in the case of charged black holes).
That is, our specific heat at constant $P$
is in fact a specific heat at constant $(P,J_i,Q)$ and coincides with $C_J$ and/or $C_Q$ considered in previous studies, e.g.,   
\cite{MonteiroSantos:2009, MonteiroEtal:2009}. When plotting the Gibbs free energy we shall plot branches with $C_P>0$ in {\em red solid lines} and
branches with $C_P<0$ in {\em dashed blue lines}.

Once the behaviour of the Gibbs free energy is known, we construct the associated {\em phase diagrams}. These are usually drawn in the $P-T$ plane 
and display the {\em coexistence lines} of various black hole phases, inter-related by the first-order phase transitions, as well as reveal possible 
{\em critical  points} where the coexistence lines terminate/merge together.    

A renowned example of a transition in black hole spacetimes is the radiation/black hole first-order phase transition of Hawking and Page observed for Schwarzschild-AdS black holes immersed in a bath of radiation \cite{HawkingPage:1983}.  Such a phenomenon (discussed further  in the next section) has a dual interpretation for a boundary quantum 
field theory via the AdS/CFT correspondence and is related to a confinement/deconfinement phase transition in the dual quark gluon plasma \cite{Witten:1998b}. 
For charged or rotating black holes one likewise observes  a small/large black hole first order phase transition 
reminiscent of the liquid/gas transition of the {\em Van der Waals} fluid  \cite{ChamblinEtal:1999a,ChamblinEtal:1999b,CveticGubser:1999a, CaldarelliEtal:2000, TsaiEtal:2012, Dolan:2011a, KubiznakMann:2012, Hristov:2013sya, Johnson:2013, NiuEtal:2011, Dolan:2012, GunasekaranEtal:2012, Chen:2013ce, Cai:2013qga,Poshteh:2013pba, Wei:2012ui, BelhajEtal:2012, HendiVahinidia:2012, Belhaj:2013cva, Dutta:2013dca, Zou:2013owa, MoLiu:2014}.

Interesting new phenomena appear in higher dimensions.  Recently it has been shown  \cite{AltamiranoEtal:2013a} that in all dimensions $d\geq 6$  singly spinning Kerr-AdS black holes demonstrate the peculiar behaviour of large/small/large 
black hole transitions reminiscent of   {\em reentrant phase transitions} (RPT) observed for multicomponent fluid systems, gels, ferroelectrics, liquid crystals, and binary gases, e.g., \cite{NarayananKumar:1994}. 
A system undergoes an RPT if a monotonic variation of any thermodynamic quantity results in two
(or more) phase transitions such that the final state is
macroscopically similar to the initial state. 
For singly spinning Kerr-AdS black holes it was found \cite{AltamiranoEtal:2013a} that for 
a fixed pressure within a certain range
(and for a given angular momentum) a monotonic lowering of the temperature yields
a large/small/large black hole transition. The situation is accompanied by a discontinuity
in the global minimum of the Gibbs free energy, referred
to as a {\em zeroth-order phase transition}, a phenomenon seen
in superfluidity and superconductivity \cite{Maslov:2004}. 
The RPT was also found for the four-dimensional Born--Infeld-AdS black hole spacetimes, deep in the non-linear regime of the Born--Infeld theory \cite{GunasekaranEtal:2012}, and for the black holes of third-order Lovelock gravity \cite{FrassinoEtal:2013}.
Remarkably, here we show (see Sec.~\ref{sec:MP}) that a similar phenomenon is observed for the asymptotically flat doubly-spinning Myers--Perry black holes of vacuum Einstein gravity. Hence, neither exotic matter nor a cosmological constant (and hence AdS/CFT correspondence) are required for this phenomena to occur in black hole spacetimes.

More intriguingly, other qualitatively different interesting phenomena known from the `every day thermodynamics' emerge when multiply rotating Kerr-AdS black holes are considered  \cite{AltamiranoEtal:2013b}.  Depending on the ratios of the angular momenta, we find  an analogue of a solid/liquid phase transition, multiple first-order 
{\em small/intermediate/large} black hole phase transitions with one {\em tricritical} (triple) and two critical points reminiscent of the solid/liquid/gas phase transition, and  the `standard' liquid/gas behaviour of a Van der Waals fluid.  
We review and provide more detail concerning these phenomena in the following sections.

\subsection{Thermodynamic volume}
A conjugate quantity to the thermodynamic pressure is the {\em thermodynamic volume} $V$. 
For AdS black hole spacetimes this can formally be obtained from the {\em first law} of black hole thermodynamics by
allowing the pressure, given by \eqref{PLambda}, (and hence the cosmological constant) to vary
\be\label{1st}
\delta M=T\delta S+\sum_i\Omega_i \delta J_i+\Phi \delta Q+V\delta P\,.
\ee
Here, $\Omega_i$ stand for the horizon angular velocities and $\Phi$ for the horizon electrostatic potential. 
The extended first law \eqref{1st} is consistent with the following {\em Smarr--Gibbs--Duhem relation}, an integrated formula relating the black hole mass to other thermodynamic quantities:
\be\label{Smarr}
\frac{d-3}{d-2}M=TS+\sum_i\Omega_i J_i+\frac{d-3}{d-2}\Phi Q-\frac{2}{d-2} VP\,,
\ee
the two being related by a dimensional (scaling) argument \cite{KastorEtal:2009}. 
Either of these relations can be used for calculating the thermodynamic volume $V$ of AdS black holes.
As we shall see in each case of interest, the final expression for $V$ has a smooth limit for $P\to 0$, which defines a thermodynamic volume for asymptotically flat black holes. Alternatively, one can calculate $V$ using the method of Killing co-potentials \cite{KastorEtal:2009, CveticEtal:2010}.\footnote{This consists of finding a co-potential 2-form whose divergence gives the Killing vector that is a horizon generator of the  black hole spacetime under consideration. Once known, such a co-potential is integrated to give the thermodynamic volume. However finding the co-potential is  not easy. For spherical rotating black hole spacetimes this was possible due to the presence of a special hidden symmetry of a Killing--Yano tensor 
\cite{KubiznakFrolov:2007}. This symmetry does not exist for black rings or black saturns.}

The thermodynamic volume has dimensions of (length)$^{d-1}$ and describes a spatial volume characterizing
the black hole spacetime.\footnote{For alternative definitions of various volumes associated with black hole spacetimes, see 
\cite{Hayward:1997jp, Parikh:2006, BallikLake:2010, BallikLake:2013}.}
In particular, for the Schwarzschild black hole in $d=4$ one obtains the intriguing relation
\be
V=\frac{4}{3}\pi r_+^3\,,
\ee
which is the volume of a ball of radius $r_+$ in the Euclidean space, with $r_+$ being the horizon radius in Schwarzschild coordinates.
For a wide variety of asymptotically flat and AdS black hole spacetimes with spherical horizon topology the thermodynamic volume has been calculated
in \cite{CveticEtal:2010}.  Although much more complicated in the presence of angular momenta and various additional charges, the thermodynamic volume in all cases studied so far was found to obey the so called {\em reverse isoperimetric inequality}, an inequality between the horizon area (black hole entropy) and the thermodynamic volume. 

In Euclidean space ${\mathbb E}^{d-1}$, the isoperimetric inequality for the volume ${\cal V}$ of a connected domain 
whose area is ${\cal A}$ states that the ratio 
\be\label{ratio}
{\cal R}= \Bigl( \frac{(d-1){\cal V}\,}{\omega_{d-2} } \Bigr )^{\frac{1}{d-1}}\,
  \Bigl(\frac{\omega_{d-2}}{\cal A}\Bigr)^{\frac{1}{d-2}}\,,
\ee
where 
\be\label{omega}
\omega_{d}=\frac{2\pi^{\frac{d+1}{2}}}{\Gamma\Bigl(\frac{d+1}{2}\Bigr)}\,
\ee
is the volume of the unit $d$-sphere, obeys ${\cal R}\leq 1$, with equality if and only if the domain is a standard round ball.
It was {\em conjectured} in \cite{CveticEtal:2010} that a {\em reverse isoperimetric inequality},
\be\label{ISO}
{\cal R}\geq 1\,,
\ee
holds for any asymptotically AdS black hole, where  ${\cal A}$ is the black hole horizon area and ${\cal V}$ 
is the thermodynamic volume $V$,  
the bound being saturated for Schwarzschild-AdS black holes. In other words, 
for a fixed thermodynamic volume the entropy of the black hole is maximized for the Schwarzschild-AdS spacetime.
Up to now this conjecture has been verified for a variety of (charged rotating) black holes with the horizon of spherical topology. 
In Sec.~\ref{sec:rings} we show that it remains true for (thin) black rings with toroidal horizon topology.

\subsection{Equation of state}

Once the thermodynamic volume is known and the cosmological constant identified with the thermodynamic pressure, one can, for a given black hole, write down the corresponding `fluid' {\em equation of state} relating the pressure, temperature, volume and  other external parameters characterizing the black hole, $P=P(V,T,J_i,Q)$. Similar to references \cite{KubiznakMann:2012, GunasekaranEtal:2012}, instead of $V$ we employ
a new quantity $v$, corresponding to the  ``{\em specific volume}'' of the fluid, defined by\footnote{%
We note an interesting coincidence pointed to us by Jennie Traschen in private communications. Similar to the specific volume of the fluid which as per usual is taken to be the total volume of the fluid divided by the
number of particles, for non-rotating AdS black holes one may define their specific volume to be
\be\label{VN}
\tilde v=V/N\,,
\ee
with $V$ being the thermodynamic volume and $N$ being identified with the number of degrees of freedom  associated with the black hole, which is proportional to its horizon area in Planck units, i.e., $N=\frac{1}{4}\frac{d-2}{d-1}\frac{A}{l_P^2}$. Although the two definitions agree for non-rotating black holes, they differ significantly when the rotation is taken into account. In particular, for the ultraspinning black holes (discussed in Sec.~\ref{sec:KerrAdS}) $v$ defined by \eqref{specific} remains finite whereas $\tilde v$ diverges to infinity since $N$ goes to zero in this limit. 
}
\be\label{specific}
v=\frac{4}{d-2}\Bigl(\frac{(d-1)V}{\omega_{d-2}}\Bigr)^{\frac{1}{d-1}}\,,
\ee
where in physical units one has to replace $v\to v l_P^{d-2}$ with $l_P$ being the Planck length.
Hence we consider the following {\em equation of state}:
\be\label{state}
P=P(v,T,J_i,Q)\,.
\ee

The equation of state allows one to employ  standard thermodynamic machinery and for instance calculate the critical exponents associated with the critical point.
For example, the `$P-v$' criticality of charged AdS black holes in four dimensions was studied in \cite{KubiznakMann:2012}. The system exhibits 
a first-order phase transition ala Van der Waals, with the coexistence line terminating at a critical point where the phase transition becomes  second-order \cite{MoLiu:2013}  and is characterized by the following (mean field theory) critical exponents:
\be\label{MFT}
\alpha=0\,,\quad \beta=\frac{1}{2}\,,\quad \gamma=1\,,\quad \delta=3\,.
\ee
 The same remains true for charged AdS black holes in higher dimensions \cite{GunasekaranEtal:2012, BelhajEtal:2012, Cai:2013qga}. In what follows we extend the analysis to singly spinning Kerr-AdS black holes in all dimensions. Since in the rotating case the equation of state \eqref{state} cannot be constructed analytically for any value of the rotation parameter, we investigate it numerically and develop
a systematic analytic expansion in the ultraspinning and slow rotation regimes. The {\em slow rotation expansion} allows us to approximately calculate the critical exponents and verify that they remain as predicted in \eqref{MFT} by  mean field theory.
The {\em ultraspinning expansion} approximates the properties of black membranes and relates to the 
effective blackfold description \cite{Emparan:2009at}
and the Gregory--Laflamme instability \cite{GregoryLaflamme:1993}.

The plan of our paper is as follows. In the next section we recapitulate the relevant thermodynamic results for four-dimensional black holes. Sec.~\ref{sec:KerrAdS} deals with the thermodynamics and various phase transitions of higher-dimensional Kerr-AdS black holes, whereas Sec.~\ref{sec:MP} studies their asymptotically flat cousins. Sec.~\ref{sec:5d} is devoted to analytically known 5-dimensional vacuum black holes with non-spherical horizon topology. In Sec.~\ref{sec:rings}  AdS black rings in all dimensions are studied in the thin (ultraspinning) approximation. The aforementioned thermodynamic phase transitions are put in context with various classical instabilities in Sec.~\ref{sec:Instab}. Sec.~\ref{sec:Discussion} is devoted to the final discussion.

\section{Black holes in 4d}\label{sec:4d}

To set  the stage for higher-dimensional results in this section we review some interesting features about the thermodynamics of four-dimensional black holes. We concentrate our attention on plotting the Gibbs free energy and discussing the associated phase diagrams.
For asymptotically flat black holes we have $P=0$, whereas from \eqref{PLambda}, $P=3/(8\pi l^2)$ for asymptotically AdS black holes.

\subsection{Asymptotically flat black holes}

\subsubsection{Schwarzschild solution}
To start with let us consider a spherically symmetric (four-dimensional) metric element 
\be\label{ss}
ds^2=-f(r)dt^2+\frac{dr^2}{f(r)}+r^2d\Omega_{(k)}^2\,,
\ee
where $d\Omega_{(k)}^2=d\theta^2+\frac{1}{k}\sin^2\!\bigl(\sqrt{k}\theta\bigr) d\varphi^2$ is a standard metric on a compact 2-surface, with $k=\{1,0,-1\}$ denoting spherical, planar, and hyperbolic geometries.   Setting
$k=1$, the horizon radius is the largest $r_+$ solving
\be
f(r_+)=0\,.
\ee
In our examples, limited to vacuum spacetimes with possibly a negative cosmological constant
and Maxwell field, one gets the following expressions for the black hole temperature, horizon entropy, and thermodynamic volume:
\be\label{SSgeneral}
T=\frac{f'(r_+)}{4\pi}\,,\quad
S=\frac{A}{4}=\pi r_+^2\,,\quad V=\frac{4}{3}\pi r_+^3\,.
\ee
Once the mass $M$ is determined the Gibbs free energy is given by \eqref{G}, $G=M-TS$.

\begin{figure}
\begin{center}
\includegraphics[width=0.4\textwidth,height=0.3\textheight]{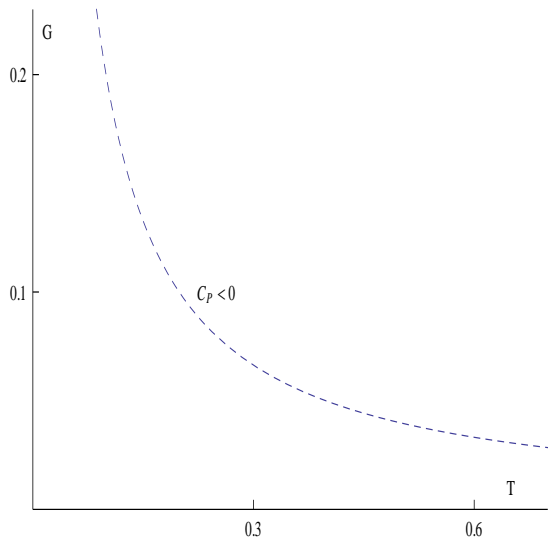}
\caption{{\bf Gibbs free energy: Schwarzschild black hole.}
The dashed blue line corresponds to a negative specific heat; 
for an asymptotically flat Schwarzschild black hole this quantity is negative for any temperature.
}
\label{Fig:Gschflat}
\end{center}
\end{figure}

In particular, for the {\em Schwarzschild black hole} one has
\be
f=1-\frac{2M}{r}\,,
\ee
and the corresponding thermodynamic quantities are 
\be
M=\frac{r_+}{2}\,,\quad T=\frac{1}{4\pi r_+}\,, \quad G=\frac{r_+}{4}=\frac{1}{16\pi T}\,,
\ee
and it is clear the Smarr relation \eqref{Smarr}, $M=2TS$, is satisfied.
It is not difficult to show that $C_P=-2 \pi r_+^2<0$, which corresponds to the well known fact that the Schwarzschild black hole is
locally thermodynamically unstable. This is related to the Gregory--Laflamme instability
of the corresponding Schwarzschild black string \cite{GregoryLaflamme:1993}.\footnote{
The thermodynamic instability of a black hole can be related to the corresponding classical instability, see, e.g.,  
\cite{GubserMitra:2000a, GubserMitra:2000b, Reall:2001, Figueras:2011he, Hollands:2012sf}. 
}
The Gibbs free energy, displayed in fig.~\ref{Fig:Gschflat}, has no interesting features and indicates no phase transitions. 

\subsubsection{Charged black hole: Reissner--Nordstr\"om solution}
\begin{figure}
\begin{center}
\includegraphics[width=0.4\textwidth,height=0.3\textheight]{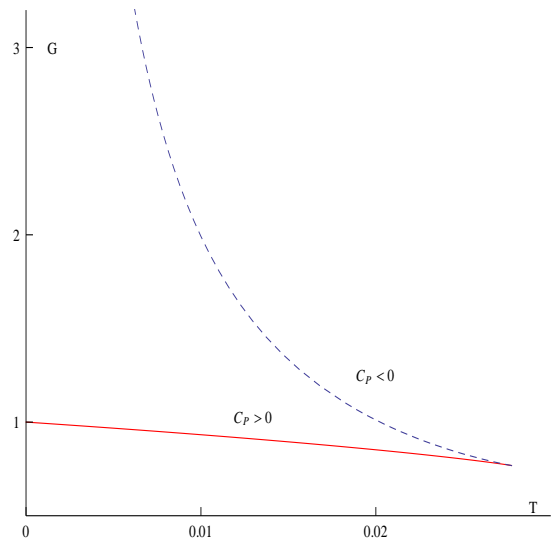}
\caption{{\bf Gibbs free energy: RN black hole.}
The Gibbs free energy of $Q=1$ RN black hole is displayed. The horizon radius $r_+$ increases from left to right and then up; 
$T=0$ corresponds to the extremal black hole with $r_+=M=Q=1$.
For a fixed temperature there are two branches of RN black holes. The lower thermodynamically preferred branch corresponds to small strongly charged nearly extremal black holes with positive $C_P$. The upper branch of weakly charged RN (almost Schwarzschild-like) black holes has higher Gibbs free energy and negative specific heat and hence is thermodynamically unstable. 
Its Euclidean action also possesses a negative zero mode. 
The situation for the Kerr-AdS black hole is qualitatively similar, with fixed $J$ replacing fixed $Q$.
}
\label{Fig:RNGQfixed}
\end{center}
\end{figure}

When the charge $Q$ is added to the Schwarzschild black hole, we obtain the Reissner--Nordstr\"om (RN) solution, for
which the metric is \eqref{ss} with
\be
f(r)=1-\frac{2M}{r}+\frac{Q^2}{r^2}\,.
\ee
The thermodynamic quantities are
\ba
T&=& \frac{r_+^2-Q^2}{4\pi r_+^3}\,,\quad M=\frac{r_+^2+Q^2}{2r_+}\,,\quad 
\Phi=\frac{Q}{r_+}\,,\nonumber\\
G&=&\frac{r_+^2+3Q^2}{4r_+}\,,\quad C_P=2\pi r_+^2\frac{r_+^2-Q^2}{3Q^2-r_+^2}\,,
\ea
and the Smarr relation \eqref{Smarr} is again satisfied.
The specific heat capacity is positive for small strongly charged black holes,
\be\label{Qrange}
\sqrt{3}|Q|> r_+>|Q|\,,
\ee 
or, equivalently, $\sqrt{3}M/2<|Q|<M$. This range corresponds to a thermodynamically stable branch of near extremal 
black holes which globally minimize the Gibbs free energy, see fig.~\ref{Fig:RNGQfixed}. 
Consequently and counter-intuitively in a fixed charge canonical ensemble
strongly charged small RN black holes are thermodynamically preferred to weakly charged (almost Schwarzschild-like)
large black holes. [Despite the fact that the entropy is larger on the upper branch in fig.~\ref{Fig:RNGQfixed}, since it increases quadratically with $r_+$, the  Gibbs free energy in the lower branch is the global minimum at fixed $Q$ due to the larger enthalpy contribution in the upper branch.] 
A directly related result  is that in the range \eqref{Qrange} there are no negative modes in the Euclidean action, whereas the negative mode appears for $r_+\geq \sqrt{3}|Q|$, indicating the onset of  Schwarzschild-like behaviour  \cite{MonteiroSantos:2009} .

\subsubsection{Rotating black hole: Kerr solution}
The rotating black hole metric described by the Kerr solution is written
\ba
ds^2&=&-dt^{2}+\frac{2Mr}{\Sigma}\left(dt-a\sin^{2}\!\theta d\phi\right)^{2}
 +\frac{\Sigma}{\Delta}dr^{2}\quad\nonumber\\
&+& \Sigma d\theta^{2}+\left(r^{2}+a^{2}\right)\sin^{2}\!\theta d\phi^{2}\,,
\ea
where 
\be
\Sigma=r^2+a^2 \cos^2\!\theta\,, \quad \Delta=r^2+a^2-2Mr\,.
\ee
The thermodynamic quantities are
\ba
T&=&\frac{1}{2\pi}\bigg[\frac{r_+}{a^2+r_+^2}-\frac{1}{2r_+}\bigg]\,,\quad 
S=\pi(a^2+r_+^2)=\frac{A}{4}\,,\nonumber\\
J&=&\frac{a}{2r_+}(a^2+r_+^2)\,,\quad \Omega=\frac{a}{r_+^2+a^2}\,,
\nonumber\\
G&=&\frac{3a^2+r_+^2}{4r_+}\,,\quad C_P=\frac{2\pi(r_+^2-a^2)(r_+^2+a^2)^2}{3a^4+6r_+^2a^2-r_+^4}\,,\nonumber\\
M&=&\frac{r_+^2+a^2}{2r_+}\,,\quad V=\frac{r_+A}{3}\Bigl(1+\frac{a^2}{2r_+^2}\Bigr)\,,
\ea
and satisfy \eqref{Smarr} .
The behaviour of the Gibbs free energy is qualitatively very similar to that of the charged black hole \cite{KubiznakMann:2012}, given by fig.~\ref{Fig:RNGQfixed}, with fixed $J$ replacing fixed $Q$.\footnote{In fact the thermodynamic behaviour remains qualitatively same even in charged rotating Kerr--Newman case.} 
As for the RN black hole we observe two branches of the Gibbs free energy. The lower one occurs for 
\be
\sqrt{3+2\sqrt{3}}|a|>r_+>|a|\,,
\ee
or, equivalently, $M>|a|>M\sqrt{2\sqrt{3}-3}$. It is thermodynamically preferred and corresponds to small fast rotating nearly extremal black holes with positive $C_P$. The upper branch of slowly rotating (almost Schwarzschild-like) black holes has higher Gibbs free energy and negative specific heat and is thermodynamically unstable. Its Euclidean action also possesses a negative zero mode  \cite{MonteiroEtal:2009}.  

We conclude that in the canonical ensemble of fixed $J$, the nearly extremal fast spinning black holes are thermodynamically preferred to the slowly rotating ones.

\subsection{AdS black holes}

\subsubsection{Schwarzschild-AdS}
The simplest (spherical) asymptotically AdS black hole is described by the Schwarzschild-AdS spacetime. The metric takes the form \eqref{ss} with
\be
f=1-\frac{2M}{r}+\frac{r^2}{l^2}\,.
\ee
The thermodynamic quantities are
\ba\label{SchaAdS}
T&=&\frac{1}{4\pi r_+ l^2}(l^2+3r_+^2)\,,\quad S=\pi r_+^2\,,\quad V=\frac{4}{3}\pi r_+^3\,,\nonumber\\
M&=&\frac{r_+}{2}\Bigl(1+\frac{r_+^2}{l^2}\Bigr)\,,\quad G=\frac{r_+}{4}\Bigl(1-\frac{r_+^2}{l^2}\Bigr)\,,\nonumber\\
C_P&=&2\pi r_+^2\frac{3r_+^2+l^2}{3r_+^2-l^2}\,,
\ea
 and satisfy \eqref{Smarr}, $M=2TS-2VP$, due to the presence of the $PV$ term.
\begin{figure}
\begin{center}
\rotatebox{-90}{
\includegraphics[width=0.4\textwidth,height=0.3\textheight]{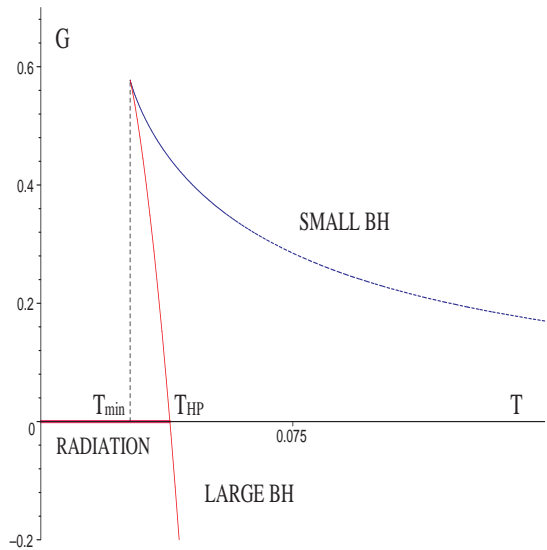}
}
\caption{{\bf Gibbs free energy: Schwarzchild-AdS black hole.}
When compared to the asymptotically flat Schwarzschild case (fig.~\ref{Fig:Gschflat}) for $P>0$ the Gibbs free energy acquires a new thermodynamically stable branch 
of large black holes. For $T>T_{\mbox{\tiny  HP}}$ this branch has negative Gibbs free energy and the corresponding black holes represent the 
globally thermodynamically preferred state. 
}
\label{Fig:Gschads}
\end{center}
\end{figure}

We observe a qualitatively different thermodynamic behaviour when compared to the asymptotically flat Schwarzschild case. Specifically $C_P$ is no longer always negative: it becomes positive for large (when compared to the AdS radius) black holes
\be
r_+>r_{\mbox{\tiny  min}}=\frac{l}{\sqrt{3}}\,,
\ee
while it is negative for $r_+<r_{\mbox{\tiny  min}}$ and ill defined at $r_+=r_{\mbox{\tiny  min}}$.

The behaviour of the Gibbs free energy $G$ is displayed in fig.~\ref{Fig:Gschads}.
We observe a minimum temperature $T_{\mbox{\tiny  \tiny min}}=2\sqrt{3}/(4\pi l)$, corresponding to $r_{\mbox{\tiny  min}}$, below which no black holes can exist. Above this temperature we have two branches of black holes. The upper one describes small (Schwarzschild-like) black holes with negative specific heat; these are thermodynamically unstable and  cannot be in a thermal equilibrium with a thermal bath of radiation. The large ($r_+>r_{\mbox{\tiny  \tiny min}}$) black holes at lower branch have positive specific heat and hence are locally thermodynamically stable. However, just above $T_{\mbox{\tiny  \tiny min}}$ the Gibbs free energy of such black holes is positive and the thermal AdS space with approximately zero Gibbs free energy represents a globally preferred thermodynamic state.\footnote{We refer to a thermal AdS space to be a (global) AdS space coupled to a bath of thermal radiation. Since the number of particle quanta representing the radiation is relatively small, the corresponding Gibbs free energy approximately equals zero.}     
This continues until temperature $T_{\mbox{\tiny  HP}}\approx 1/(\pi l)$ for which the black hole Gibbs free energy becomes negative, with the corresponding black hole radius given by 
\be
r_{\mbox{\tiny  HP}}=l\,.
\ee
Black holes with $r_+> r_{\mbox{\tiny  HP}}$ have negative Gibbs free energy and represent the globally preferred state. 
This means that at $T=T_{\mbox{\tiny  HP}}$ there is a first order {\em Hawking--Page} \cite{HawkingPage:1983} phase transition between thermal radiation and large black holes. This phase transition 
can be interpreted as a confinement/deconfinement phase transition in the dual quark
gluon plasma \cite{Witten:1998b}.

\begin{figure}
\begin{center}
\rotatebox{-90}{
\includegraphics[width=0.4\textwidth,height=0.3\textheight]{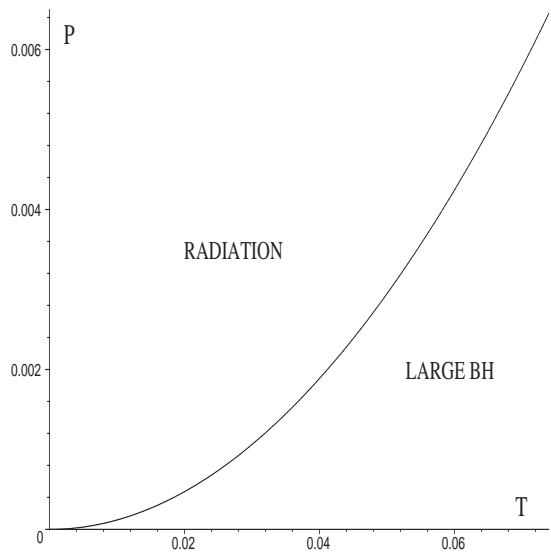}
}
\caption{{\bf Hawking--Page transition} is a first-order phase transition between thermal radiation in AdS and large stable Schwarzschild-AdS black hole. It occurs when $G$ of the Schwarzschild-AdS black hole approximately vanishes. Considering various pressures $P$ gives the radiation/large black hole coexistence line \eqref{HPcoexistence}  displayed in this figure. Similar to a ``solid/liquid'' phase transition, this line continues all the way to infinite pressure and temperature. 
}
\label{Fig:HPtrans}
\end{center}
\end{figure}

Considering an extended phase space, the coexistence line of thermal radiation/large black hole phases, determined from $G=0$,
reads 
\be\label{HPcoexistence}
P|_{\mbox{\tiny  coexistence}}=\frac{3\pi}{8} T^2\,.
\ee
The corresponding $P-T$ phase diagram is displayed in fig.~\ref{Fig:HPtrans}.    

\begin{figure}
\begin{center}
\rotatebox{-90}{
\includegraphics[width=0.4\textwidth,height=0.3\textheight]{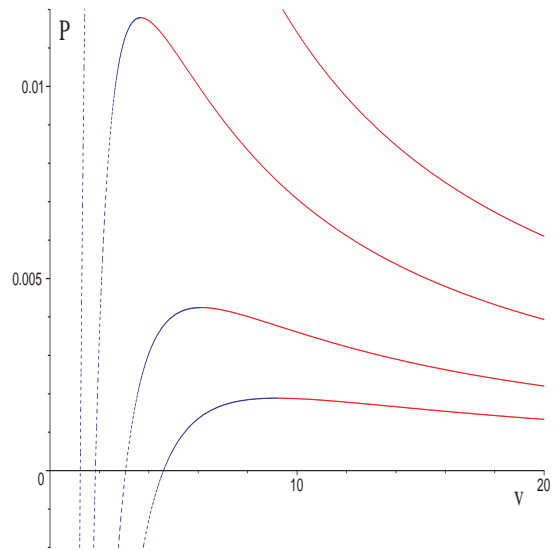}
}
\caption{{\bf Equation of state: Schwarzschild-AdS black hole}. The equation of state \eqref{HPstate} is displayed for various temperatures. 
For a given temperature the maximum occurs at $v=2r_0$. The dashed blue curves correspond to small unstable black holes. The red curves depict the stable large black hole branch; we observe   `ideal gas' behaviour for large temperatures. 
}
\label{Fig:PVSchwAdS}
\end{center}
\end{figure}
By rewriting the temperature equation \eqref{SchaAdS} while using \eqref{PLambda}, we get a corresponding `fluid equation of state' 
for the Schwarzschild-AdS black hole, given by 
\be\label{HPstate}
P=\frac{T}{v}-\frac{1}{2\pi v^2}\,,\quad v=2\Bigl(\frac{3V}{4\pi}\Bigr)^{1/3}=2r_+\,.
\ee     
 The behaviour of this equation is displayed in the $P-V$ diagram in fig.~\ref{Fig:PVSchwAdS}. For each isotherm there is a maximum which occurs for $v=1/(\pi T)$. For a given temperature this precisely corresponds to $r_+=r_{\mbox{\tiny  \tiny min}}$; 
the dashed blue curves with positive slope (and possibly negative pressures) to the left of the maximum correspond to small black holes with negative $C_P$ which are thermodynamically unstable, whereas solid red curves correspond to large black holes with positive $C_P$ that are locally thermodynamically stable.

\begin{figure}\label{fig:Fig1}
\begin{center}
\rotatebox{-90}{
\includegraphics[width=0.39\textwidth,height=0.32\textheight]{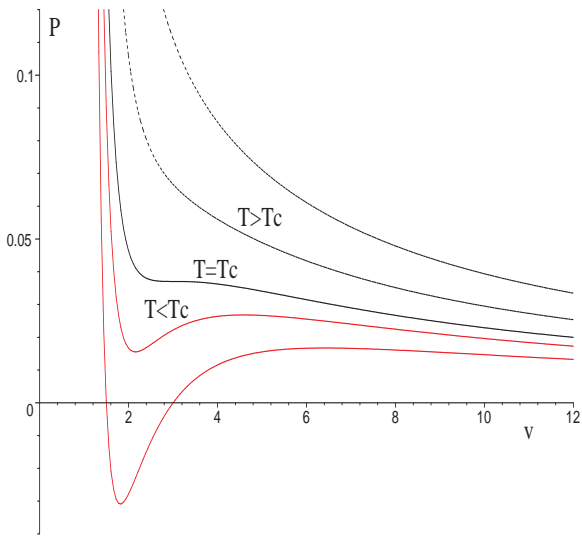}
}
\caption{{\bf $P-v$ diagram of Van der Waals fluid.} The temperature of isotherms decreases from top to bottom. The two upper dashed lines correspond to the ``ideal gas'' phase for $T>T_c$, the critical isotherm $T=T_c$ is denoted by the thick solid line, lower solid lines correspond to temperatures smaller than the critical temperature; for $T<T_c$ parts of the isotherms are actually unphysical, and must be replaced by a constant pressure line according to the Maxwell's equal area prescription \cite{KubiznakMann:2012}. Constants $a$ and $b$ were set equal to one.  
}  \label{fig:PVVdWstate}
\end{center}
\end{figure}

Let us compare the fluid equation of state of the Schwarzschild-AdS black hole \eqref{HPstate} with the famous {\em Van der Waals (VdW) equation}, e.g. \cite{Goldenfeld:1992},  which is a popular two parameter closed form modification of the ideal gas law that approximates the behaviour of real fluids. The VdW equation takes into account the nonzero size of fluid molecules 
(described by a constant $b>0)$ and the attraction between them (described by a constant $a>0$) and is often used to capture basic qualitative features of the liquid--gas phase transition. The equation reads (setting the Boltzmann constant $k_B=1$)
\be\label{VdW}
P=\frac{T}{v-b}-\frac{a}{v^2}\,,
\ee
where $v=V/N$ is the specific volume of the fluid, $P$ its pressure, $T$ its temperature, and the fluid parameters $a$ and $b$ are characteristics 
of a given fluid. The characteristic VdW $P-v$ diagram is displayed in fig.~\ref{fig:PVVdWstate}.
The equation admits a critical point, described by $\{T_c, v_c, P_c\}$, with universal critical ratio 
\be\label{universalVdWratio}
\rho_c=\frac{P_c v_c}{T_c}=\frac{3}{8}\,,
\ee
and the mean field theory critical exponents \eqref{MFT}.

We pause to consider topological Schwarzschild-AdS black holes.  Making use of the metric \eqref{ss} for arbitrary $k$, the 
formulae \eqref{SchaAdS} take the following more general form:
\ba
M &=& \frac{r_+ A_k}{8}\Bigl(k + \frac{r^2_+}{\ell^2}\Bigr)\,, \quad S=\frac{\pi A_k}{4} r^2_+\,, \nonumber\\
T &=& \frac{k \ell^2 + 3r^2_+}{4\pi\ell^3 r_+}\,,\quad V = \frac{\pi A_k}{3} r^3_+ \,,
\ea
where  $A_k$ is the area of the constant-curvature space divided by $\pi$; for a sphere, $A_{k=1} = 4$, 
for a torus, $A_{k=0} = A B$, where $A$ and $B$ and the sides of the torus, and there is no nice simple formula for $A_{k=-1}$.
In all cases the Smarr formula \eqref{Smarr} and first law \eqref{1st} hold.
It is then straightforward to show that the fluid corresponding to the  Schwarzschild-AdS black hole is characterized by
\be
a=\frac{k}{2\pi}\,,\quad b=0\,.
\ee
That is, its equation of state differs from the ideal gas law by the presence of a nontrivial parameter $a=k/(2\pi)$. This is directly related to the  topology of the horizon. For {\em planar} Schwarzschild-AdS black holes, $k=0$ and we recover the ideal gas law characterized by $a=0=b$, 
\be
Pv=T\,,
\ee
whereas for the $k=-1$ hyperbolic case we get a peculiar `repulsion' feature $a=-\frac{1}{2\pi}<0$, with $b=0$, the volume of molecules still vanishing.

We note one more interesting fact  for the planar ($k=0)$ AdS black holes.   In this case the following specific Smarr-like relation has been employed  e.g., \cite{Bertoldi:2010ca,Berglund:2011cp}:
\be\label{differentSmarr}
3 M=  2 TS\,,
\ee
[or $(d-1) M=(d-2) TS$ in  $d$-dimensions] without any reference to the $PV$ term. 
The value of our  Smarr relation \eqref{Smarr} is that it follows from the general geometric argument and 
applies to a wide class of (even possibly unknown) solutions that satisfy the assumptions of the theorem, whereas \eqref{differentSmarr} is a `phenomenological' observation valid for a particular sub-class of solutions. We also note that this relation is not related to the corresponding 
first law, $dM=TdS$, by a dimensional scaling argument.

\subsubsection{Charged AdS black hole}
The charged AdS black hole metric is described by \eqref{ss} with
\be
f=1-\frac{2M}{r}+\frac{Q^2}{r^2}+\frac{r^2}{l^2}\,.
\ee
The thermodynamic quantities are 
\ba
T&=&\frac{1}{4\pi r_+^3 l^2}\Bigl(l^2(r_+^2-Q^2)+3r_+^4\Bigr)\,,\quad S=\pi r_+^2\,,\nonumber\\
V&=&\frac{4}{3}\pi r_+^3\,,\quad \Phi=\frac{Q}{r_+}\,,\quad 
G=\frac{l^2r_+^2-r_+^4+3Q^2l^2}{4l^2r_+}\,,\nonumber\\
C_P&=&2\pi r_+^2\frac{3r_+^4+l^2r_+^2-Q^2l^2}{3r_+^4-l^2r_+^2+3Q^2l^2}\,.
\ea
The specific heat is negative for 
\be
\frac{l \sqrt{1-\sqrt{1-36Q^2/l^2}}}{\sqrt{6}}<r_+<\frac{l \sqrt{1+\sqrt{1-36Q^2/l^2}}}{\sqrt{6}}\,,
\ee
and positive otherwise.

\begin{figure}
\begin{center}
\rotatebox{-90}{
\includegraphics[width=0.39\textwidth,height=0.34\textheight]{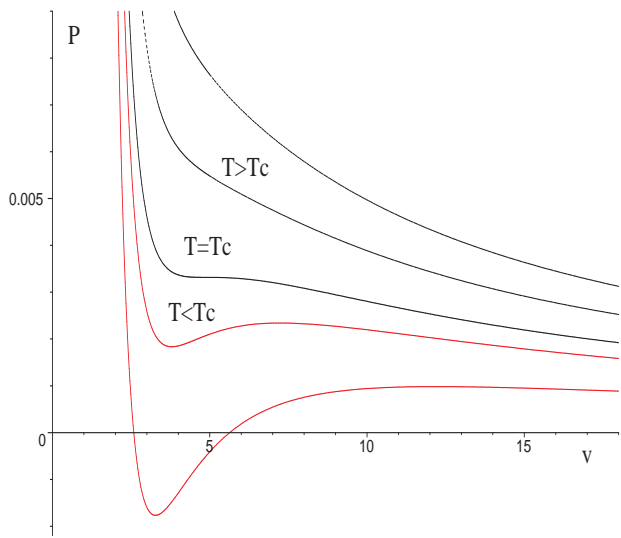}
}
\caption{{\bf Equation of state: charged AdS black hole.}
The temperature of isotherms decreases from top to bottom. The two upper dashed lines correspond to the ``ideal gas'' one-phase behaviour for $T>T_c$, the critical isotherm $T=T_c$ is denoted by the thick solid line, lower solid lines correspond to temperatures smaller than the critical temperature.  We have set $Q=1$.  
$P-v$ diagram for the Kerr-AdS black hole with $J=1$ is qualitatively similar.
}\label{Fig:RNstate}
\end{center}
\end{figure}
\begin{figure}
\begin{center}
\includegraphics[width=0.4\textwidth,height=0.3\textheight]{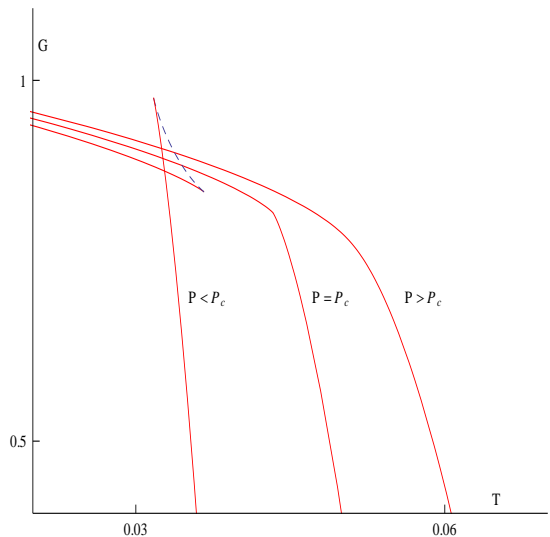}
\caption{{\bf Gibbs free energy: charged AdS black hole.}
Characteristic swallowtail behaviour is observed for $P<P_c$, corresponding to a small/large black hole phase transition.
An unstable branch of the Gibbs free energy is displayed in dashed blue line. We have set $Q=1$.
The behaviour of $G$ for Kerr-AdS black hole with $J=1$ is qualitatively similar.
}
\label{Fig:Grnads}
\end{center}
\end{figure}

\begin{figure}
\begin{center}
\rotatebox{-90}{
\includegraphics[width=0.39\textwidth,height=0.34\textheight]{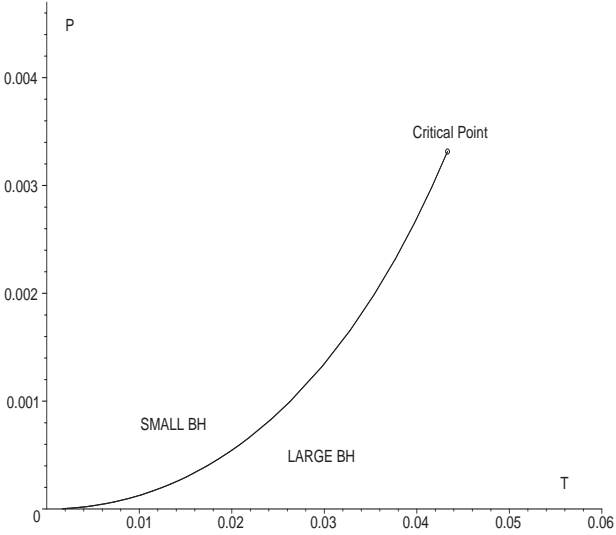}
}
\caption{{\bf Phase diagram: charged AdS black hole.} The coexistence line of the SBH/LBH phase transition of the charged AdS black hole system in $(P, T)$-plane is displayed. The critical point is highlighted by a small circle at the end of the coexistence line. The
phase diagram for a Kerr-AdS black hole with $J=1$ is qualitatively similar.
} \label{Fig:RNPT} 
\end{center}
\end{figure}

 
It was first noticed in \cite{ChamblinEtal:1999a,ChamblinEtal:1999b} that in a canonical (fixed charge) ensemble, charged AdS black holes allow for a first order {\em small-black-hole/large-black-hole} phase (SBH/LBH) transition which is in many ways reminiscent of the liquid/gas transition of a Van der Waals fluid.
This analogy becomes more complete in extended phase space \cite{KubiznakMann:2012}. 
Namely, the equation of state,
\be\label{RNstate}
P=\frac{T}{v}-\frac{1}{2\pi v^2}+\frac{2Q^2}{\pi v^4}\,,\quad v=2r_+\,,
\ee
mimics qualitatively the behaviour of the Van der Waals equation, shown in fig. \ref{fig:PVVdWstate}, with its black hole counterpart \eqref{RNstate} illustrated in fig. \ref{Fig:RNstate}.
Below $P_c$, the Gibbs free energy displays a characteristic swallowtail behaviour, depicted in fig.~\ref{Fig:Grnads}, indicating a first-order SBH/LBH phase transition. The corresponding coexistence line is displayed in fig.~\ref{Fig:RNPT}. 
It terminates at a critical critical point, characterized by 
\be
T_c=\frac{\sqrt{6}}{18\pi Q}\,,\quad v_c=2\sqrt{6} Q\,,\quad P_c=\frac{1}{96\pi Q^2}\,, 
\ee
where the phase transition becomes of the second order \cite{Banerjee:2010bx, MoLiu:2013}, and is characterized by the mean field theory critical exponents \eqref{MFT}. Remarkably the relation $\rho_c=P_c v_c/T_c=3/8$ is identical to  the Van der Waals case. The overall situation reminds one of the liquid/gas phase transition.  
A similar situation occurs for charged AdS black holes in higher dimensions \cite{GunasekaranEtal:2012}. 

We remark that the Hawking--Page phase transition as well as the SBH/LBH phase transition ala Van der Waals can be also observed for asymptotically flat or de Sitter charged black holes provided these are placed in a finite cavity \cite{Carlip:2003ne}.

We close this subsection by noting that strongly charged small AdS black holes are subject to superradiant instabilities. For a charged scalar field coupled to Einstein--Maxwell theory a similar situation occurs:  
at small temperatures  charged scalar hair forms and the resulting `hairy black holes' are thermodynamically preferred  \cite{Gubser:2008px, Hartnoll:2008vx, Hartnoll:2008kx, Maeda:2010hf, Basu:2010uz, Dias:2011tj, Horowitz:2010gk, Hartmann:2013nla}. This means that in the presence of a charged scalar field, the left-most branch of small black holes in the Gibbs free diagram \ref{Fig:Grnads}, although locally thermodynamically stable, does not globally minimize the Gibbs free energy. Rather, there is another branch of hairy black holes with 
lower Gibbs free energy. The hair/no-hair black hole phase transition is of the second order and underlies the theory of holographic superconductors.

\subsubsection{Kerr-AdS}

The thermodynamics of four-dimensional rotating AdS black holes is qualitatively similar to the charged AdS case.\footnote{In fact a qualitatively similar behaviour occurs for charged rotating AdS black holes, see \cite{CaldarelliEtal:2000}.} 
The metric is  
\begin{eqnarray}
ds^2&=&-\frac{\Delta}{\rho^2}(dt-\frac{a}{\Xi}\sin^2\!\theta d\varphi)^2+
\frac{\rho^2}{\Delta}dr^2+\frac{\rho^2}{\Sigma}d\theta^2 \nonumber \\
&+&
\frac{\Sigma \sin^2\!\theta}{\rho^2}[adt-\frac{(r^2+a^2)}{\Xi}d\varphi]^2\,,
\end{eqnarray}
where
\begin{eqnarray}
\Delta&=&(r^2+a^2)(1+\frac{r^2}{l^2})-2mr\,,\quad
\Sigma=1-\frac{a^2}{l^2}\cos^2\theta \,,\nonumber \\
\Xi&=&1-\frac{a^2}{l^2}\,, \quad
\rho^2=r^2+a^2\cos^2\theta \,. \nonumber
\end{eqnarray}
The thermodynamic quantities read 
\begin{eqnarray}
M&=&\frac{m}{\Xi^2}\,, \quad J=\frac{ma}{\Xi^2} \,, \quad  \Omega=\frac{a}{l^2}\frac{r_+^2+l^2}{r_+^2+a^2}\,, \nonumber\label{OHJ}\\
T&=&\frac{1}{2\pi r_+}\Bigr[\frac{(a^2+3r_+^2)(r_+^2/l^2+1)}{2(a^2+r_+^2)}-1\Bigr]\,, \nonumber\label{BHT} \\
S &=&\pi \frac{(a^2+r_+^2)}{\Xi}=\frac{A}{4} \,,\quad V=\frac{r_+A}{3}\Bigl(1+\frac{1+r_+^2/l^2}{2r_+^2}\frac{a^2}{\Xi}\Bigr)\,, \nonumber\label{BHS}\\
G &=& \frac{(r^2_+ + 3a^2) l^4 -(r^2_+ - a^2)^2 l^2 + (a^2 + 3 r^2_+)a^2 r^2_+}{l^4 \Xi^2 r_+}\,, \nonumber \\
\end{eqnarray}
and satisfy the Smarr relation \eqref{Smarr}; 
rather lengthy formula for $C_P$ can be found in \cite{MonteiroEtal:2009}.

The Gibbs free energy, equation of state, and the $P-T$ phase diagram are qualitatively similar to figs.~\ref{Fig:Grnads}, \ref{Fig:RNstate} and 
\ref{Fig:RNPT}---with fixed $J$ replacing fixed $Q$.  For any fixed $J$, there is a critical point, characterized by $(P_c, V_c, T_c)$, that can be determined numerically. For $P<P_c$, the Gibbs free energy displays  swallowtail behaviour and there is a corresponding SBH/LBH first order phase transition with a phase diagram similar to fig.~\ref{Fig:RNPT}. It will be shown in the next section that in the limit of slow rotation,  $a\ll l$, the equation of state can be approximated by 
\ba
P&=&\frac{T}{v}-\frac{1}{2\pi v^2}+\frac{48J^2}{\pi v^6}-\frac{384 J^4(7+8\pi Tv)}{(1+\pi Tv)^2\pi v^{10}}\nonumber\\
&&+\frac{36864(13\pi Tv\!+\!11)J^6}{\pi v^{14}(1+\pi Tv)^3}\!+\!O\bigl[(a/l)^8\bigr]\,,\nonumber \\
v&=&2\Bigl(\frac{3V}{4\pi}\Bigr)^{\!1/3}\,.
\ea
The first three terms were first obtained in \cite{GunasekaranEtal:2012}. Using this equation of state, one can approximately find 
the critical point and study its characteristics. In particular, one finds that the critical exponents 
remain as predicted by the mean field theory, given by \eqref{MFT}. 

We pause to remark that, similar to the charged AdS case, the rotating AdS black holes close to extremity are unstable with respect to   superradiant instabilities;  the resultant objects (hairy black hole, soliton, or boson star) are expected to globally minimize the Gibbs free energy \cite{Hawking:1999dp, Sonner:2009fk, Dias:2010ma, Cardoso:2013pza}.

\section{Higher-dimensional Kerr-AdS black hole spacetimes}\label{sec:KerrAdS}

\subsection{General metrics}\label{sec:GeneralMetrics}
General Kerr-AdS black hole spacetimes \cite{GibbonsEtal:2004, GibbonsEtal:2005}  are $d$-dimensional metrics that solve the Einstein equations with  cosmological constant
\be
R_{ab} =\frac{2\Lambda}{(d-2)}  g_{ab}
\ee
and generalize the $d$-dimensional asymptotically-flat rotating black hole spacetimes of Myers and Perry \cite{MyersPerry:1986}.   
In   Boyer--Lindquist  coordinates the metric takes the form
\ba \label{metric}
ds^2&=&-W\Bigl(1+\frac{r^2}{l^2}\Bigr)d\tau ^2+\frac{2m}{U} \Bigl(W d\tau -\sum_{i=1}^{N} \frac{a_i \mu_i ^2 d\varphi _i}{\Xi _i}\Bigr)^2\nonumber\\
&+&\sum_{i=1}^{N} \frac{r^2+a_i^2}{\Xi _i} \mu_i ^2 d\varphi _i^2+\frac{U dr^2}{F-2m}+\sum_{i=1}^{N+\varepsilon}\frac{r^2+a_i ^2}{\Xi _i} d\mu _i ^2 \nonumber\\
&-&\frac{l^{-2}}{W (1+r^2/l^{2})}\Bigl(\sum_{i=1}^{N+\varepsilon}\frac{r^2+a_i ^2}{\Xi _i} \mu_i d\mu_i\Bigr)^2\,,
\ea
where 
\ba\label{metrcifunctions}
W&=&\sum_{i=1}^{N+\varepsilon}\frac{\mu _i^2}{\Xi _i}\,,\quad U=r^\varepsilon \sum_{i=1}^{N+\varepsilon} \frac{\mu _i^2}{r^2+a_i^2} \prod _j ^N (r^2+a_j^2)\,,\nonumber\\
F&=&r^ {\varepsilon -2} \Bigl(1+\frac{r^2}{l^2}\Bigr) \prod_{i=1}^N (r^2+a_i^2)\,,\quad \Xi_i=1-\frac{a_i^2}{l^2}\,.\quad
\ea
To treat even ($\varepsilon=1)$  odd ($\varepsilon=0)$ spacetime dimensionality $d$ simultaneously, we have parametrized
\be
d=2N + 1 + \varepsilon\,,
\ee 
and in even dimensions set for convenience $a_{N+1}=0$.
The  coordinates $\mu_i$ are not independent, but obey the constraint
\begin{equation}\label{constraint}
\sum_{i=1}^{N+\varepsilon}\mu_i^2=1\,.
\end{equation}
In general the spacetime admits $N$ independent angular momenta $J_i$, described by
$N$ rotation parameters $a_i$, and generalizes the previously known singly-spinning case \cite{HawkingEtal:1999}.
In $d=4$ it reduces to the four-dimensional Kerr-AdS metric studied in the previous section.  
With this metric it shares a remarkable property---it possesses a hidden symmetry associated with the Killing--Yano tensor \cite{KubiznakFrolov:2007} that is responsible for integrability of geodesic motion and various test field equations in these spacetimes, see, e.g., review \cite{FrolovKubiznak:2008}.

The thermodynamic quantities associated with Kerr-AdS black holes were first calculated in \cite{Gibbons:2004ai}.
The mass $M$, the angular momenta $J_i$, and the angular velocities of the horizon $\Omega_i$ read
\ba \label{TD}
M&=&\frac{m \omega _{d-2}}{4\pi (\prod_j \Xi_j)}\bigl(\sum_{i=1}^{N}{\frac{1}{\Xi_i}-\frac{1-\varepsilon }{2}}\bigr)\,,\nonumber\\
J_i&=&\frac{a_i m \omega _{d-2}}{4\pi \Xi_i (\prod_j \Xi_j)}\,,\quad \Omega_i=\frac{a_i (1+\frac{r_+^2}{l^2})}{r_+^2+a_i^2}\,,
\ea
while the temperature $T$, the horizon area $A$, and the entropy $S$ are given by
\ba\label{TS}
T&=&\frac{1}{2\pi }\Bigr[r_+\Bigl(\frac{r_+^2}{l^2}+1\Bigr)
\sum_{i=1}^{N} \frac{1}{a_i^2+r_+^2}-\frac{1}{r_+}
\Bigl(\frac{1}{2}-\frac{r_+^2}{2l^2}\Bigr)^{\!\varepsilon}\,\Bigr]\,,\nonumber\\
A&=&\frac{\omega _{d-2}}{r_+^{1-\varepsilon}}\prod_{i=1}^N 
\frac{a_i^2+r_+^2}{\Xi_i}\,,\quad S=\frac{A}{4}\,. 
\ea
The horizon radius $r_+$ is determined as the largest root of $F-2m=0$ and $\omega_{d}$ is given by \eqref{omega}.

The thermodynamic volume reads \cite{CveticEtal:2010, Dolan:2013}
\begin{eqnarray} \label{VBHKerr}
V& =& \frac{r_+ A }{d-1}\left[1+\frac{1+{r^2_+}/{l^2}}{(d-2)r_+^2}\sum_i \frac{a_i^2}{\Xi_i}\right]  \nonumber\\ \label{VBHKerr2}
&=&	{\frac{r_+ A }{d-1}+{8\pi\over (d-1)(d-2)}\sum_i a_iJ_i }\,, 
\end{eqnarray}
and indeed is required to ensure that the Smarr formula \eqref{Smarr} holds.
It is known to obey the {\em reverse isoperimetric inequality} \eqref{ISO} provided that in \eqref{ratio} we identify ${\cal A}=A$ and ${\cal V}=V$
as given by \eqref{TS} and \eqref{VBHKerr}.
In fact the inequality is saturated for non-rotating black holes, whereas it becomes most extreme $({\cal R}\to \infty)$
in the ultraspinning limit, see discussion around Eq. \eqref{ratioInfinity}.

Note that the `naive' {\em geometric volume} \cite{Parikh:2006, BallikLake:2010, CveticEtal:2010,BallikLake:2013} (given by the spatial integral of $\sqrt{-g}$ integrated up to the horizon radius) 
\be
V'=\frac{r_+ A}{d-1}\,
\ee
and the thermodynamic volume $V$ in \eqref{VBHKerr} differ by an $\sum_i a_iJ_i$ term and behave differently in the limits of slow and fast rotation. 
Whereas the thermodynamic volume \eqref{VBHKerr} is dominated by its first geometric term for slow rotations 
\be\label{Vslow}
V_{\mbox{\tiny  slow}}\approx V'_{\mbox{\tiny  slow}}\approx\frac{\omega_{d-2}r_+^{d-1}}{d-1}\,,
\ee
in the ultraspinning regime the second (angular momentum) term dominates.
To see this let us, for simplicity, consider for a moment 
a singly spinning Myers--Perry black hole ($a_1=a$, other $a_i=0$, and $l\to \infty$). 
In this case the formula \eqref{VBHKerr} reduces to  the following expression 
\be\label{VKerr}
V=V'+\frac{4\pi}{d-1}\frac{J^2}{M}
\ee
which is also valid for the $d=4$ Kerr-AdS case.
In $d\geq 6$ there is no limit on how large the rotation parameter $a$ can be \cite{EmparanMyers:2003} and one can, in principle, take the {\em ultraspinning limit} $a\to \infty$. For fixed $M$ this implies $r_+\to 0$, and the second term in the previous formula dominates.
Hence, for the ultraspinning black holes, the thermodynamic volume differs significantly from the geometric one, being approximately given by
\be\label{VeqV'}
V\approx \frac{4\pi}{d-1}\frac{J^2}{M}\approx \frac{\omega_{d-2} a^4 r_+^{d-5}}{(d-1)(d-2)}\,,
\ee
which is to be compared with formula \eqref{Vslow}. The conclusions for multiply-spinning and/or AdS black holes are   analogous.

The Gibbs free energy is given by \eqref{G} and is related to the Euclidean action $I$ \cite{Gibbons:2004ai},  
\be\label{action}
I=\frac{\omega_{d-2}}{8\pi T\bigl(\prod_j \Xi_j\bigr)}\Bigl(m-\frac{r_+^\varepsilon}{l^2}\prod_{i=1}^N(r_+^2+a_i^2)\Bigr)\,,
\ee
by 
\be\label{GibbsKerrAdS}
G=M-TS=TI+\sum_{i}\Omega_i J_i\,,
\ee
see also Sec.~\ref{sec:Instab} where $I$ is used to estimate the onset of ultraspinning instabilities.  
In what follows we shall discuss in detail some special cases of these general metrics and the associated interesting thermodynamic features.

\subsection{Classical swallowtail}
The behaviour of the Gibbs free energy $G$, \eqref{GibbsKerrAdS}, depends crucially on the number of spacetime dimensions $d$ and the number and ratios 
of the nontrivial angular momenta. 

\begin{figure}
\begin{center}
\includegraphics[width=0.4\textwidth,height=0.28\textheight]{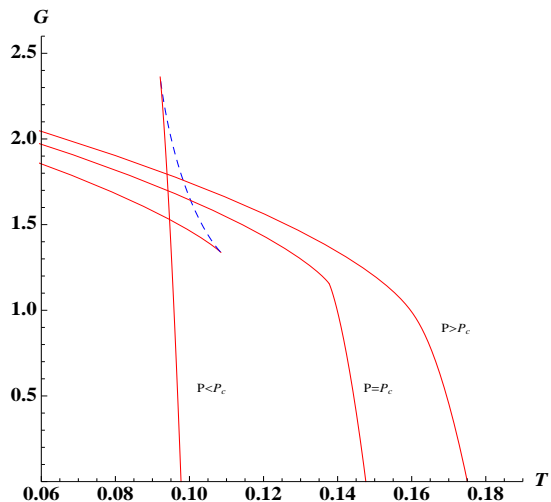}
\caption{{\bf Gibbs free energy: Kerr-AdS in $d=5$.}
The Gibbs free energy of equal spinning Kerr-AdS black hole is displayed for $J_1=J_2=1$. The situation is very similar to what happens 
for charged or rotating black holes in four dimensions.
A qualitatively similar swallowtail behaviour, indicating a
SBH/LBH phase transition, is observed for any ratio of the two angular momenta, as long as, at least one of them
is non-trivial. Equal spinning Ker-AdS black holes in all $d\geq 5$ demonstrate the qualitatively same feature.   
}  
\label{Fig:5DKerr}
\end{center}
\end{figure}

The dimension $d=5$  is an exception. 
The Gibbs free energy is in principle a function of two possible angular momenta $J_1$ and $J_2$. 
However, as long as at least one of them is nontrivial,  there exists a critical pressure $P_c$ below which we observe the qualitatively same characteristic swallowtail behaviour indicating the small/large black hole phase transition, as shown in fig.~\ref{Fig:5DKerr}. 
The situation is in fact very similar to what happens for Kerr-AdS in $d=4$. The corresponding $P-V$ and $P-T$ diagrams (not shown) are reminiscent of what was observed for charged or rotating black holes in $d=4$ and are analogous to the Van der Waals $P-V$ and $P-T$ diagrams.

The classical swallowtail behaviour is also observed for {\em equal-spinning} ($J_1=J_2=......=J_N=J$) Kerr-AdS 
black holes in all $d\geq 5$ dimensions.

\subsection{Reentrant phase transition}
\begin{figure}
\begin{center}
\includegraphics[width=0.39\textwidth,height=0.3\textheight]{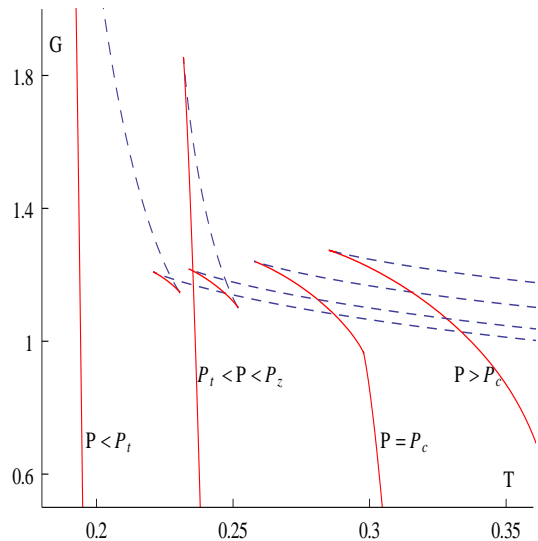}
\caption{ {\bf Gibbs free energy: singly spinning Kerr-AdS in $d=6$.}
The behaviour of $G$ for $d\geq 6$ is completely different from that of $d<6$, cf., figs.~\ref{Fig:Grnads} and \ref{Fig:5DKerr}.  
The Gibbs free energy for a singly spinning black hole is displayed for increasing pressure
(from left to right). As with Schwarzschild-AdS black holes, for $P\geq P_c$, the (lower) large black hole branch is thermodynamically stable  whereas the upper branch is unstable.  For $P=P_c$ we observe critical behaviour. 
In a range  of pressures $P\in(P_t,P_z)$, we observe a discontinuity in the global minimum of $G$ or zeroth-order phase transition, signifying the presence of a reentrant phase transition. 
This feature is qualitatively similar and present for all $d\geq6$ singly spinning Kerr-AdS black holes. For $P<P_t$ only one branch of stable large black holes exists.  
Note  that the specific heat is positive in zones where the Gibbs free energy reaches a global minimum. 
}
\label{Fig:6DKerr}
\end{center}
\end{figure}
It was observed in \cite{AltamiranoEtal:2013a} that in all $d\geq 6$ dimensions the {\em singly spinning} Kerr-AdS black holes 
admit a reentrant large/small/large black hole phase transition. This is quite exciting as similar reentrant phase transitions are 
commonly observed for multicomponent fluid systems, gels, ferroelectrics, liquid crystals, and binary gases, e.g., \cite{NarayananKumar:1994},
and the observed feature could provide a toy gravitational model for such `every day' thermodynamic phenomena. 

In general, a system undergoes a reentrant phase transition if a monotonic variation of one thermodynamic quantity results in two
(or more) phase transitions such that the final state is macroscopically similar to the initial state. 
For singly spinning Kerr-AdS black holes in a certain range of pressures (and a given angular momentum) it was found \cite{AltamiranoEtal:2013a} that a monotonic lowering of the black hole temperature yields a large/small/large black hole transition. The situation is accompanied by a zeroth-order phase transition, a discontinuity in the global minimum of the Gibbs free energy, a phenomenon seen
in superfluidity and superconductivity \cite{Maslov:2004}, and recently in four-dimensional Born--Infeld black holes 
\cite{GunasekaranEtal:2012}\footnote{Interestingly, such reentrant phase transitions 
 were recently found to be absent in higher-dimensional Born--Infeld-AdS black holes, giving the four-dimensional result a special significance \cite{Zou:2013owa}.}
and black holes of third-order Lovelock gravity \cite{FrassinoEtal:2013}.
In this subsection we provide more detail about this interesting transition. We start with a review of singly spinning Kerr-AdS black holes.

For singly spinning ($a_1=a$ and other $a_i=0$) Kerr-AdS black holes the metric \eqref{metric} significantly simplifies and reduces to 
\ba\label{Singlyspinning}
ds^2&=&-\frac{\Delta}{\rho^2}(dt-\frac{a}{\Xi}\sin^2\!\theta d\varphi)^2+
\frac{\rho^2}{\Delta}dr^2+\frac{\rho^2}{\Sigma}d\theta^2\quad \nonumber\\
&+&
\frac{\Sigma \sin^2\!\theta}{\rho^2}[adt-\frac{r^2\!+\!a^2}{\Xi}d\varphi]^2\!+\!r^2\cos^2\!\theta d\Omega_{d-4}^2\,,\qquad\ \ 
\ea
where
\ba
\Delta&=&(r^2+a^2)(1+\frac{r^2}{l^2})-2mr^{5-d}\,,\quad \Sigma=1-\frac{a^2}{l^2}\cos^2\!\theta\,,\nonumber\\
\Xi&=&1-\frac{a^2}{l^2}\,,\quad \rho^2=r^2+a^2\cos^2\!\theta\,,
\ea
and $d\Omega_{d}^2$ denotes the metric element on a $d$-dimensional sphere.
The associated thermodynamic quantities  become  
\begin{eqnarray}\label{singlespin}
M&=&\frac{\omega_{d-2}}{4\pi}\frac{m}{\Xi^2}\Bigl(1+\frac{(d-4)\Xi}{2}\Bigr)\,, \quad
J=\frac{\omega_{d-2}}{4 \pi}\frac{ma}{\Xi^2}\,,\nonumber\\
{\Omega}&=&\frac{a}{l^2}\frac{r_+^2+l^2}{r_+^2+a^2}\,,\quad S=\frac{\omega_{d-2}}{4}\frac{(a^2+r_+^2) r_+^{d-4}}{\Xi}=\frac{A}{4}\,,\nonumber\\
T&=&\frac{1}{2\pi}\Bigr[r_+\Bigl(\frac{r_+^2}{l^2}+1\Bigr) \left(\frac{1}{a^2+r_+^2}+
\frac{d-3}{2 r_+^2}\right)-\frac{1}{r_+}\Bigr]\,,\nonumber\\
G&=&\frac{\omega_{d-2}r_+^{d-5}}{16\pi \Xi^2}\Bigl(3a^2\!+\!r_+^2-\frac{(r_+^2\!-\!a^2)^2}{l^2}\!+\!\frac{3a^2r_+^4\!+\!a^4r_+^2}{l^4}\Bigr)\,,\nonumber\\
V&=&\frac{r_+A}{d-1}\Bigl[1+\frac{a^2}{\Xi}\frac{1+r_+^2/l^2}{(d-2)r_+^2}\Bigr]\,,
\end{eqnarray}
where $r_+$ is the largest positive real root of $\Delta=0$.

To analyze the behaviour of $G=G(P,T,J)$ for fixed $J$ and $P$ we proceed as follows. 
We first express $a=a(J,r_+, P)$ from the (quartic) $J$ equation \eqref{singlespin}. Inserting the (well-behaved) solution into the expressions for $G$ and $T$, consequently expressed  as functions of $P$ and $r_+$, we can then plot $G$ vs. $T$ parametrically.
Similarly, we  obtain an expression for the specific heat $C_P(r_+, P, J)$ and plot this function for fixed $P$ and $J$ (not displayed) to see for which regions of $r_+$ it becomes negative, thereby finding which branches of $G$ are thermodynamically unstable.

\begin{figure*}
\centering
\begin{tabular}{cc}
\rotatebox{-90}{
\includegraphics[width=0.4\textwidth,height=0.33\textheight]{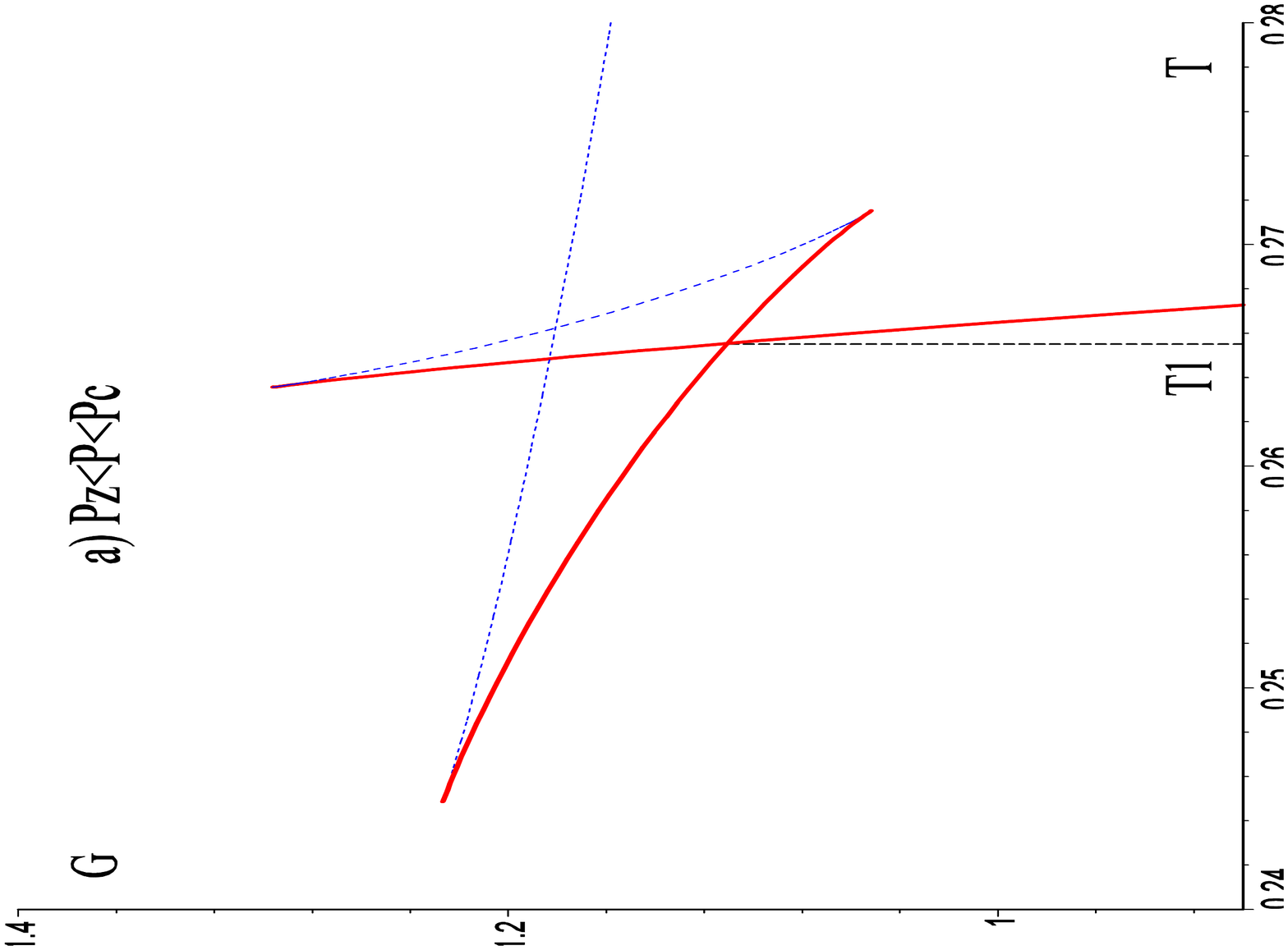}}&
\rotatebox{-90}{
\includegraphics[width=0.4\textwidth,height=0.33\textheight]{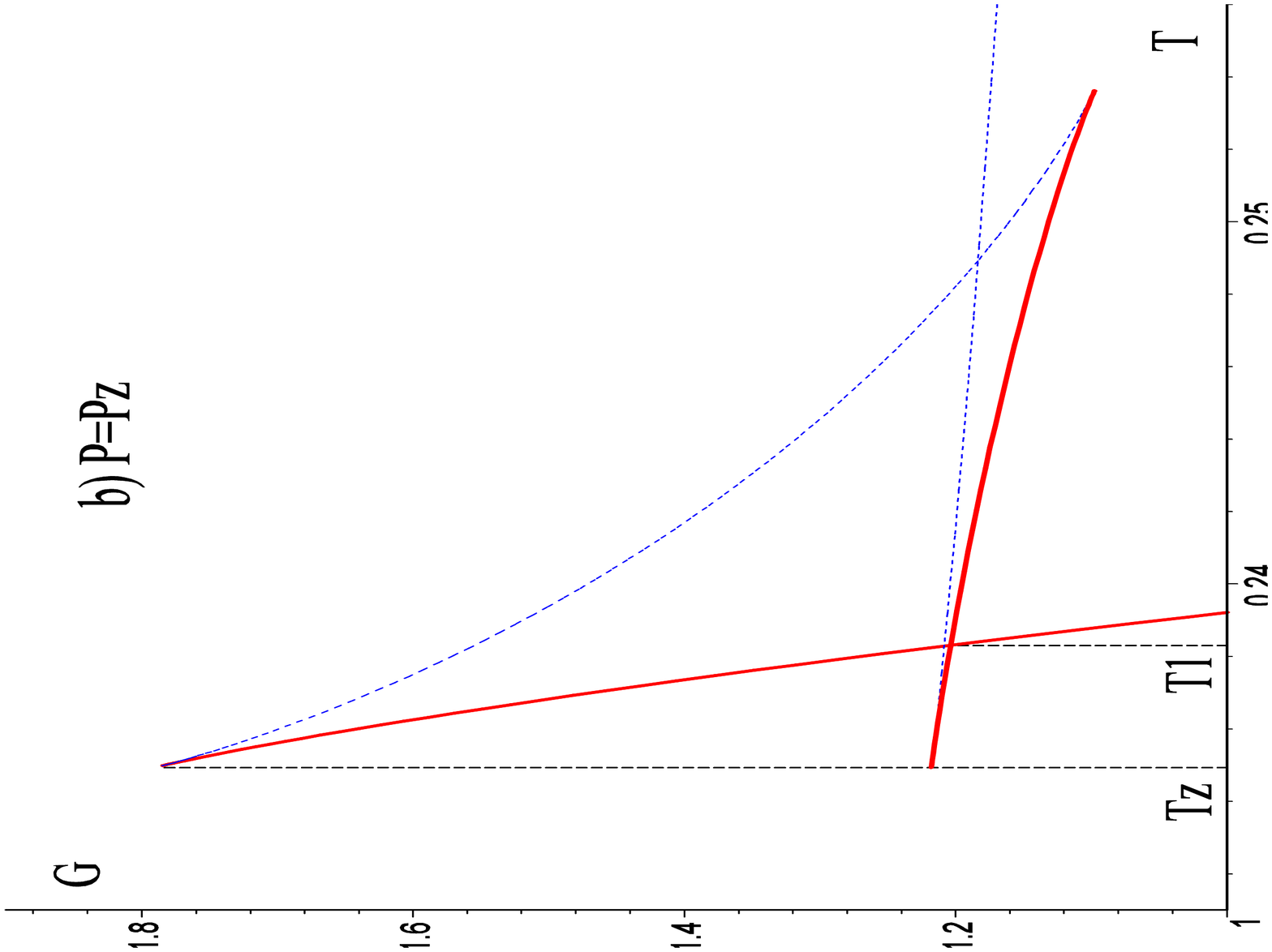}}\\
\rotatebox{-90}{
\includegraphics[width=0.4\textwidth,height=0.33\textheight]{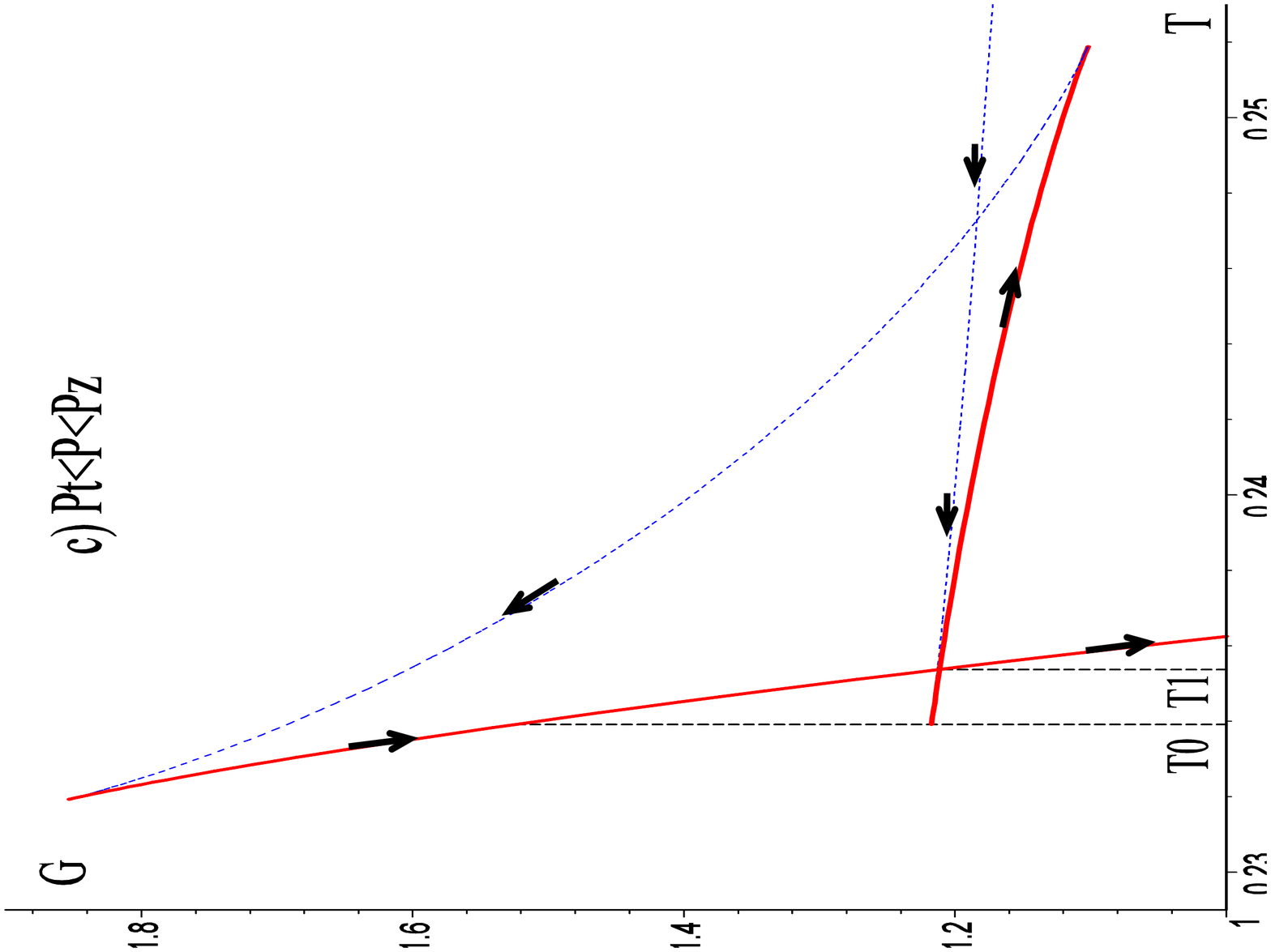}}&
\rotatebox{-90}{
\includegraphics[width=0.4\textwidth,height=0.33\textheight]{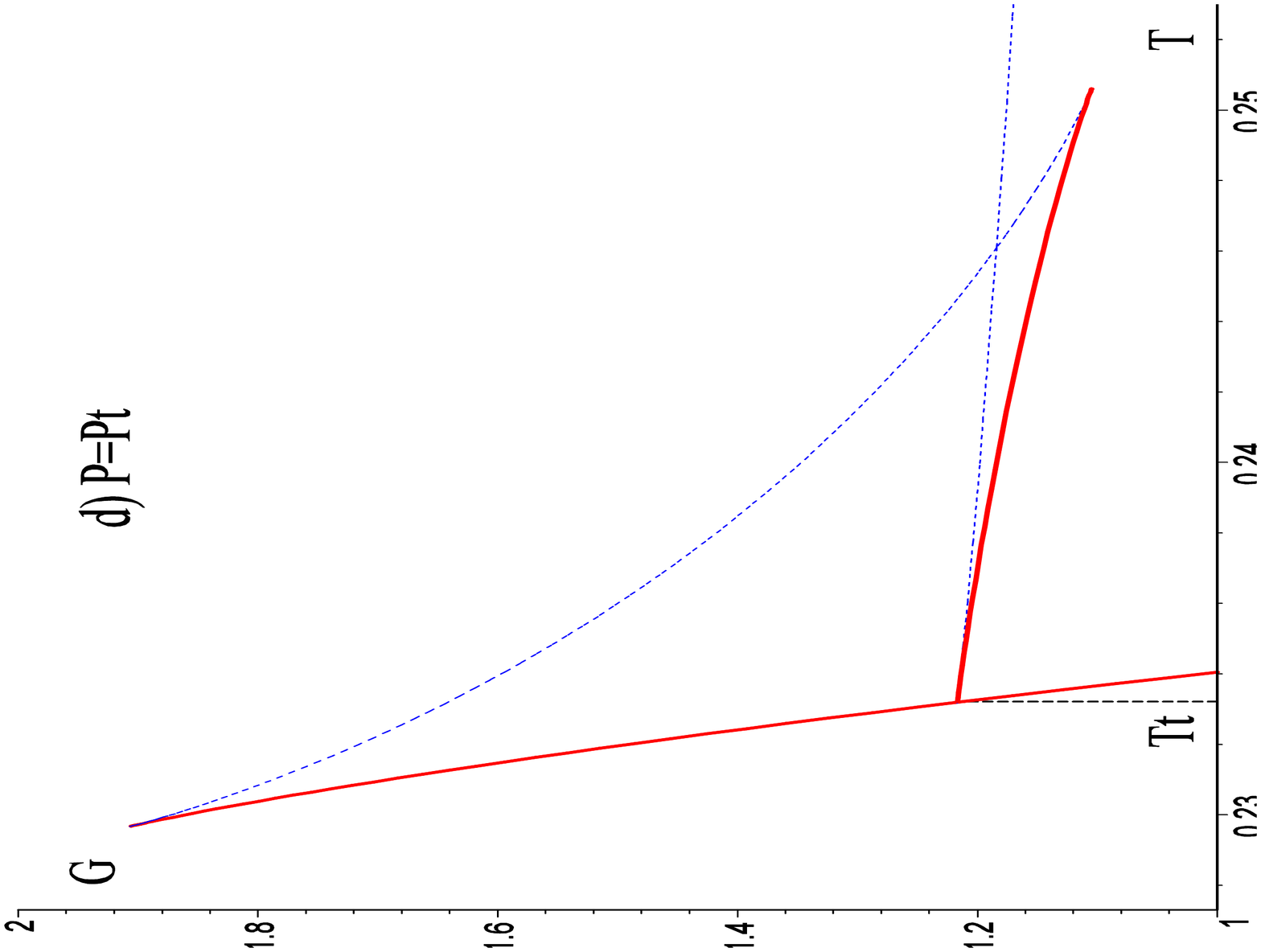}}
\end{tabular}
\caption{{\bf Reentrant phase transition: singly spinning Kerr-AdS black holes in $d=6$.}
The figure (a close up of fig.~\ref{Fig:6DKerr}) illustrates the behaviour of the Gibbs free energy for various 
pressures, $P=\{0.073, Pz=0.0579, 0.0564, P_t=0.0553\}$ (from top to bottom), 
when the reentrant phase transition is present. We have set $J=1$, for which $P_c=0.0958$. Solid-red/dashed-blue lines correspond to $C_P$ positive/negative respectively.
{\em Picture a)} describes a LBH/SBH first order phase transition at temperature $T_1$ as indicated by a dashed vertical line. 
{\em Picture b)} displays the behaviour of $G$ for $P=P_z$ for which the `upper peak' occurs at the same temperature $T=T_z$ as the 
lower peak---indicating the `entrance' of the reentrant phase transition. The LBH/SBH first order phase transition is 
still present and occurs at $T=T_1>T_z$.  
{\em Picture c)} displays a typical behaviour of $G$ when 
the reentrant phase transition is present, $P\in(P_t,P_z)$. Black arrows indicate increasing $r_+$. 
If we start decreasing the temperature from, say $T=0.24$, the system follows the lower vertical solid 
red curve until it joins the upper horizontal solid red curve---this corresponds to a first order SBH/LBH
phase transition at $T=T_1$. As $T$ continues to decrease the system follows this upper curve until  
$T= T_0$, where $G$ has a discontinuity at its global minimum. Further decreasing $T$, the system jumps to the 
uppermost vertical red line---this corresponds to the zeroth order phase transition between small and intermediate black holes.  
We note that the blue dashed curve with the smallest $r_+$ admits black holes subject to the ultraspinning instabilities, see Sec.~\ref{sec:Instab}, whereas all other branches are stable with respect to this instability.
Finally, {\em picture d)} shows the behaviour for 
$P=P_t$ when the lower peak merges with the vertical red line. For this pressure temperature of the 
zeroth-order phase transition $T_0$ coincides with the temperature of the first-order phase transition $T_1$, 
we call it $T_t=T_0=T_1$. 
}  
\label{Fig:6DKerrZorder}
\end{figure*} 
The resulting Gibbs free energy in $d=6$ is illustrated in fig.~\ref{Fig:6DKerr}; the behaviour in higher dimensions is qualitatively similar.
As with the Schwarzschild-AdS case discussed in Sec.~\ref{sec:4d}, for a given pressure $P$ there is a minimum temperature 
$T_{\mbox{\tiny  \tiny min}}$ for which the black holes can exist. Above this temperature we observe two (or more) branches of black holes. The branch corresponding to the largest black holes has positive $C_P$ and is locally thermodynamically stable. Its Gibbs free energy becomes negative above a certain temperature, analogous to $T_{\mbox{\tiny  HP}}$. The new feature present in six and higher dimensions is that between $T_{\mbox{\tiny  \tiny min}}$ and $T_{\mbox{\tiny  HP}}$ the branch of largest black holes may not correspond to a global minimum of the Gibbs free energy.

The complicated behaviour of $G$ between $T_{\mbox{\tiny  \tiny min}}$ and $T_{\mbox{\tiny  HP}}$ stems from the presence of a critical point occurring in the large black hole branch at $(T_c,P_c)$.
For $P>P_c$, $G$ resembles the curve characteristic for the Schwarzschild-AdS black hole seen in Sec.~\ref{sec:4d}, fig.~\ref{Fig:Gschads}. However, below $P_c$ the situation is more subtle, as  displayed in fig.~\ref{Fig:6DKerrZorder}.
The reentrant phase transition occurs for a range of pressures $P\in(P_t,P_z)$. The pressure $P_z$, see fig.~\ref{Fig:6DKerrZorder} b), is a pressure for which $T_{\mbox{\tiny  \tiny min}}(P_z)=T_z$ occurs for two locally thermodynamically stable branches of black holes. Slightly below this pressure we observe a reentrant phase transition, fig.~\ref{Fig:6DKerrZorder} c). It is connected with the discontinuity of the global minimum of the Gibbs free energy at $T=T_0$,  where we observe a zeroth order phase transition between small and large black holes. The reentrant phase transition terminates at $P=P_t$ below which the global minimum of the Gibbs free energy is again continuous and occurs for a branch of large black holes.  
We also note a presence of the first order SBH/LBH phase transition at $T=T_1$ in the range of pressures $P\in (P_t,P_c)$. 
As displayed in fig.~\ref{Fig:6DKerrZorder} d), at $P=P_t$ temperatures $T_0$ and $T_1$ coincide to give $T_0=T_1=T_t$. 
 At $T_t$ both the zeroth order and the first order phase transitions terminate and we observe a ``virtual triple point'',  characterized by $(P_t,T_t)$, where three black hole phases (small, large and intermediate) coexist together.\footnote{We use the term ``virtual triple point'' to stress a distinction from the standard triple point where three first-order phase transition coexistence lines join together. In a virtual triple point a zeroth-order phase transition coexistence line joins the first order phase transition coexistence lines where `three types' of black holes meet.}

Fig. \ref{Fig:6DKerrZorder} c) illustrates in detail the typical behaviour of $G$ when the reentrant phase transition is present, $P\in(P_t,P_z)$. 
Black arrows indicate increasing $r_+$. If we start decreasing the temperature from, say $T=0.24$, the system follows the lower vertical solid 
red curve of large stable black holes until it joins the upper horizontal solid red curve of small stable black holes---this corresponds to a first order LBH/SBH phase transition at $T=T_1$. As $T$ continues to decrease the system follows this upper curve until  
$T=T_0$, where $G$ has a discontinuity at its global minimum. Further decreasing $T$, the system jumps to the 
uppermost vertical red line of large stable black hole---this corresponds to the zeroth order phase transition between small and large black holes.
In other words, as $T$ continuously decreases, we observe LBH/SBH/LBH  reentrant phase transition.

\begin{figure}
\vspace{-0.5cm}
\begin{center}
\includegraphics[width=0.47\textwidth,height=0.32\textheight]{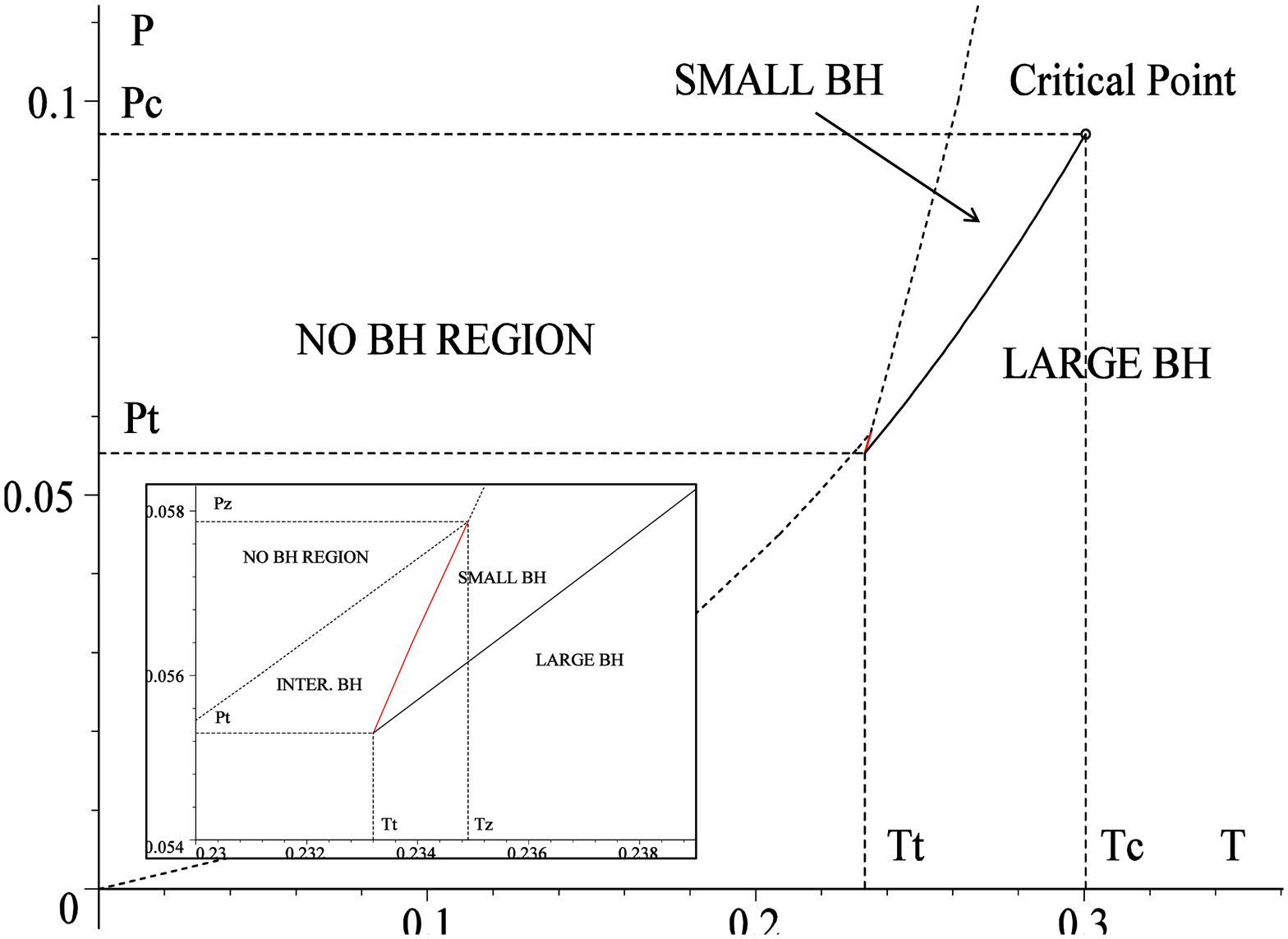}
\caption{{\bf Reentrant phase transition in $P-T$ plane.} 
The coexistence line of the first order phase transition between small and large black holes is depicted by a thick black solid line for $J=1$ and 
$d=6$. It initiates from the critical point $(P_c, T_c)$ and terminates at virtual triple point at $(P_t, T_t)$. The red solid line (inset) indicates the `coexistence line' of small and intermediate black holes, separated by a finite gap in $G$, indicating the 
reentrant phase transition.  It commences from  $(T_z, P_z)$ and terminates at $(P_t, T_t$).
 The ``No BH region",  given by $T_{\mbox{\tiny  \tiny min}}$, is to the left of the dashed oblique curve, containing the $(T_z, P_z)$ point.
A similar figure is valid for any $d\geq 6$. 
}
\label{Fig:PTZeroth1}
\end{center}
\end{figure}
This interesting situation is clearly illustrated in the $P-T$ diagram in fig.~\ref{Fig:PTZeroth1}.
There is the expected SBH/LBH line of coexistence corresponding to the liquid/gas Van der Walls case, ending in a
critical point at $(T_c,P_c)$.  On the other end this line terminates at $(T_t,P_t)$, where there is a virtual triple point where the large, small, and large again black holes coexist. For $T\in(T_t, T_z)$ a new LBH/SBH line of coexistence emerges (see inset of fig.~\ref{Fig:PTZeroth1}) that terminates in another ``critical point" at $(T_z,P_z)$. 
We note that the range for the reentrant phase transition is quite narrow and must be determined numerically. For example  for $J=1$ and $d=6$ we obtain
 $(T_t, T_z, T_c)\approx (0.2332, 0.2349, 0.3004)$ and $(P_t, P_z, P_c) \approx (0.0553, 0.0579, 0.0958)$.

\begin{figure}
\vspace{-0.7cm}
\begin{center}
\rotatebox{-90}{
\includegraphics[width=0.4\textwidth,height=0.32\textheight]{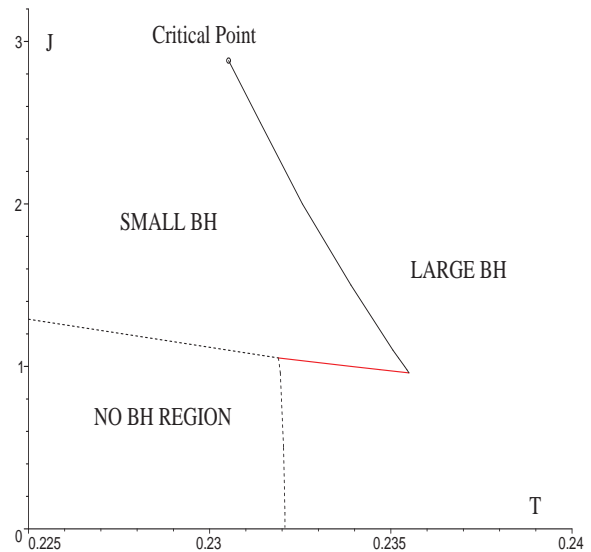}
}
\caption{{\bf Reentrant phase transition in $J-T$ plane.}
The coexistence line of the first order phase transition between small and large black holes is depicted by a thick black solid line. 
The red curve denotes a zeroth-order phase transition between small and intermediate (large) black holes. The dashed line outlines the `no black hole region'. The diagram is displayed for fixed $l\approx 2.656$ and $d=6$.
}
\label{Fig:PTZerothJ-14}
\end{center}
\end{figure}
Several remarks are in order:
\begin{enumerate}[a)]
\item
It is well known that in $d\geq 6$ dimensions there is no ``kinematic'' limit on how fast the singly spinning Kerr-AdS black holes can rotate. 
However, fast spinning black holes are subject to various dynamical instabilities, such as ultraspinning instability, superradiant instability,
or bar mode instability; these will be discussed in greater detail in Sec.~\ref{sec:Instab}. 
It turns out that black holes which participate in the reentrant phase transition are stable with respect to the ultraspinning instability:
in fig.~\ref{Fig:6DKerrZorder} c) only the blue dashed curve with the smallest $r_+$ admits black holes subject to this instability.
Unfortunately, this is no longer true for the superradiant and bar mode instability, which `compete' with the reentrant phase transition.
\item
One may wonder why the reentrant phase transition, which is characteristic for multicomponent systems where various phenomena compete among each other to result in reentrance, should occur at all in a `homogeneous' system of one black hole. What are the competing phenomena in our case?
A possible explanation is related to the ultraspinning regime.\footnote{If so, this would also explain why we see reentrance in $d\geq 6$ dimensions but not in $d=4$ or $5$ where such a regime does not exist.} It is well known that as we spin the spherical black hole faster and faster, its 
horizon flattens and the resulting object is in many respects similar to a black brane, see the next subsection. However, the thermodynamic behaviour of black branes is completely different from that of spherical black holes. It happens that small black holes that participate in the reentrant phase transition are `almost ultraspinning' and hence possess almost black brane behaviour. For this reason it may be the competition between the 
black brane thermodynamic behaviour and the black hole thermodynamic behaviour which causes the `multicomponency' and results in the reentrant phase transition.    
\item
We note that all the interesting behaviour leading to the reentrant phase transition occurs for a positive Gibbs free energy, i.e., below temperature $T_{\mbox{\tiny  HP}}$. For this reason, one may expect that the thermal AdS (see Sec.~\ref{sec:4d}) is actually preferred thermodynamic state in this region and the various black holes participating in the reentrant phase transition are actually metastable. 
If so, the reentrant phase transition may actually be destroyed and one would simply observe a Hawking--Page transition between thermal radiation and large black holes at  $T=T_{\mbox{\tiny  HP}}$.\footnote{We stress that similar arguments also apply to the four-dimensional charged AdS black hole discussed in Sec.~\ref{sec:4d} and the corresponding `Van der Waals' phase transition.}
\item
The observed reentrant phase transition is well suited for the AdS/CFT interpretation.
Although first observed \cite{AltamiranoEtal:2013a} in the context of extended phase space thermodynamics, 
the existence of the reentrant phase transition does not require a variable cosmological constant. 
For any fixed value of $\Lambda$ within the allowed range of pressure, the reentrant phase transition will take place. 
This opens up a possibility for an AdS/CFT interpretation---in particular in the dual CFT there will
be a corresponding reentrant phase transition within the allowed range of $N$. In fact, we can fix the pressure and
construct a phase diagram plotting $J$ vs. $T$ (fig.~\ref{Fig:PTZerothJ-14}) showing that reentrant phase behaviour occurs.
Hence in the dual CFT at this fixed pressure there will be a corresponding reentrant transition as the relative values
of the quantities dual to the angular momenta are adjusted.  
\item
The existence of reentrant phase transitions in the context of black hole thermodynamics seems quite general. Similar phenomena have been 
observed in Born--Infeld black hole spacetimes \cite{GunasekaranEtal:2012} and (quite recently) in the context of third-order 
Lovelock gravity \cite{FrassinoEtal:2013}.
We shall also see in Sec.~\ref{sec:MP}, that reentrant phase transitions are observed for the asymptotically flat doubly-spinning Myers--Perry black holes of vacuum Einstein gravity. Hence, neither exotic matter nor a cosmological constant (and hence AdS/CFT correspondence) are required for this phenomenon to occur in black hole spacetimes. 
\end{enumerate}

\subsection{Equation of state}
Let us now look at the equation of state. For simplicity  we shall first study 
the case of singly spinning Kerr-AdS black holes, given by \eqref{Singlyspinning}--\eqref{singlespin}; the case of 
equal spinning Kerr-AdS black holes is discussed towards the end.

\begin{figure}
\begin{center}
\includegraphics[width=0.46\textwidth,height=0.3\textheight]{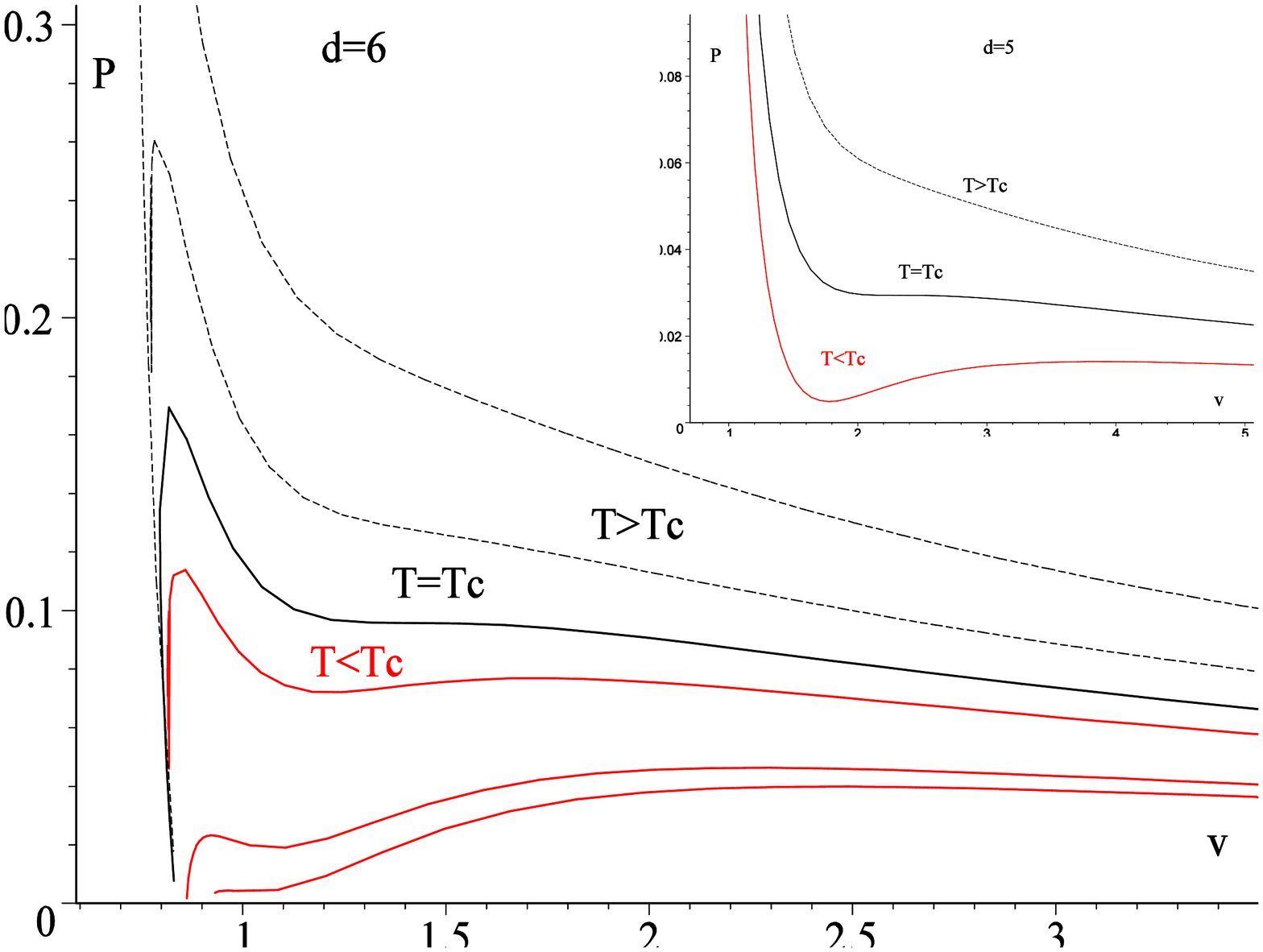}
\caption{{\bf Exact $P-v$ diagram: Kerr-AdS black holes.}
In $d=5$ (inset) we observe standard Van der Waals like behaviour, similar to $d=4$ in fig.~\ref{Fig:RNstate}.
In $d\geq 6$ the $P-v$ diagram is more complex than that
of the standard Van der Waals, reflecting the interesting
behaviour of the Gibbs free energy and a possible reentrant phase transition. Namely, whereas for 
large $v$ the isotherms approximate the Van der Waals behaviour, for small $v$, 
similar to Schwarzschild-AdS in fig.~\ref{Fig:PVSchwAdS}, the isotherms turn and lead back to a region with negative pressures. 
}
\label{Fig:6DKerrExact}
\end{center}
\end{figure}

The aim is to express the pressure $P$ as a function of  the specific volume $v$, the black hole temperature $T$, and the angular momentum $J$,
\be\label{PvT}
P=P(v,T,J)\,,
\ee
which is the equation of state \eqref{state}.
Finding this relation for general $J$ is algebraically involved. For this reason we plot the exact solution numerically:
we solve the quartic $J$ equation \eqref{singlespin} for $a=a(r_+, J, P)$,
obtain the resulting expressions $T=T(r_+, J, P)$ and
$v=v(r_+, J,P)$, and then solve them for fixed $J$ and $T$. 
The result in $d=5$ is qualitatively similar to what was observed in four dimensions, whereas in $d\geq 6$ the situation is more interesting, as shown in fig.~\ref{Fig:6DKerrExact}.

In two cases of physical interest---the slowly rotating
case and the ultraspinning regime---the  equation of state \eqref{PvT} can be approximated analytically.  In the former case, as with $d=4$ \cite{GunasekaranEtal:2012},  we consider an $\epsilon=a/l$ expansion. This slow rotation expansion allows us to approximate the equation of state demonstrating the Van der Waals behaviour and to study the corresponding critical point. In the ultraspinning regime, $\Xi=1-a^2/l^2\to 0$,
black hole geometry approaches that of a black membrane \cite{EmparanMyers:2003},
and is characterized by   completely different thermodynamic behaviour. We also construct the Kerr-AdS equation of state at the very ultraspinning limit and show in Sec.~\ref{sec:rings} that it coincides with the equation of state for ultraspinning black rings.

\subsubsection{Slow rotation expansion}
\begin{figure}
\vspace{-0.7cm}
\begin{center}
\rotatebox{-90}{
\includegraphics[width=0.39\textwidth,height=0.34\textheight]{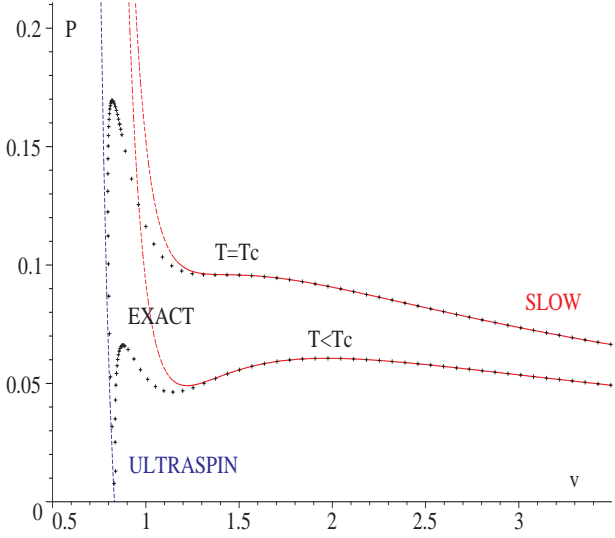}
}
\caption{{\bf The two approximations.}
Exact critical and subcritical isotherms, depicted by black crosses, are compared to the slow spinning expansion \eqref{Pbh} denoted by red curves and the ultraspinning expansion \eqref{Pultra} denoted by a blue curve; value of $r+$ decreases right to left. 
Obviously, the slow rotation approximation is valid for $r_+\to \infty$ whereas the ultraspinning one for $r_+\to 0$.
In the regime of very small $P$ (not displayed), $v$ grows arbitrarily large for any temperature $T$,  according to \eqref{PVultra}.
Note that the ultraspinning black holes correspond to the upper branch in the Gibbs free energy in fig.~\ref{Fig:6DKerr}, which is unstable. We have set $d=6$ and  $J=1$.}
\label{Fig:6DKerr1st}
\end{center}
\end{figure}

Here we demonstrate that  in the regime of slow rotation, where the equation of state can be well
approximated, a Van der Waals type of phase transition takes place. Within the validity of the slow rotation expansion we can analytically study 
the properties of the critical point as well as the critical exponents.

The slow rotation expansion is an expansion in the parameter
\be
\epsilon=\frac{a}{l}\to 0\,,
\ee
where we keep the pressure $P$ and the mass $M$ of the black hole fixed.

To obtain the expansion for the equation of state and the Gibbs free energy we proceed as
follows. First we solve the equation
\be\label{m=M}
\frac{4\pi M\Xi^2}{\omega_{d-2}\Bigl(1+\frac{d-4}{2}\Xi\Bigr)}=m=\frac{1}{2}r_+^{d-5}(r_+^2+a^2)\Bigl(1+\frac{r_+^2}{l^2}\Bigr)\,,
\ee
for $r_+$ in terms of $M$ and  $l$, while we express $a=\epsilon l$. 
The l.h.s. of this equation comes from $M$ equation \eqref{singlespin} and the r.h.s. from $\Delta=0$.  We expand
 \be\label{solr+}
r_+=\sum_{I=0}^k r_I \epsilon^I\,,
\ee
to some given order $k$ and solve Eq.~\eqref{m=M} order by order. The first term yields the relation
\be\label{r0M}
M=\frac{\omega_{d-2}(d-2)r_0^{d-3}}{16\pi}\Bigl(1+\frac{r_0^2}{l^2}\Bigr)\,,
\ee
which determines $r_0$ in terms of $M$; higher-order $r_I$ are obtained from subsequent terms and are
complicated expressions depending on $M$ and $l$, or alternatively
on $r_0$ and $l$.

Inserting the expansion \eqref{solr+} and the solution \eqref{r0M} into the expressions for $J$, $T$ and $V$, in 
\eqref{singlespin}, we find 
\ba
J&=&\frac{\omega_{d-2}r_0^{d-3}(l^2+r_0^2)}{8\pi l}\epsilon+\sum_{{\small I}=2} J_I\epsilon^{\small I}\,,\nonumber\\
T&=&\frac{r_0^2(d-1)+l^2(d-3)}{4\pi r_0 l^2}+\sum_{{\small I}=1}T_{\small I} \epsilon^{\small I}\,,\nonumber\\
v&=&\frac{r_0}{\kappa}+\sum_{{\small I}=1}v_{\small I} \epsilon^{\small I}\,,
\ea
while $P$ is given in terms of $l$ through \eqref{PLambda}.
Here $\{J_{\small I}, T_{\small I}, v_{\small I}\}$ are some concrete functions of $l$ and $r_0$.
Note that as $\epsilon\to 0$,  $J\to 0$, whereas all other quantities remain finite.

In order to get $P=P(v,T,J)$ we proceed perturbatively in $\epsilon$, successively  eliminating $r_0$ and $\epsilon$ from the resultant expressions. To zeroth order we find
\be
P=\frac{T}{v}-\frac{d-3}{\pi (d-2)v^2}+O(\epsilon^2)\,. 
\ee
To obtain the first correction we note that $J\propto \epsilon$ and hence the first correction is proportional
to $J^2$, while we fix the dependence on $T$ and $v$ using  dimensional analysis:
\ba\label{Pbh}
P&=&\frac{T}{v}-\frac{d-3}{\pi(d-2)v^2}+\frac{\pi (d-1)16^d}{4\omega_{d-2}^2 (d-2)^{2(d-1)}}\frac{J^2}{v^{2(d-1)}}\nonumber \quad\\
&&+O(\epsilon^4)\,.
\ea
When $d=4$ this expression reduces to the one presented in \cite{GunasekaranEtal:2012}, albeit obtained in a different fashion.
In principle by expanding to a sufficiently high order $k$, one can algorithmically
find the equation of state to an arbitrary order in $\epsilon$.  However  in the vicinity of the critical point the exact $P-v$ relation is 
``well approximated'' by the first three terms in \eqref{Pbh};  fig.~\ref{Fig:6DKerr1st} illustrates the situation for $d=6$. This allows us to study the properties of the critical point using the truncation \eqref{Pbh}.  We shall comment on the possible deviation
beyond this approximation in the end of this subsection.
The Gibbs free energy can likewise be obtained in the slow rotating expansion.

\subsubsection{Critical point}
The critical point $\{P_c, v_c, T_c\}$ of the truncated equation of state \eqref{Pbh} is obtained
from solving
\be
\frac{\partial P}{\partial v}=0\,,\quad \frac{\partial ^2P}{\partial v^2}=0\,,
\ee
which yields
\ba
v_c&=&\frac{4}{d-2}\Bigl[\frac{2^6\pi^2(2d-3)(d-1)^2J^2}{(d-2)(d-3)\omega_{d-2}^2}\Bigr]^{\frac{1}{2(d-2)}}\!\!,\qquad\nonumber\\
T_c&=&\frac{4(d-3)}{\pi(2d-3)}\frac{1}{v_c}\,,\quad
P_c=\frac{d-3}{\pi(d-1)}\frac{1}{v_c^2}\,.
\ea
We note the critical ratio [cf. Eq.~\eqref{universalVdWratio}]
\be\label{rhoc}
\rho_c=\frac{P_c v_c}{T_c}=\frac{2d-3}{4(d-1)}\,,
\ee
which reduces to $\rho_c=5/12$  for $d=4$ \cite{GunasekaranEtal:2012}.

Introducing new variables
\be
p=\frac{P}{P_c}\,,\quad \nu=\frac{v}{v_c}\,,\quad \tau=\frac{T}{T_c}\,,
\ee
the parameter $J$ can be `scaled away' and we obtain the
{\em law of corresponding states} of the form
\be\label{p_form}
p=\frac{1}{\rho_c}\frac{\tau}{\nu}+h(\nu)\,,
\ee
with $\rho_c$ given by \eqref{rhoc} and 
\be
h(\nu)=-\frac{d-1}{\nu^2(d-2)}+\frac{1}{(d-2)(2d-3)\nu^{2(d-1)}}\,.
\ee
It was shown in \cite{GunasekaranEtal:2012} that when the law of corresponding states
takes the form \eqref{p_form} and $1-\frac{1}{6}\rho_c h^{(3)}(1)>0$, which in our case is satisfied,
we recover the mean field theory critical exponents 
\be
\beta=\frac{1}{2}\,,\quad \gamma=1\,,\quad \delta=3\,.
\ee
Moreover, since
\be
S(T,v)=\frac{\omega_{d-2}\bigl[v(d-2)\bigr]^{d-2}}{4^{d-1}}+O(\epsilon^4)\,,
\ee
which implies that to  cubic order in $\epsilon$ we have $C_v=T\left(\frac{\partial S}{\partial T}\right)_v=0$, and so  the critical exponent 
$\alpha$ is
\be 
\alpha=0\,, 
\ee
which also coincides with the mean field theory prediction \eqref{MFT}.

\subsubsection{Remark on exact critical exponents in $d=4$}

So far we have studied the properties of the critical point using the truncated equation of state \eqref{Pbh}. Although this provides a good approximation to the exact equation of state, the exact critical point will be shifted from the preceding values and one may wonder whether the critical exponents will be modified. We  argue  that this is unlikely to happen. For simplicity, we limit ourselves to $d=4$.

For this purpose we approximate the exact equation of state by the slow rotating expansion to  a higher order than \eqref{Pbh},  writing
\ba
P_{(d=4)}&=&\frac{T}{v}-\frac{1}{2\pi v^2}+\frac{48J^2}{\pi v^6}-\frac{384 J^4(7+8\pi Tv)}{(1+\pi Tv)^2\pi v^{10}}\nonumber\\
&&+\frac{36864(13\pi Tv+11)J^6}{\pi v^{14}(1+\pi Tv)^3}+O(\epsilon^8)\,. \label{Pbh3}
\ea
Near a critical point, it can be shown numerically that this approximation to high precision reproduces the exact equation of state.
Since  equation \eqref{Pbh3} does not give rise to the law of the corresponding states of the form \eqref{p_form}  the critical exponents may,
in principle, be different. Similarly we have 
\be
S_{(d=4)}=\frac{\pi v^2}{4}-\frac{192 \pi J^4}{v^6(1+\pi Tv)^2}\!+\!\frac{36864 \pi J^6}{v^{10}(1+\pi Tv)^3}\!+\!O(\epsilon^8)
\ee
to this same level of approximation. 
Using these two equations, we have found numerically the critical point, with $\rho_c$ slightly different from $5/12$, and (within the numerical accuracy) confirmed it to be characterized by the same critical exponents $\alpha=0, \beta=\frac{1}{2}\,,\gamma=1\,,\delta=3\,.$

\subsubsection{Ultraspinning expansion}

In $d\geq 6$  singly spinning spherical black holes admit an interesting feature, the so called {\em ultraspinning regime}, which is present already for  asymptotically flat Myers--Perry black holes \cite{EmparanMyers:2003}.
This limit consists of  rotating the black hole faster and faster, while keeping its mass and the cosmological constant fixed. Consequently, the black hole horizon flattens, becomes pancake-like, and the black hole enters the regime of black membrane-like behaviour.
For asymptotically flat Myers--Perry black holes, the ultraspinning limit is achieved when the rotation parameter $a\to \infty$. In the presence of a negative cosmological constant one instead considers $a\to l$ (for an alternative see \cite{Armas:2010hz}), or, equivalently,
\be\label{Xi0}
\epsilon=\Xi^{\frac{1}{d-5}}=\Bigl(1-\frac{a^2}{l^2}\Bigr)^{\frac{1}{d-5}}\to 0\,.
\ee
In this subsection we derive the form of the equation of state and the Gibbs free energy employing the ultraspinning expansion. As a by-product we obtain the thermodynamic volume of the ultraspinning membrane-like black hole and demonstrate that it remains finite.

Similar to the slow rotating case discussed in the previous section, in the
ultraspinning expansion we expand in small parameter $\epsilon$, \eqref{Xi0}, while we keep the pressure $P$ and the mass $M$ of the black hole fixed.

We again solve the equation
\be
\frac{4\pi M\Xi^2}{\omega_{d-2}\Bigl(1+\frac{d-4}{2}\Xi\Bigr)}=m=\frac{1}{2}r_+^{d-5}(r_+^2+a^2)\Bigl(1+\frac{r_+^2}{l^2}\Bigr)\,,
\ee
to find the expansion
\be\label{solr}
r_+=r_2\epsilon^2+\sum_{{\small I}=3}^k r_{\small I} \epsilon^{\small I}\,,  \quad r_2=\left(\frac{8\pi M}{l^2\omega_{d-2}}\right)^{\frac{1}{d-5}}\,,
\ee
up to some order $k$\,; the $r_I$ are calculable functions of $M$ and $l$.
Plugging this solution into the expressions for $J$, $T$ and $V$ in \eqref{singlespin}, we find the following expansions:
\ba\label{JTvUltra}
J&=&Ml+\sum_{{\small I}=1} J_{\small I}\epsilon^{\small I}\,,\nonumber\\
T&=&\frac{d-5}{4\pi r_2\epsilon^2}+\sum_{{\small I}=-1}T_{\small I} \epsilon^{\small I}\,,\nonumber\\
v&=&\frac{4}{d-2}\left[\frac{8\pi Ml^2}{\omega_{d-2}(d-2)}\right]^{\frac{1}{d-1}}+\sum_{{\small I}=1}v_{\small I} \epsilon^{\small I}\,,
\ea
while $P$ is given by \eqref{PLambda}.
Here $\{J_{\small I}, T_{\small I}, v_{\small I}\}$ are calculable functions of $l$ and $M$. 
We note that in the $\epsilon\to 0$ limit, the radius of the horizon   $r_+\to 0$ and the temperature diverges, whereas both the (specific) volume as well the angular momentum remain finite. We also find
\be\label{Aultra}
A=\omega_{d-2} l^2\epsilon^{d-3}r_2^{d-4}+o(\epsilon^{d-3})\,,
\ee
and hence the number of degrees of freedom associated with the horizon, $N\propto A/l_P^2$ vanishes.

To find the equation of state $P=P(v,T,J)$, as well as other quantities, to any order in $\epsilon$, we proceed algorithmically, as in the case of the slow rotating expansion.
In particular, to the zeroth order we find
\ba\label{PVultra}
P&=&\frac{4\pi}{(d-1)(d-2)}\frac{J^2}{V^2}+\dots\nonumber\\
&=&\frac{\pi(d-1)16^d}{4\omega_{d-2}^2(d-2)^{2d-1}}\frac{J^2}{v^{2(d-1)}}+\dots\,.
\ea
We note the interesting coincidence that this expression coincides with the 3rd term in slow rotation expansion \eqref{Pbh}.  
Higher-order  terms depend crucially on the  number of dimensions. For example, the  next term in the expansion \eqref{solr} reads
\ba
d<9:&& \quad r_{d-3}=r_2\frac{6-d}{2(d-5)}\,,\nonumber\\
d=9:&& \quad r_6=-r_2\Bigl(\frac{3}{8}+\frac{r_2^2}{2l^2}\Bigr)\,,\nonumber\\
d>9:&& \quad r_6=-\frac{2r_2^3}{l^2(d-5)}\,,
\ea
and we observe dimension-dependent `branching'.  

For this reason, we give up on finding a general expression
and provide rather the expanded equations of state and the Gibbs free energies in selected dimensions $(d=6,7,8,9$ and $10)$. The first three non-trivial terms in the expansion of equation of state read
\ba\label{Pultra}
P_6&=&\frac{\pi J^2}{5 V^2}-\frac{\sqrt{5\omega_4}}{16\sqrt{\pi^3 VT}}
+\frac{9\sqrt{5\omega_4}J^2}{3200\sqrt{\pi^3V^5T^5}}\,,\nonumber\\
P_7&=&\frac{2\pi J^2}{15V^2}-\frac{\sqrt{30\omega_5}}{32 \sqrt{\pi^4VT^2}}+
\frac{\sqrt{30\omega_5}J^2}{300\sqrt{\pi^4 V^5T^6}}\,,\nonumber\\
P_8&=&\frac{2\pi J^2}{21V^2}-\frac{9\sqrt{14\omega_6}}{128\sqrt{\pi^5VT^3}}+
\frac{135\sqrt{14\omega_5}J^2}{12544\sqrt{\pi^5V^5T^7}}\,,\nonumber\\
 P_9&=& \frac{\pi  J^2}{14 V^2}
-\frac{\sqrt{7{\omega _7}}}{4 \sqrt{2\pi^6VT^4}}
+\frac{9 \sqrt{\omega _7}J^2}{28\sqrt{14 \pi^6 V^5T^8}}\,, \ \\
   P_{10}&=&
\frac{\pi  J^2}{18 V^2}-\frac{75\sqrt{{5\omega _8}}}{128 \sqrt{2\pi^7VT^5}}+
\frac{4375 \sqrt{5\omega _8}J^2}{36864 \sqrt{2\pi^7V^5T^9}}\,.\qquad \nonumber
\ea
Similarly, for the Gibbs free energy we find 
\ba
G_6&=&\frac{2\sqrt{5\pi P} J}{5}+\frac{5^{\frac{1}{4}}\sqrt{\omega_4 J}}{8\pi\sqrt{T}(\pi P)^{\frac{1}{4}}}-\frac{5\omega_4}{1024 \pi^3 TP} \,,\nonumber\\
G_7&=&\frac{2\sqrt{30\pi P}J}{15}+\frac{2^{3/4}15^{1/4}\sqrt{\omega_5 J}}{16\pi^{7/4}P^{1/4}T}
-\frac{15\omega_5}{2048\pi^4PT^2}\,,\nonumber\\
G_8&=&\frac{2\sqrt{42\pi P}J}{21}+\frac{6^{3/4}7^{1/4}3\sqrt{\omega_6 J}}{64\pi^{9/4}P^{1/4}T^{3/2}}-\frac{567\omega_6}{32768\pi^5PT^3}\,, \nonumber\\
G_9&=&\frac{\sqrt{2\pi P}J}{\sqrt{7}}+
\frac{14^{1/4}\sqrt{J\omega_{7}}}{4\pi^{11/4}P^{1/4}T^2}
+\frac{3\sqrt{\omega_7 J}{14}^{1/4} P^{3/4}}{14\pi^{15/4} T^4}\nonumber\\
&&-\frac{7\omega_7}{128\pi^6P T^4}\,,
\nonumber\\
G_{10}&=&\frac{1}{3} \sqrt{2 \pi P} J
+\frac{25 \sqrt{15} 2^{1/4}\sqrt{J \omega_{8}}}{128\pi^{13/4} P^{1/4} T^{5/2}}\nonumber\\
&&+\frac{4375 \sqrt{15} 2^{1/4}P^{3/4} \sqrt{J \omega_{9}}}{18432\pi ^{17/4} T^{9/2}}\,. 
\ea

We display the two expansions, the slow rotating one and the ultraspinning one together with the exact quantities in fig.~\ref{Fig:6DKerr1st}.

\subsubsection{Ultraspinning limit: black membranes}\label{sec:Ultraspin}
The full ultraspinning limit, in which $\epsilon$ given by \eqref{Xi0} actually equals zero, is obtained by keeping only the dominant terms in the ultraspinning expansion \eqref{JTvUltra}--\eqref{Aultra}. For Kerr-AdS black holes this was studied in \cite{CaldarelliEtal:2008} (see also \cite{Armas:2010hz} for an alternative approach). As noted above, in this limit the black hole horizon radius shrinks to zero, the black hole temperature becomes infinite, the angular momentum and the (specific) thermodynamic volume remain finite, and the horizon area (and hence also the entropy) shrinks to zero; the angular velocity of the horizon goes as $\Omega_Hl\to 1$.

These ultraspinning black holes share  many properties with black membranes, including the Gregory--Laflamme instability \cite{EmparanMyers:2003}.
In fact, when the limit is accompanied with the following change of coordinates:
\be
\tau=\frac{t}{\epsilon^2}\,,\quad \rho=\frac{r}{\epsilon^2}\,,\quad \sigma=\epsilon^{\frac{7-d}{2}}l\sin\theta\,,
\ee
together with defining $\mu=\frac{8\pi M}{\omega_{d-2}l^2}$,
the ultraspinning black hole metric becomes (up to a constant conformal prefactor) that of a black membrane \cite{CaldarelliEtal:2008}:
\ba
ds^2_{M}&=&-fd\tau^2+\frac{d\rho^2}{f}+d\sigma^2+\sigma^2d\varphi^2+\rho^2d\Omega_{d-4}^2\,,\nonumber\\
f&=&1-\frac{\mu}{\rho^{d-5}}\,.
\ea
We emphasize again that the thermodynamic volume remains finite and reads
\be
V=\frac{8\pi Ml^2}{(d-1)(d-2)}\,.
\ee
At the same time the area of the black hole vanishes according to \eqref{Aultra}. This implies that in the full ultraspinning limit 
the   isoperimetric ratio \eqref{ratio} diverges to infinity:
\be\label{ratioInfinity}
{\cal R}\to \infty\,.
\ee 
Since $TS=O(\epsilon^{d-5})$, the Gibbs free energy is
\be
G=M=\sqrt{\frac{16\pi P}{(d-1)(d-2)}}J\,,
\ee
while the equation of state takes  interesting form
\be
P=\frac{4\pi}{(d-1)(d-2)}\frac{J^2}{V^2}\,.
\ee
We shall see in Sec.~\ref{sec:rings} that exactly the same equation of state is valid for the ultraspinning black rings.

\subsubsection{Equal spinning AdS black holes}
Let us briefly mention how the equation of state gets modified for the equal spinning Kerr-AdS black holes. 
Equal spinning Kerr-AdS black holes are characterized by $N=(d-\varepsilon-1)/2$ equal angular momenta, 
\be
J_1=J_2=\dots=J_N=J\,.
\ee  
In this case the symmetry of the spacetime is significantly enhanced and the general metric \eqref{metric} and its thermodynamic 
characteristics \eqref{TD}--\eqref{VBHKerr} simplify considerably. For example, the thermodynamic volume now reads 
\be\label{V1BH}
V=\frac{r_+A}{d-1}\Bigl[1+\frac{N a^2}{\Xi}\frac{1+r_+^2/l^2}{(d-2)r_+^2}\Bigr]\,.
\ee
As mentioned above, the Gibbs free energy exhibits a classical swallowtail, see fig.~\ref{Fig:5DKerr}, with a critical point at $(T_c,P_c)$ and
the corresponding equation of state mimics the Van der Waals behaviour. 

Repeating the procedure for the slow rotation expansion, $\frac{a}{l}\to 0$, as we did for the singly spinning case, we recover  
\ba\label{eq}
P&=&\frac{T}{v}-\frac{d-3}{\pi (d-2)v^2}+\frac{\pi (d-1)^{2-\varepsilon}16^d}{8\omega_{d-2}^2 (d-2)^{2(d-1)-\varepsilon}}\frac{J^2}{v^{2 (d-1)}}\nonumber \quad\\
&&+O\bigl[(a/l)^4\bigr]\,.
\ea
The equation of state \eqref{eq} admits a critical point at 
\ba
v_c&=&\frac{4}{d-2}\Bigl[\frac{2^6\pi^2}{\omega_{d-2}^2}\frac{2d-3}{d-3}\frac{(d-1)^{3-\varepsilon}}{(d-2)^{1-\varepsilon}}J^2\Bigr]^\frac{1}{2(d-2)}\!\!,\qquad\nonumber\\
T_c&=&\frac{4(d-3)}{\pi(2d-3)}\frac{1}{v_c}\,,\quad
P_c=\frac{d-3}{\pi(d-1)}\frac{1}{v_c^2}\,.
\ea
The critical ratio reads $\rho_c=\frac{P_c v_c}{T_c}=\frac{2d-3}{4(d-1)}$\,,
and is the same as for singly spinning case, \eqref{rhoc}. Also the critical exponents 
of \eqref{eq} remain as predicted by the mean field theory, \eqref{MFT}.

\subsection{An analogue of triple point and solid/liquid/gas phase transition}
We now turn to a study of the thermodynamic behaviour of multi-spinning $d=6$ Kerr-anti de Sitter  black holes with two fixed (non-trivial) angular momenta  
$J_1$ and $J_2$. Following \cite{AltamiranoEtal:2013b}, depending on the ratio
\be
q=J_2/J_1\,,
\ee
we find qualitatively different interesting phenomena known from the `every day thermodynamics' of simple substances.
In addition to the large/small/large reentrant phase transition seen for $q=0$ in previous subsections, we find for  $0<q< 0.00905$   an analogue of a `solid/liquid' phase transition.  Furthermore,   the system exhibits a large/intermediate/small black hole phase transition with one triple and two critical points for $q\in (0.00905, 0.0985)$. This  behaviour is reminiscent of the solid/liquid/gas phase transition although the coexistence line of small and intermediate black holes does not continue for an arbitrary value of pressure (similar to the solid/liquid coexistence line) but rather terminates at one of the critical points. Finally, for $q>0.0985$ we observe the `standard liquid/gas behaviour' of a Van der Waals fluid.

Specifying to $d=6$, we consider the Gibbs free energy $G=G(P,T,J_1,J_2)$, obtained from the expression \eqref{GibbsKerrAdS} by eliminating parameters $(l, r_+, a_1, a_2)$ in favor of $(P, T,J_1,J_2)$ using Eqs.~\eqref{TD}, \eqref{TS} and \eqref{PLambda}.  An analytic solution is not possible
since higher-order polynomials are encountered and so we proceed numerically: for a given $P, r_+, J_1$ and $J_2$, we solve $J_i$ equations in \eqref{TD} for $a_1$ and $a_2$  and thence calculate the values of $T$ and $G$ using \eqref{TS} and \eqref{GibbsKerrAdS}, yielding a $G-T$ diagram.
Once the behaviour of $G$ is known, we compute the corresponding phase diagram,  coexistence lines, and critical points in the $P-T$ plane. We display our results in figs. \ref{fig:3}--\ref{fig:6}. 
Since the qualitative behaviour of the system depends only on  $q=J_2/J_1$ we set everywhere $J_1=1$.

\subsubsection{Solid/liquid analogue}
\begin{figure}
\vspace{-0.5cm}
\begin{center}
\rotatebox{-90}{
\includegraphics[width=0.37\textwidth , height=0.34\textheight]{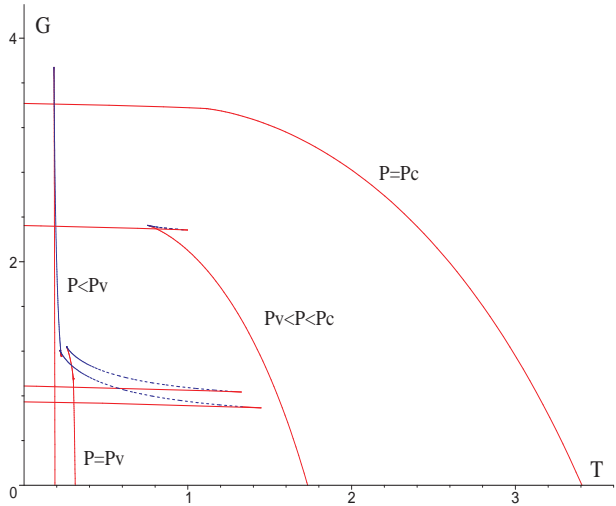}
}
\caption{{\bf Gibbs free energy for $q=0.005$} Kerr-AdS black hole in $d=6$ is displayed for  decreasing pressures (from top to bottom). 
The horizon radius $r_+$ increases from left to right. The uppermost isobar corresponds to $P=P_{c}=4.051$; for higher pressures only one branch of (stable) black holes with positive $C_P$ exists. The second uppermost isobar displays the swallowtail behaviour and implies the existence of a first order phase transition. For $P=P_v\approx 0.0958$ another critical point emerges but occurs for a branch that does not minimize $G$ globally. Consequently, out of the two swallowtails for $P<P_v$ only one occurs in the branch globally minimizing $G$ and describes a `physical' first order phase transition.
} \label{fig:3}
\end{center}
\end{figure}

For $0<q<q_1\approx 0.00905$ the behaviour of $G$, displayed for $q=0.005$ in fig.~\ref{fig:3},  completely changes to the $q=0$ behaviour discussed in the previous subsection, cf. figs.~\ref{Fig:6DKerr} and \ref{Fig:6DKerrZorder}. Namely the unstable branch of tiny hot black holes on the right of the $G-T$ diagram in fig.~\ref{Fig:6DKerr} disappears and a new branch of (locally) stable tiny cold black holes appears to the left. The $q=0$ `no black hole region' is eliminated and the situation is very similar to what happens when a small charge is added to a Schwarzschild black hole, cf. Sec.~\ref{sec:4d}. The zeroth-order phase transition  is `replaced' by a `solid/liquid'-like phase transition of small to large black holes.
Although in this range of angular momenta $G$ admits two critical points (one at $P=P_c$ and another at $P=P_v<P_c$), only the one at $P=P_c$ occurs for stable black holes that minimize $G$ globally. Consequently we observe one phase transition between small and large black holes.  
The corresponding coexistence line (not displayed) terminates at a critical point characterized by $(T_c,P_c)$ and is qualitatively similar to fig.~\ref{Fig:RNPT}. As $q$ decreases, the critical point occurs for larger and larger pressures $P_c$ and temperatures $T_c$ and in the limit $q\to 0$ we find `an infinite' coexistence line, similar to what happens for a solid/liquid phase transition. This is indicated in fig.~\ref{Fig:varyq}.

\subsubsection{Triple point and solid/liquid/gas analogue}
\begin{figure}
\vspace{-0.7cm}
\rotatebox{-90}{
\includegraphics[width=0.37\textwidth , height=0.34\textheight]{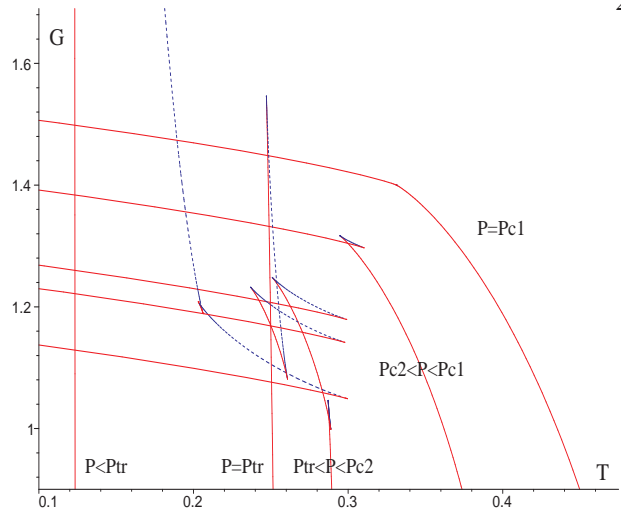}
}
\caption{ {\bf Gibbs free energy for $q=0.05$} Kerr-AdS black hole in $d=6$ is displayed for  various pressures (from top to bottom) $P=0.260, 0.170, 0.087, 0.0642, 0.015$. The horizon radius $r_+$ increases from left to right. The uppermost isobar corresponds to $P=P_{c_1}=0.260$; for higher pressures only one branch of stable black hole with positive $C_P$ exists. The second uppermost isobar displays the swallowtail behaviour, implying a first order phase transition.
The third isobar corresponds to $P$ between $P_{c_2}=0.0957$ and $P_{c_1}$ for which we have `two swallowtails'. For such pressures there are two first order phase transitions. The fourth isobar displays the tricritical pressure $P_{tr}=0.0642$ where the two swallowtails `merge' and the triple point occurs. Finally the lower-most isobar corresponds to $P<P_{tr}$.
} \label{fig:5}
\end{figure}
\begin{figure}
\vspace{-0.7cm}
\rotatebox{-90}{
\includegraphics[width=0.37\textwidth , height=0.34\textheight]{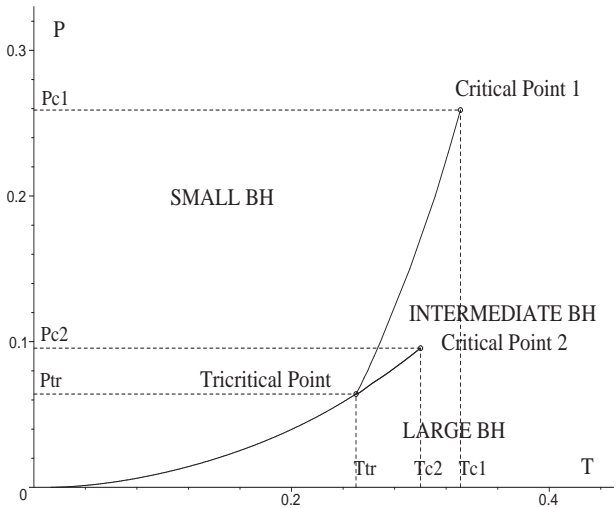}
}
\caption{ {\bf  $P-T$ diagram for $q=0.05$.} The diagram is analogous to the solid/liquid/gas phase diagram.
Note however that there are two critical points. That is, small/intermediate black hole coexistence line does not extend to infinity but rather terminates, similar to the ``liquid/gas'' coexistence line, in a critical point. }
\label{fig:6}
\end{figure}

For fixed $q_1<q<q_2\approx 0.0985$ a new phenomenon occurs: for a certain range of pressures we observe two swallowtails in the stable branch of black holes, see fig.~\ref{fig:5}, corresponding to a small/intermediate/large black hole phase transition. 
As pressure decreases, the two swallowtails move closer together until, at $P=P_{tr}$,
the corresponding first order phase transitions coincide---giving a triple point. Below $P_{tr}$, only one of the first order phase transitions 
continues to occur in a stable branch. The whole situation is reminiscent of the solid/liquid/gas phase transition.    

Let us describe in more detail what happens for fixed $q=0.05$ as pressure decreases, illustrated in figs.~\ref{fig:5} and \ref{fig:6}. For very large pressures there are no swallowtails and only one branch of locally stable black holes with positive $C_P$ exists. Decreasing the pressure, 
at $P=P_{c_1}\approx 0.260$ a critical point occurs; slightly below $P_{c_1}$ we observe a first order phase transition between large and small black holes. This continues to happen until at $P=P_{c_2}\approx 0.0957$ we observe another critical point. Slightly below $P_{c_2}$ the system exhibits 
two swallowtails in the stable branch of black holes and thence two first order phase transitions between various size black holes are present: from small to intermediate and from intermediate to large.
As $P$ decreases even further, the two swallowtails move closer to each other, until at $P=P_{tr}\approx 0.0642$ the corresponding first order phase transitions coincide---we observe a triple point. Below $P_{tr}$ only one of the first order phase transitions continues to occur in the stable branch of black holes. The corresponding $P-T$ phase diagram displaying all the features is depicted in fig.~\ref{fig:6}; in many respects it is reminiscent of  a solid/liquid/gas phase diagram.

\begin{figure}
\vspace{-0.2cm}
\rotatebox{-90}{
\includegraphics[width=0.37\textwidth , height=0.34\textheight]{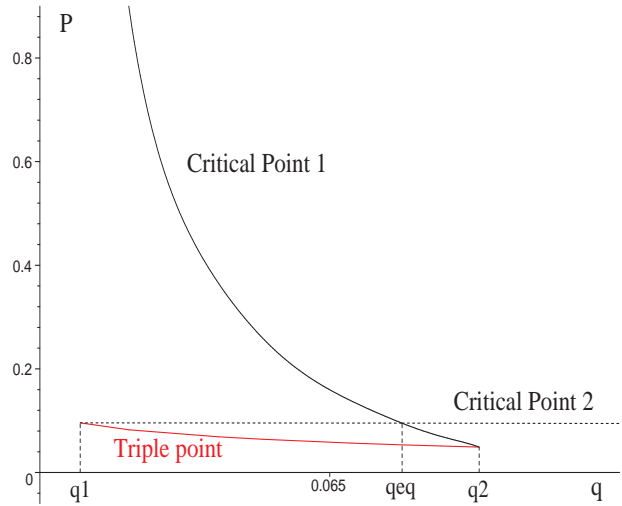}
}
\caption{ {\bf  Critical pressures and variable $q$.}
Depending on the value of $q$, we observe one critical point or two critical points and one triple point.
This figure displays the corresponding $P_{c_1}$, $P_{c_2}$, and $P_{tr}$.
For $q<q_1$ only one critical point  occurs: as $q\to 0$, $P_{c_1}$ rapidly diverges to infinity and the coexistence line 
becomes infinitely long---this is an analogue of a solid/liquid phase transition. 
Between $q_1$ and $q_2$ we observe two critical points at $P_{c_1}$ and $P_{c_2}$, displayed by solid thick black line and thin dashed line respectively, and a triple point displayed by the red solid curve. As $q$ increases, $P_{c_1}$ rapidly decreases until it meets $P_{tr}$ at 
$q=q_{2}$ where it terminates; in between there exists $q=q_{eq}$ where $P_{c_1}=P_{c_2}$. Above $q_2$ only the second critical point at $P_{c_2}$ exists.  
}  \label{Fig:varyq}
\end{figure}

So far we considered a fixed parameter $q=0.05$. Let us now look at what happens as we allow $q$ to vary, see fig.~\ref{Fig:varyq}. 
We already know that for $q<q_1$ we observe one first order phase transition terminating at a critical point characterized by $(T_{c_1}, P_{c_1})$, as described in the previous subsection. 
Qualitatively new feature occurs at $q=q_1$: a triple point and a second critical point emerge from the coexistence line at $P_{tr}=P_{c_2}\approx 0.09577$ and $T_{tr}=T_{c_2}\approx 0.30039$. 
As $q$ increases, the triple point moves away from the second critical point; the values of $T_{tr}$ and $P_{tr}$ decrease, while the values of $T_{c_2}$ and $P_{c_2}$ almost do not change. A small/intermediate/large black hole phase transition may occur and the situation resembles that of the solid/liquid/gas phase transition.
At the same time as $q$ increases, the first critical point moves towards the triple point ($P_{c_1}$ rapidly decreases). At $q=q_{eq}\approx 0.08121$ both critical points occur at the same pressure
$P_{c_1}=P_{c2}\approx 0.0953$, whereas $T_{c_1}\approx 0.2486<T_{c_2}\approx 0.2997$. Increasing $q$ even further, the first critical point moves closer and closer to the triple point and finally for $q=q_2\approx 0.0985$ the two merge at  $P_{tr}=P_{c_1}\approx 0.049$. Above $q_2$ only the second critical point remains. 

\begin{figure}
\begin{center}
\includegraphics[width=0.52\textwidth,height=0.29\textheight]{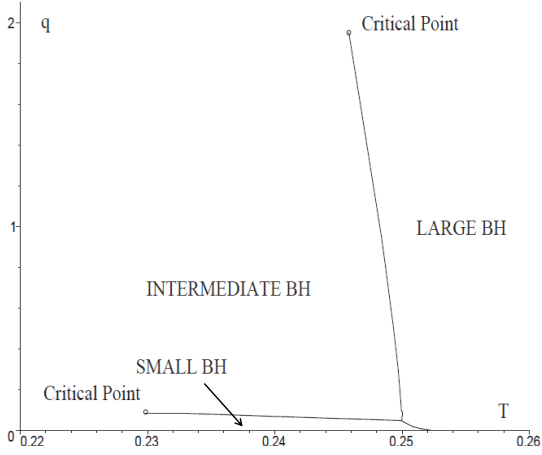}
\caption{{\bf Kerr-AdS analogue of solid/liquid/gas phase transition in $q-T$ plane.}
The diagram is displayed for fixed pressure $l\approx 2.491$.
}
\label{Fig:PTZerothJ-21}
\end{center}
\end{figure}

We remark that, similar to the reentrant phase transition, the analogue of a solid/liquid/gas phase transition as well as of the triple point occurs for any fixed $P$ in the allowed range. Once again, the possibility for an AdS/CFT interpretation emerges: a corresponding phase transition may occur in the dual CFT within the allowed range of $N$. To stress this point we plot in fig.~\ref{Fig:PTZerothJ-21} a $q-T$ phase diagram, showing the existence of small/intermediate/large black hole phase transition and the corresponding triple point for a fixed pressure. 

\subsubsection{Van der Waals behaviour}
For $q>q_2$ only one small/large black hole coexistence line exists and terminates at a corresponding
critical point. We observe an analogue of  a Van der Waals  `liquid/gas' phase transition.

In fact, the situation is a little more subtle. For $q_2<q<q_3\approx 0.1274$, similar to the `solid/liquid analogue', the second critical point still exists, but occurs for the locally unstable branch of black holes with negative heat capacity that does not globally minimize the Gibbs free energy. Hence, only one phase transition is physical. For $q>q_3$ this second critical point completely disappears.
In both cases we observe a `standard liquid/gas' Van der Waals phase transition, qualitatively similar to $d=4, 5$ case and the case of equal spinning Kerr-AdS black holes in any dimension $d\geq 5$.

\section{Myers--Perry solutions}\label{sec:MP}
We now turn to consider the $d$-dimensional multi-rotating black holes with spherical horizon topology for which the cosmological constant and hence the pressure vanishes, described by the Myers--Perry solution  \cite{MyersPerry:1986}. The most interesting observation is that even in this asymptotically flat case we find that reentrant phase transitions can take place.

The Myers-Perry (MP) solutions are the most general  rotating black hole solutions (with zero NUT charge) to the vacuum Einstein equations $(R_{ab}=0)$ 
 of spherical horizon topology in $d$-dimensions \cite{MyersPerry:1986}.  The metric as well as 
its associated thermodynamic quantities, particularly thermodynamic volume, can be obtained as the 
$\Lambda\to 0$ (or $l\to\infty$) limit of the general Kerr-AdS case discussed in previous section.
In Boyer--Lindquist coordinates the metric reads 
\ba \label{metricmp}
ds^2&=&-d\tau ^2+\frac{2m}{U} \Bigl(d\tau -\sum_{i=1}^{N}a_i \mu_i ^2 d\varphi _i\Bigr)^2+\frac{U dr^2}{F-2m}\nonumber\\
&+&\sum_{i=1}^{N} (r^2+a_i^2)\mu_i ^2 d\varphi _i^2+\sum_{i=1}^{N+\varepsilon}(r^2+a_i ^2)d\mu _i ^2\,,
\ea
where as before $\varepsilon=(1,0)$ refers to (even, odd)  spacetime dimensionality, 
\be\label{metrcifunctions}
F=r^{\varepsilon -2}\prod_{i=1}^N (r^2+a_i^2)\,,\quad
U=F\Bigl(1-\sum_{i=1}^N\frac{a_i^2\mu_i^2}{r^2+a_i^2}\Bigr)\,,
\ee
and the thermodynamic quantities are given by \cite{MyersPerry:1986}
\ba \label{TDsMP}
M&=&\frac{m \omega _{d-2}(d-2)}{8\pi}\,,\quad
J_i=\frac{a_i m \omega _{d-2}}{4\pi}\,,\quad \Omega_i=\frac{a_i}{r_+^2+a_i^2}\,,\nonumber\\
A&=&\frac{\omega _{d-2}}{r_+^{1-\varepsilon}}\prod_{i=1}^N
(a_i^2+r_+^2)\,,\quad V=\frac{r_+ A}{d-1}\Bigl(1\!+\!\frac{1}{d-2}\sum_{i=1}^N\frac{a_i^2}{r_+^2}\Bigr)\,,\nonumber\\
T&=&\frac{1}{2\pi }\Bigl(
\sum_{i=1}^{N} \frac{r_+}{a_i^2+r_+^2}-\frac{1}{2^\varepsilon r_+}\Bigr)\,,\quad S=\frac{A}{4}\,.
\ea
The thermodynamic volume   is  the $l\to \infty$ limit of \eqref{VBHKerr}. It inherits the property of obeying the reverse isoperimetric inequality \eqref{ISO},  provided in \eqref{ratio} we
identify ${\cal A}=A$ and ${\cal V}=V$.

\subsection{Five-dimensional case}
\begin{figure}
\begin{center}
\rotatebox{-90}{
\includegraphics[width=0.39\textwidth,height=0.34\textheight]{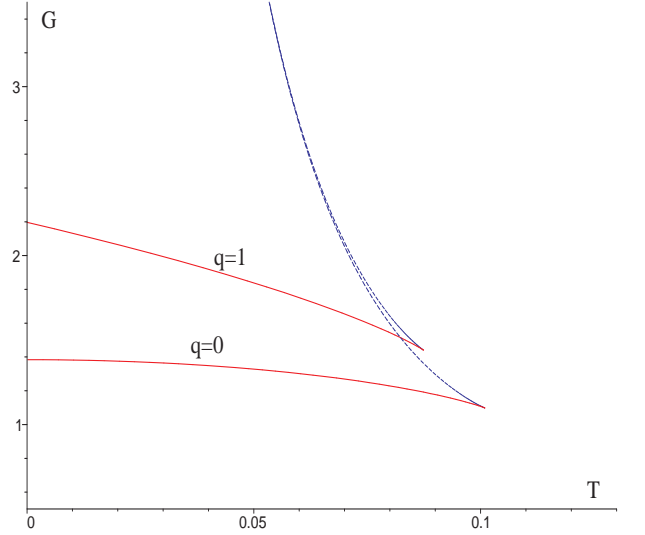}
}
\caption{{\bf Gibbs free energy: $d=5$ MP black holes.}
The lower curve corresponds to the singly-spinning $(q=0)$ MP black hole whereas the upper curve is for equal spinning  ($q=1$) one. 
The qualitative behaviour remains the same irrespective of the value of $q$ and is similar to $d=4$ Kerr case, cf. fig.~\ref{Fig:RNGQfixed}.
}
\label{fig:5dKerr}
\end{center}
\end{figure}
Let us first consider the $d=5$ MP black hole solution, which in general admits
two angular momenta $J_1$ and $J_2$. As with the Kerr-AdS case, we consider the ratio
\be\label{qrat}
q=J_2/J_1\,,
\ee
and set $J_1=1$. As shown in  fig.~\ref{fig:5dKerr}, we find that the Gibbs free energy
shares qualitatively the same behaviour with the $d=4$ Kerr metric for any value of $0\leq q \leq 1$.
Namely, two branches of black holes exist below the critical temperature at the cusp in $G$ and no black holes
are possible above this temperature. The lower, ``fast spinning'' branch of small black holes
corresponds to the global minimum of the Gibbs free energy and possess positive $C_P$.
The upper, ``Schwarzschild-like'', branch of slowly spinning
large black holes is thermodynamically unstable.

\subsection{Reentrant Phase Transition}
\begin{figure*}
\centering
\begin{tabular}{ccc}
\rotatebox{-90}{
\includegraphics[width=0.25\textwidth,height=0.22\textheight]{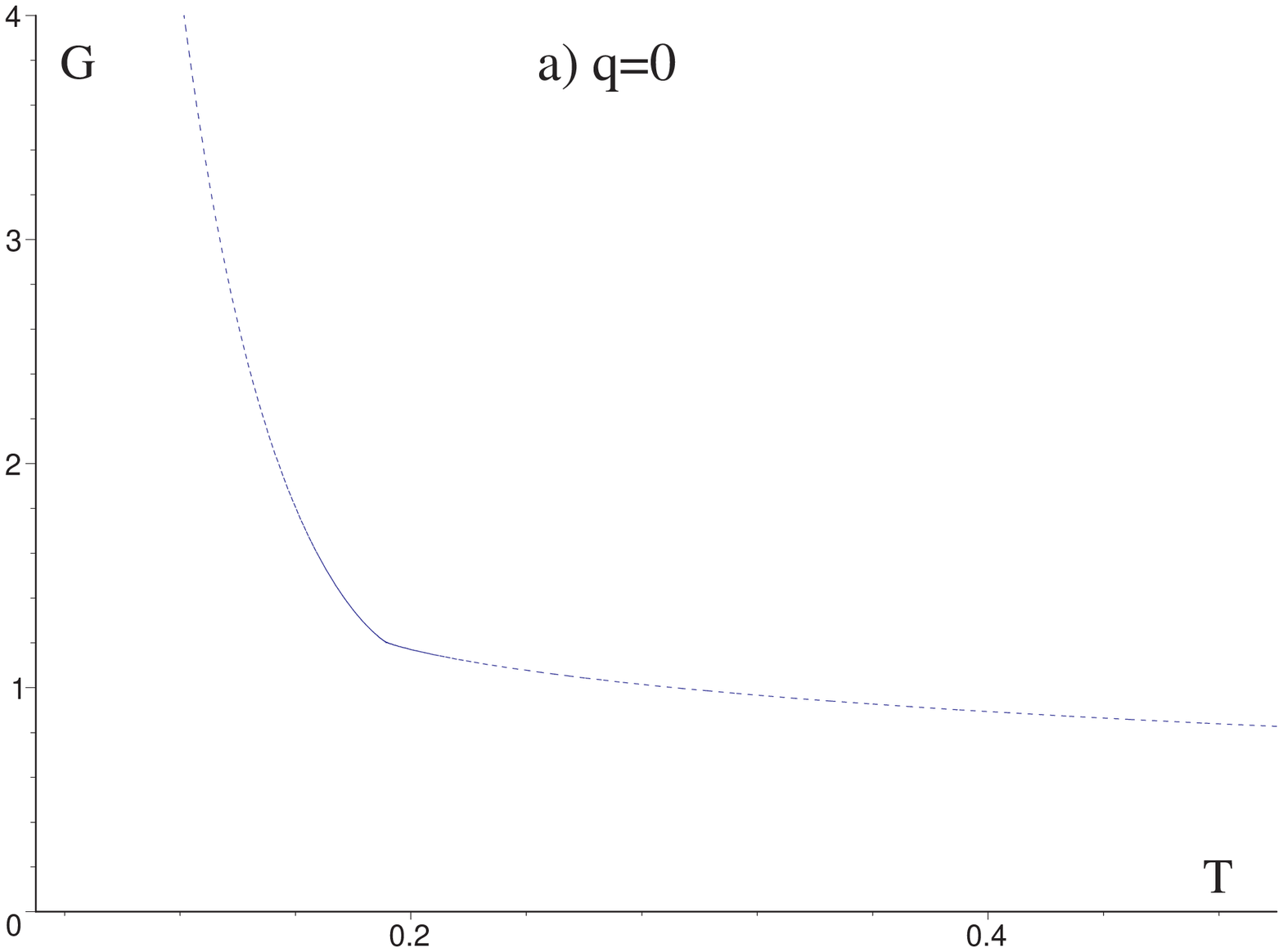}}&
\rotatebox{-90}{
\includegraphics[width=0.25\textwidth,height=0.22\textheight]{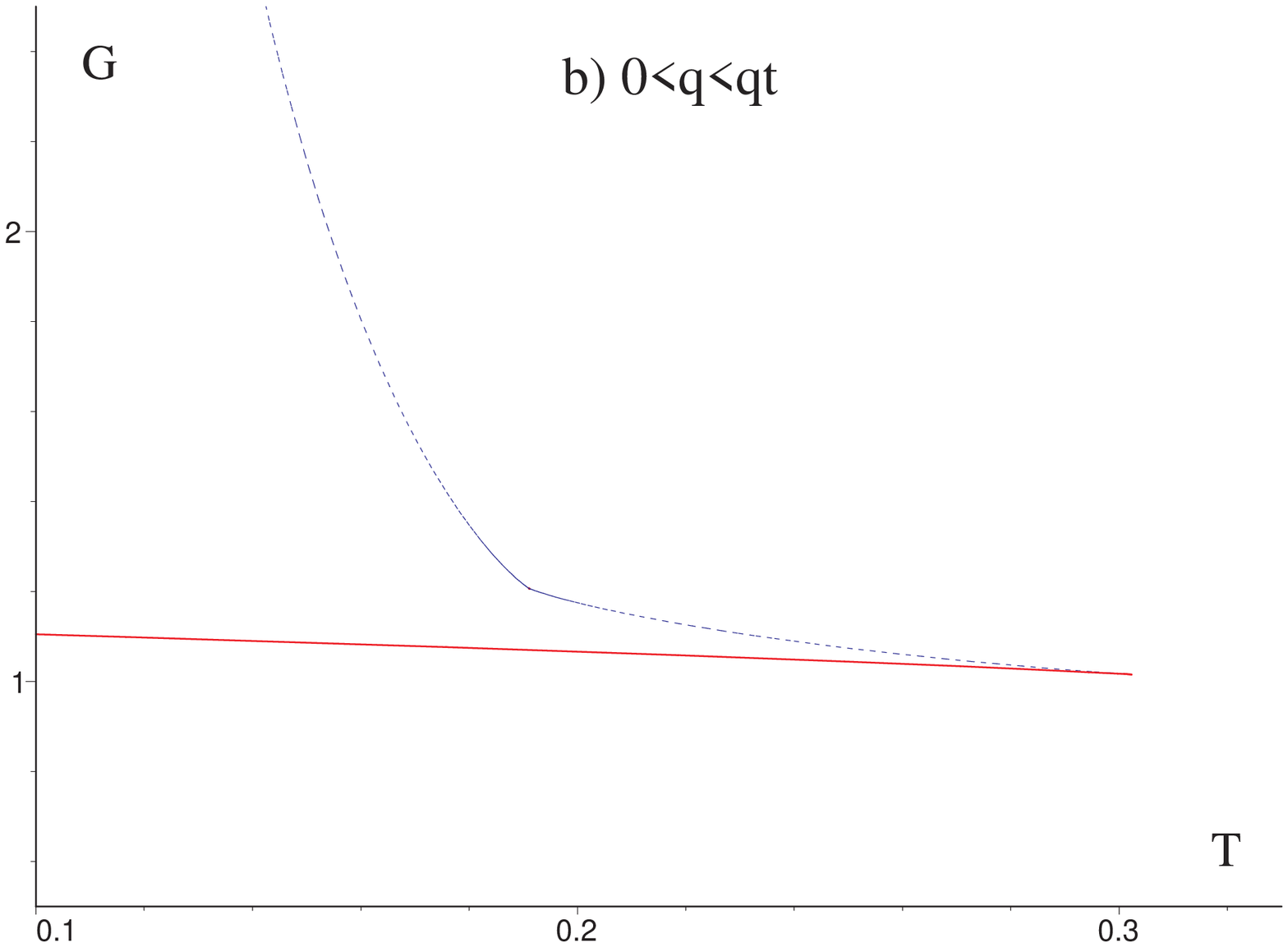}}&
\rotatebox{-90}{
\includegraphics[width=0.25\textwidth,height=0.22\textheight]{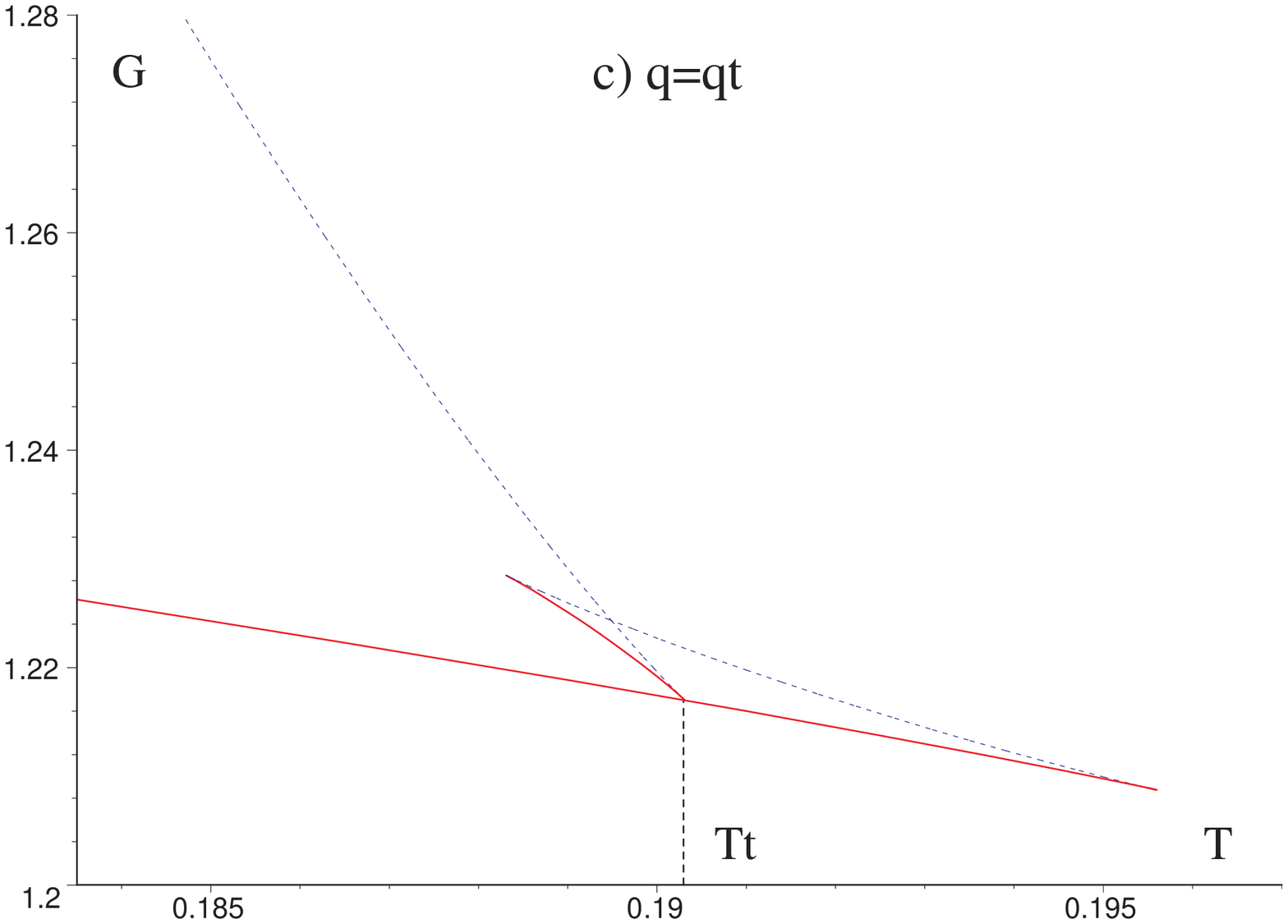}}\\
\rotatebox{-90}{
\includegraphics[width=0.25\textwidth,height=0.25\textheight]{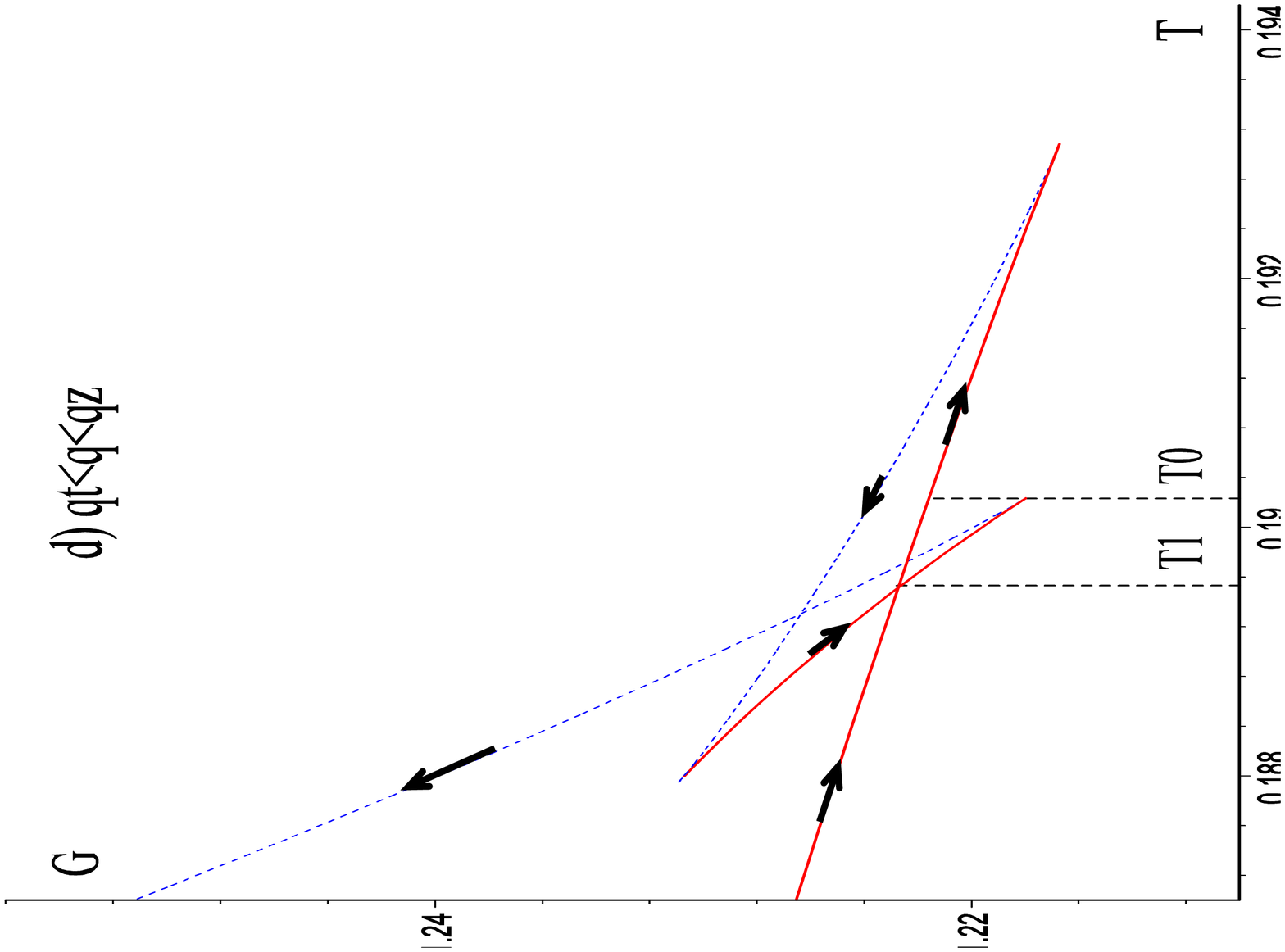}}&
\rotatebox{-90}{
\includegraphics[width=0.25\textwidth,height=0.22\textheight]{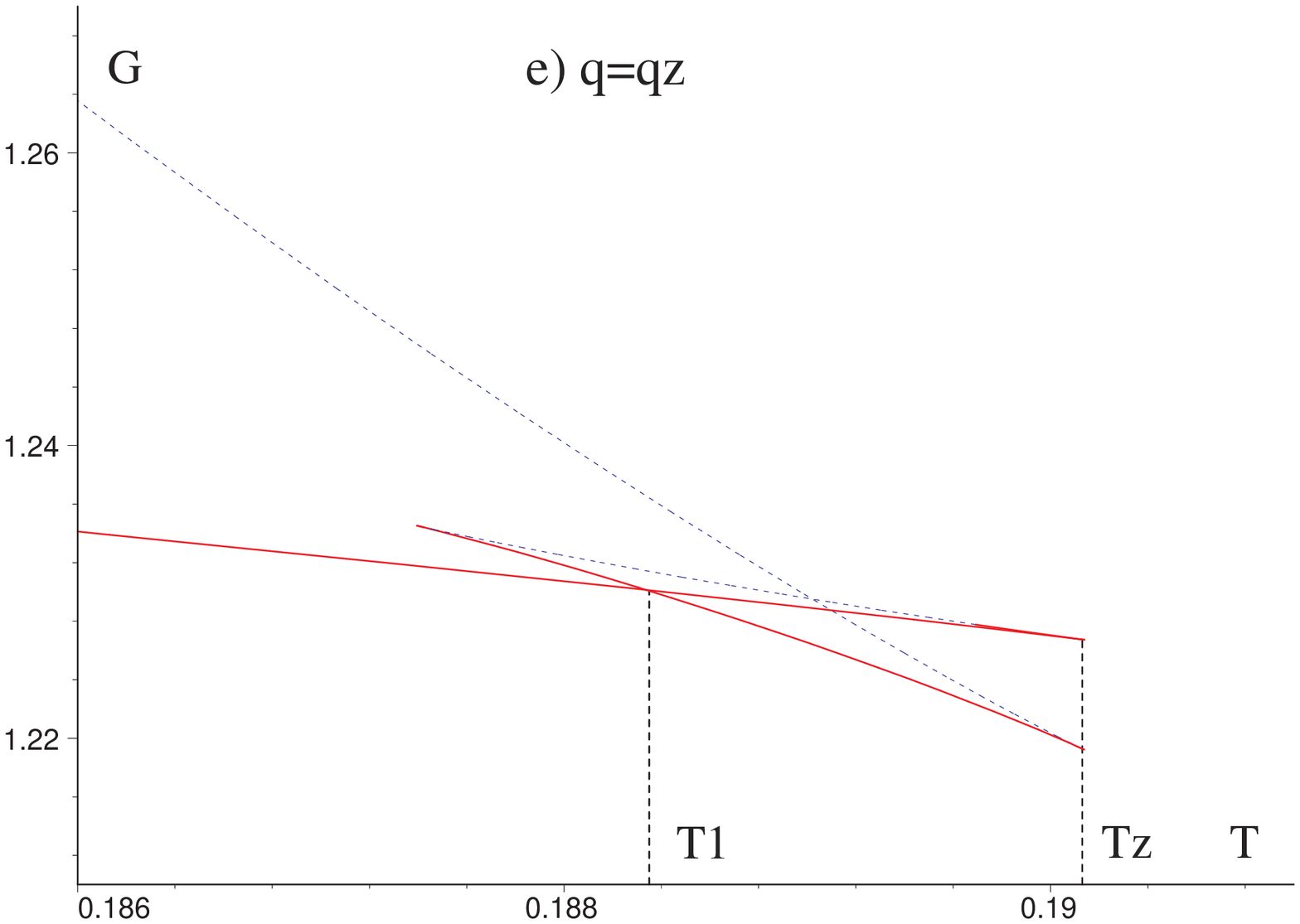}}&
\rotatebox{-90}{
\includegraphics[width=0.25\textwidth,height=0.22\textheight]{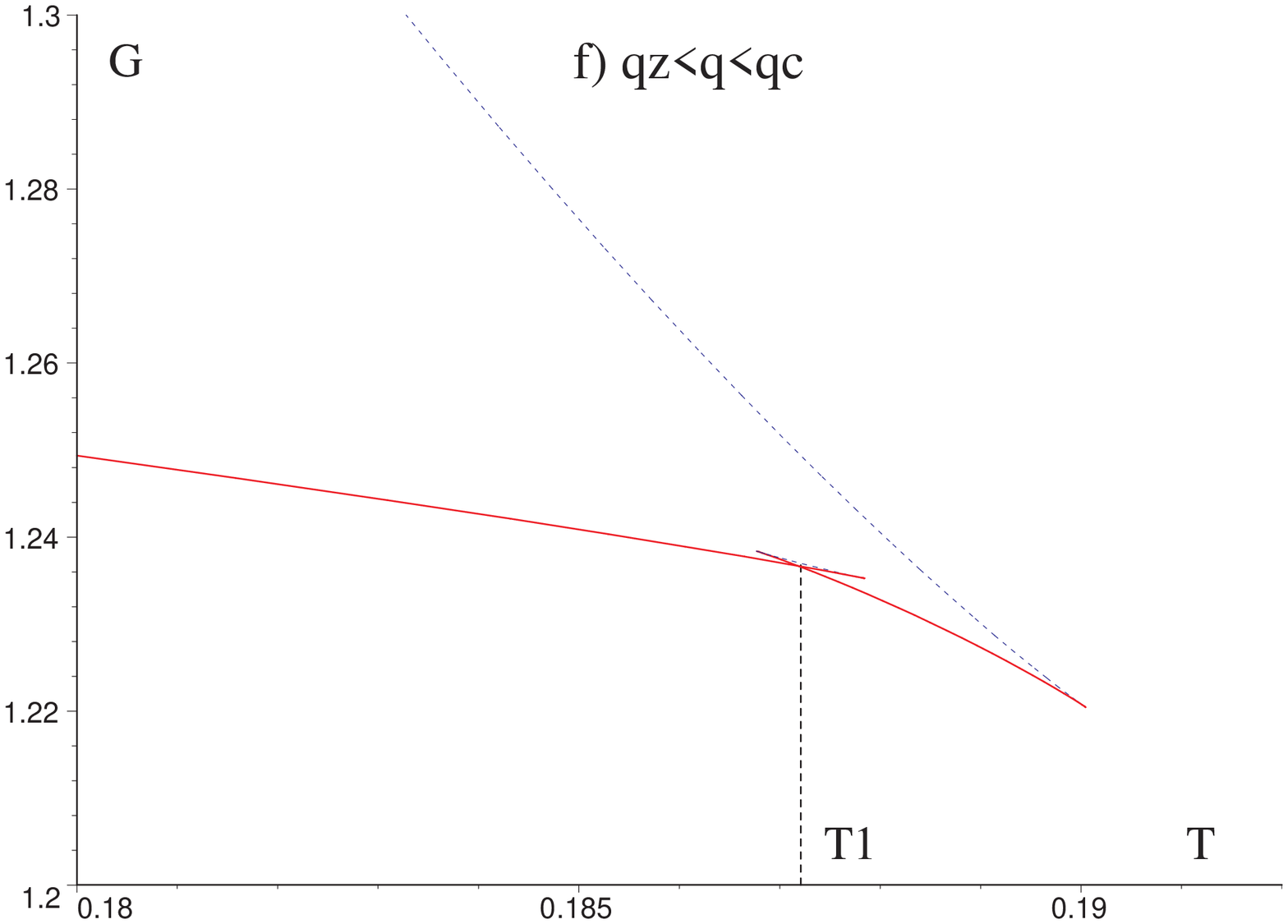}}\\
\rotatebox{-90}{
\includegraphics[width=0.25\textwidth,height=0.22\textheight]{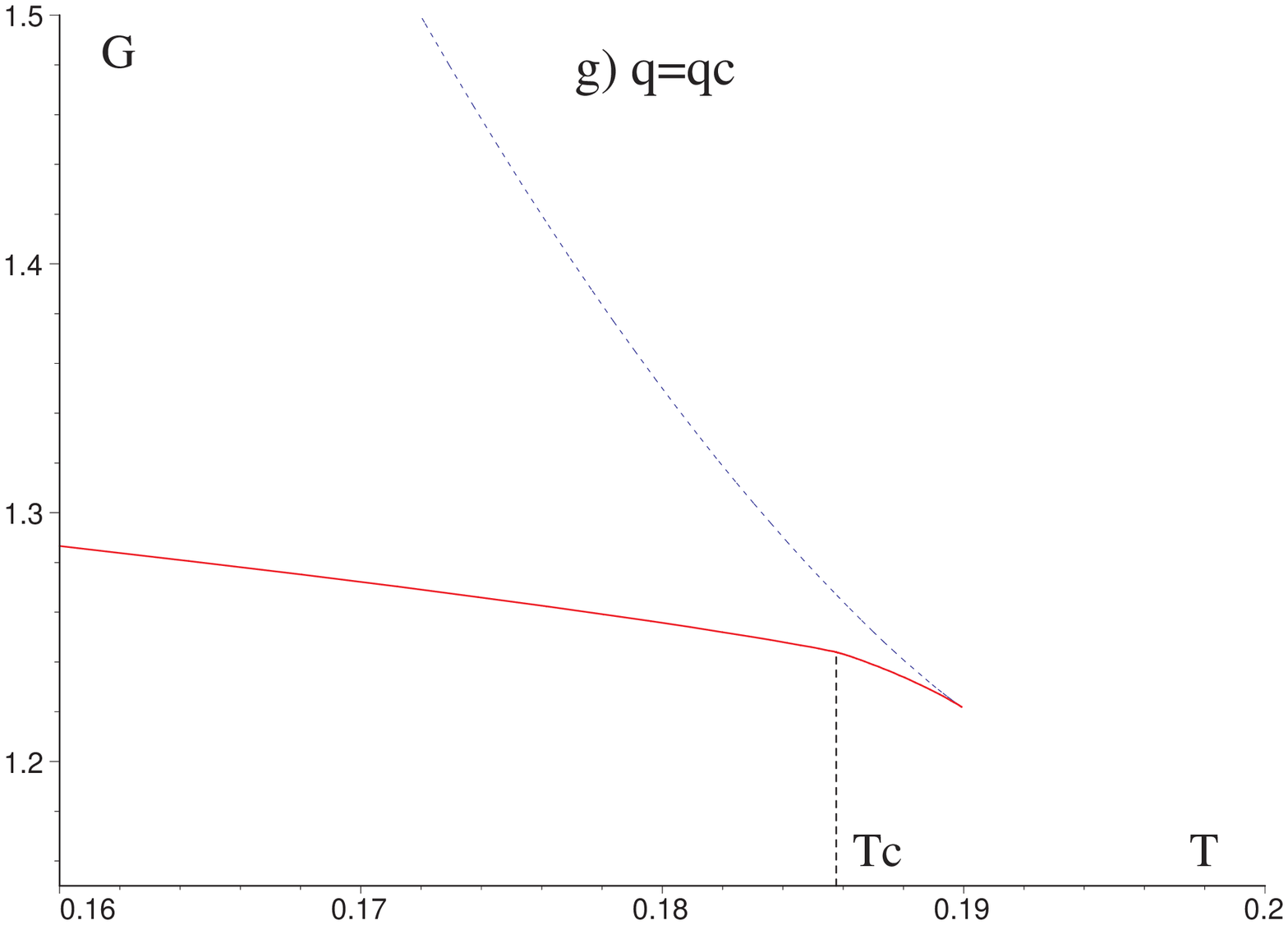}}&
\rotatebox{-90}{
\includegraphics[width=0.25\textwidth,height=0.22\textheight]{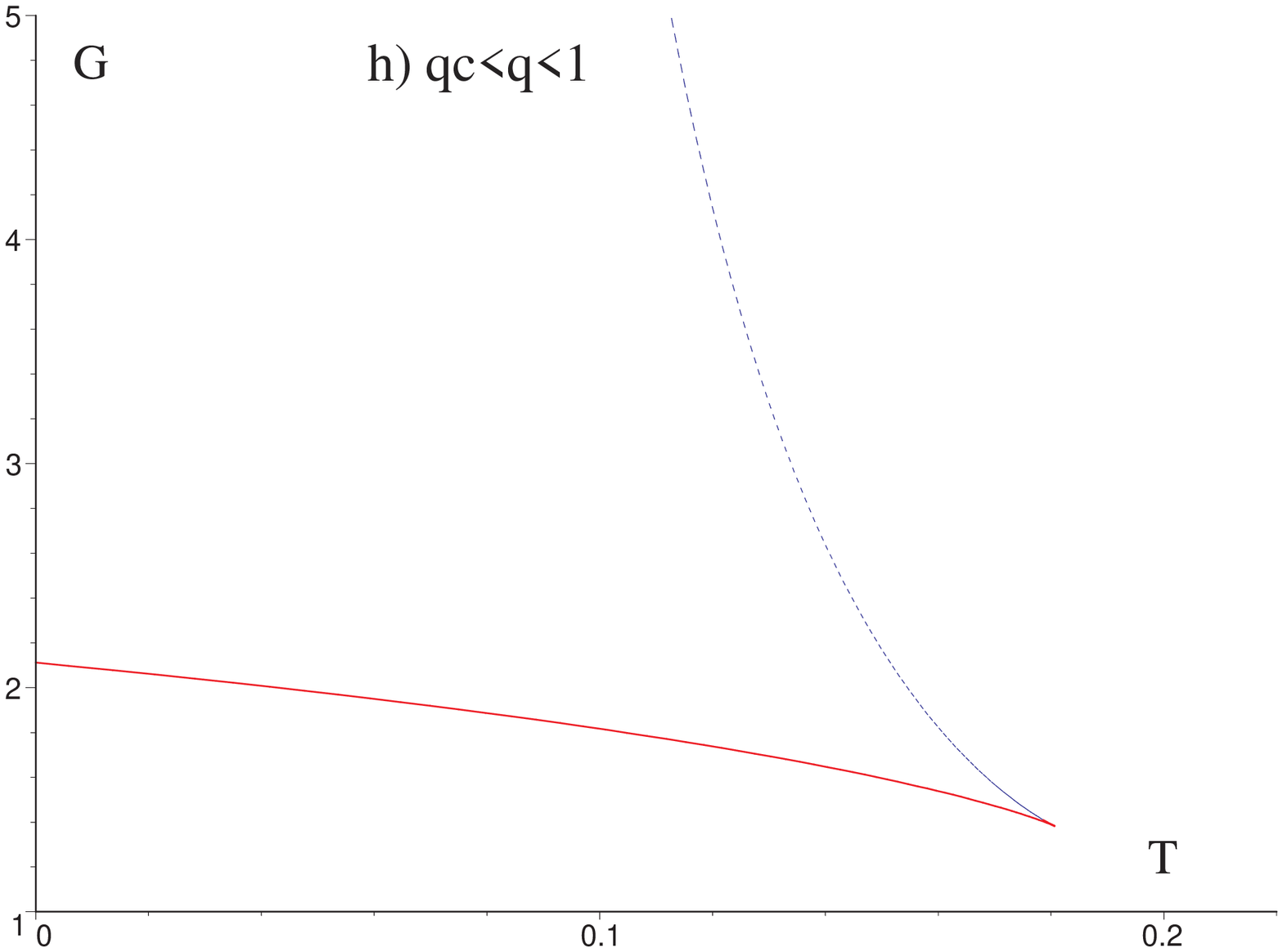}}& 
\rotatebox{-90}{
\includegraphics[width=0.25\textwidth,height=0.22\textheight]{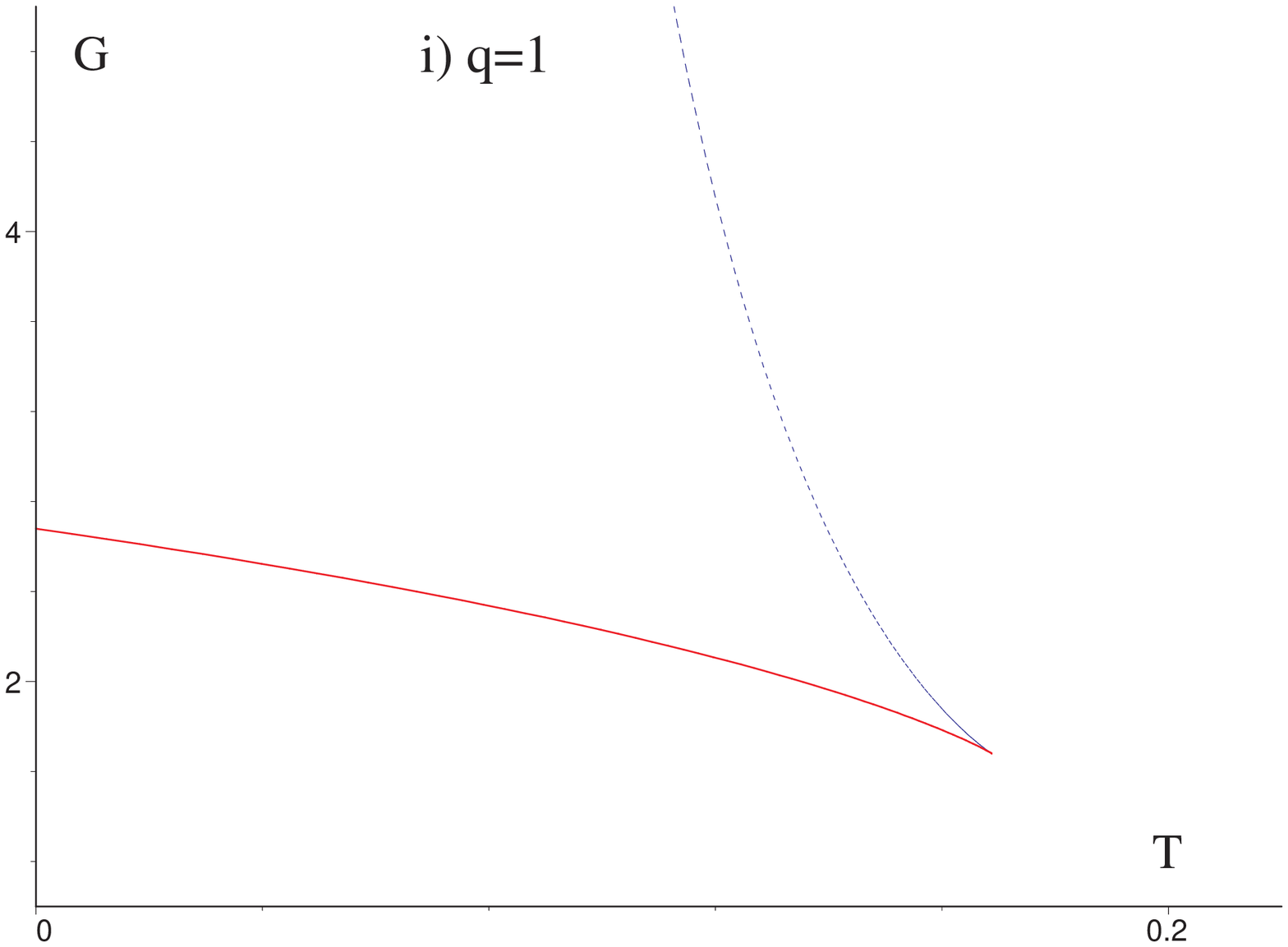}}
\\
\end{tabular}
\caption{{\bf Reentrant phase transition: $d=6$ doubly spinning MP black hole.} The Gibbs free energy is displayed for various ratios of angular momenta $q=J_2/J_1=\{0, 0.05, 0.1082, 0.112, 0.11715, 0.122, 0.1278, 0.5, 1\}$ (from top to bottom). 
Solid-red/dashed-blue lines correspond to $C_P$ positive/negative respectively.
{\em Picture a)} shows the thermodynamically unstable branch of $q=0$ black holes with negative $C_P$; qualitatively similar behaviour occurs for singly spinning MP black holes in all $d\geq 6$. When a second angular momentum is switched on, {\em picture b)}, a new branch of small thermodynamically stable black holes appears, similar to the $d=4$ Kerr case. At $q=q_t\approx  0.1082$, {\em picture c)}, the swallowtail in the upper unstable branch touches the lower stable branch, and we observe the appearance of a reentrant phase transition. {\em Picture d)}
displays  typical behaviour of $G$ when the reentrant phase transition is present, 
$q\in(q_t,q_z)$, $q_z\approx 0.11715$. Black arrows indicate increasing $r_+$. 
If we start increasing the temperature from, say $T=0.186$, the system follows the lower left horizontal solid 
red curve, describing small black holes. At $T=T_1$ it undergoes a first order phase transition and follows the lowest stable branch of large black holes until, at $T=T_0$, this branch terminates. At this point the system `jumps' to the former upper stable branch of small black holes. In other words, we observe an SBH/LBH/SBH reentrant phase transition. 
{\em Picture e)} shows $q=q_z$ at which the reentrant phase transition terminates, while the first order phase transition at $T=T_1$ is still present. 
The first order phase transition continues to occur, {\em pictures f) and g)}, until at $q=q_c\approx 0.1278$ we observe a critical point at $T_c\approx 0.18577$. Above $q_c$ the system displays `Kerr-like' behaviour with no phase transitions.     
}\label{fig:q}
\end{figure*}

Next we consider the $d=6$ MP solution with two angular momenta $J_1$ and $J_2$. As before, 
we consider the ratio \eqref{qrat} and without loss of generality set $J_1=1$.  We now  observe various phenomena reflected in the behaviour of the Gibbs free energy for various values of $q$, as shown in  fig.~\ref{fig:q}. 
For $q=0$ we observe a singly branch of thermodynamically unstable black holes with negative $C_P$; qualitatively same behaviour occurs for singly spinning MP black holes in all $d\geq 6$. For larger values of $q$ more interesting phase behaviour occurs. 

\begin{figure}
\begin{center}
\rotatebox{-90}{
\includegraphics[width=0.39\textwidth,height=0.34\textheight]{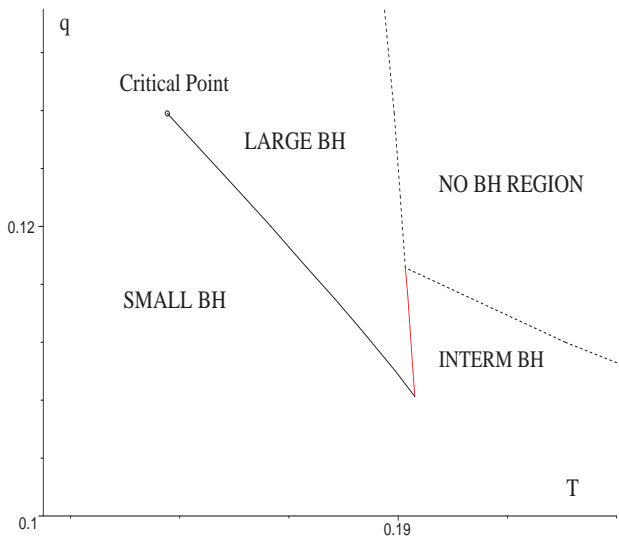}
}
\caption{{\bf $q-T$ phase diagram: $d=6$ MP black holes.}
The first order phase transition between small and large black holes is displayed by solid black curve; it terminates at a critical
point characterized by $(T_c,q_c)$. The solid red curve corresponds to the 0th-order phase transition between large and intermediate (small) black holes; it occurs for $q\in(q_t,q_z)$ and $T\in(T_t,T_z)$. In this range of $q$'s  reentrant phase transition is possible. The `no BH region' is outlined by thin dashed black curve. 
}
\label{fig:qTMP}
\end{center}
\end{figure}

For $0<q<q_t\approx 0.1082$ a new branch of small thermodynamically stable black holes appears, similar to the Kerr case in $d=4$, except there is
a swallowtail in the upper unstable branch of large black holes. In this range of $q$'s the global minimum of $G$ corresponds to small fast spinning black holes that exist up to some maximum temperature $T_{\mbox{\tiny max}}$. Above this temperature no black holes are possible; this `no black hole region' is shown in fig.~\ref{fig:qTMP}. 
At $q=q_t\approx  0.1082$,  fig.~\ref{fig:q} c), the swallowtail in the upper unstable branch reaches the lower stable branch and we observe the entrance of reentrant phase transition. 

The typical behaviour of $G$ for $q\in(q_t,q_z)$ is displayed in fig.~\ref{fig:q} d) and indicates the presence of the  reentrant phase transition. 
Following this figure, if we start increasing the temperature from, say $T=0.186$, the system follows the lower left horizontal solid 
red curve, describing small black holes. At $T=T_1$ it undergoes a first order phase transition and follows the lowest stable branch of large black holes until, at $T=T_0$, there is a cusp in $G$ and this branch terminates. Beyond the cusp, as the temperature increases even further, there is a discontinuity in the global minimum of $G$, yielding a  zeroth-order phase transition to the upper red branch of small black holes.
In other words, we observe a SBH/LBH/SBH reentrant phase transition. This reentrant phase transition, however, stands in contrast to the situation in the Kerr-AdS case discussed in the previous section, cf. graphs \ref{Fig:6DKerrZorder}, \ref{Fig:PTZeroth1}, \ref{Fig:PTZerothJ-14}. In the Kerr-AdS case the black holes undergoing the reentrant phase transition were singly spinning, the phase transition was of the LBH/SBH/LBH type with $T_0<T_1$ and the `no black hole region' occurred for small temperatures, whereas in the MP case the  black holes must be doubly spinning, the  phase transition is of the SBH/LBH/SBH type with $T_1<T_0$, and the `no black hole region' occurs for high temperatures.

The reentrant phase transition terminates at $q=q_z\approx 0.11715$ as shown in fig.~\ref{fig:q} e). In between $q_z$ and $q_c\approx 0.1278$ there is a first order SBH/LBH phase transition illustrated in fig.~\ref{fig:q} f). This transition terminates at a critical point at $T_c\approx 0.18577$, fig.~\ref{fig:q} g). Above $q_c$, figs. h) and i), the system displays `Kerr-like' behaviour with no phase transitions. 
The overall  situation is summarized in the $q-T$ phase diagram in fig.~\ref{fig:qTMP}.   

The important lesson of this section is that we showed that the reentrant phase transition  
can take place in black hole spacetimes without the cosmological constant, that is, when the system has zero pressure. A gauge/gravity interpretation of this phenomenon, similar to the asymptotically AdS case, remains to be explored.

\section{Five-dimensional black rings and black saturns}\label{sec:5d}

In this section we look at the thermodynamics of other analytically known $d=5$ black hole objects with
horizon topologies more complicated than spherical, specifically  singly spinning black rings and black saturns. Unfortunately, due to the lack of a generating technique for $\Lambda\neq 0$, only vacuum solutions are known. For this reason, there is no effective thermodynamic way to calculate the thermodynamic volume and the fluid equation of state is trivial; we limit ourselves to studying the Gibbs free energy for these objects and comparing it to the Gibbs free energy of the $d=5$ MP solution. In the next section we shall try to circumvent the lack of knowledge of exact asymptotically AdS black ring solutions by considering the approximate thin AdS black rings constructed via the blackfold approach. In the given (fast spinning approximation) this will allow us to calculate the thermodynamic volume, verify the reverse isoperimetric inequality, and construct the corresponding equation of state. 

\subsection{Singly Spinning Black Ring}
\begin{figure}
\begin{center}
\rotatebox{-90}{
\includegraphics[width=0.39\textwidth,height=0.34\textheight]{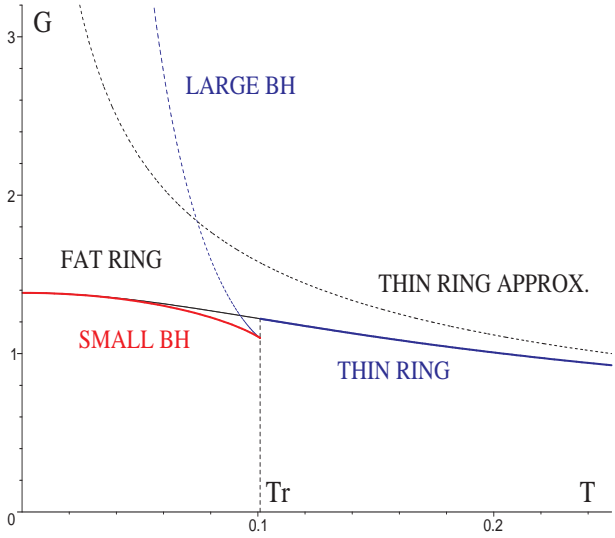}
}
\caption{{\bf Gibbs free energy: black ring vs. black hole.}
The Gibbs free energy of black rings is compared to the
Gibbs free energy of an MP black hole for $J=1$.
For temperatures below $T_r$ we observe three branches of black holes: {\em small MP} black holes displayed by thick solid red curve have positive $C_P$ and the lowest Gibbs free energy, {\em fat black rings} (depicted by the thick black solid curve slightly above the red curve) have positive $C_P$ and are locally thermodynamically stable, and {\em large MP} black holes (upper branch depicted by the thin blue dashed curve) with negative $C_P$. 
Above $T_r$ black holes are no longer possible ($G$ has a cusp there) and only a branch of thin black rings (displayed by a thick solid blue curve) with $C_P< 0$ exists. The approximate thin black ring solution, constructed using the blackfold approach in the next section, is displayed by the thin dashed black curve. We observe that the approximation works well for high temperatures, whereas it is completely off for low temperatures. We also note that the exact Gibbs free energy is actually lower than the one given by the blackfold approximation.  
}
\label{fig:GRing}
\end{center}
\end{figure}

The exact solution for a singly spinning black ring in five dimensions was constructed by Emparan and Reall \cite{EmparanReall:2002a} and reads 
\ba
ds^2&=&-\frac{F(y)}{F(x)}\Bigl(dt-R\sqrt{\lambda(\lambda-\nu)\frac{1+\lambda}{1-\lambda}}\frac{1+y}{F(y)}d\phi_2\Bigr)^2\nonumber\\
&+&\frac{R^2F(x)}{(x-y)^2}\Bigl(\frac{dx^2}{G(x)}+\frac{G(x)}{F(x)}d\phi_1^2
-\frac{G(y)}{F(y)}d\phi_2^2-\frac{dy^2}{G(y)}\Bigr)\,,\nonumber\\
\ea
where
\be
F(\xi)=1+\lambda\xi\,,\quad G(\xi)=(1-\xi^2)(1+\nu\xi)\,.
\ee
The dimensionful constant $R$ sets a scale for the solution. The absence of conical singularities fixes $\lambda$ and the periodicity 
of $\phi_1$ and $\phi_2$,
\be\label{brcon}
\lambda=\frac{2\nu}{1+\nu^2}\,,\quad 0\leq \phi_1,\phi_2\leq 2\pi \frac{\sqrt{1-\lambda}}{1-\nu}\,,
\ee
while the remaining dimensionless parameter
$\nu$ lies in the range $\nu\in(0,1)$. The coordinates $x$ and $y$ are restricted to the ranges
$-1\leq x\leq 1$ and $-1/\nu\leq y<-1$; the event horizon is located at $y=-1/\nu$.

The thermodynamic quantities are given by
\ba\label{RdRingExact}
M&=&\frac{3\pi R^2\nu}{2(1-\nu)(1+\nu^2)}\,,\nonumber\\
J&=&\frac{\pi \nu R^3}{\sqrt{2}}\!\left(\frac{1+\nu}{(1-\nu)(1+\nu^2)}\right)^{3/2}\!\!\!\!\!\!\,,\quad 
\Omega=\frac{1}{R}\sqrt{\frac{\lambda-\nu}{\lambda(1+\lambda)}}\,,
\nonumber\\
T&=&\frac{(1-\nu)\sqrt{1+\nu^2}}{4\sqrt{2}\pi \nu R}\,,\quad 
S=\frac{2\sqrt{2}\pi^2R^3\nu^2}{(1-\nu)(1+\nu^2)^{3/2}}=\frac{A}{4}\,,
\nonumber\\
C_P&=&\frac{12\sqrt{2}\pi^2(\nu-1/2)\nu^2\sqrt{1+\nu^2}}{(1-\nu)(2+\nu^2)(1+\nu^2)^2}R^3\,,
\ea
and using \eqref{brcon} it is straightforward to show that the Smarr relation \eqref{Smarr} holds.
From the behaviour of the heat capacity we see there is a clear distinction between
{\em `fat' black rings} with $\nu>1/2$ and $C_P >0$,  and
{\em `thin' black rings} with $\nu<1/2$ and $C_P<0$.

It is not by any means obvious that the two kinds of black objects, the spherical MP black hole and the toroidal black ring, may be treated as ``two phases'' of the same system. Notwithstanding this concern, we plot the Gibbs free energy for the black ring
\be
G=M-TS=\frac{\pi R^2\nu(2+\nu)}{2(1+\nu^2)(1-\nu)}
\ee
parametrically in  fig.~\ref{fig:GRing}, for $J=1$, comparing it to  
the Gibbs free energy of the $d=5$ singly spinning asymptotically flat MP black hole.
We observe the following interesting behaviour. 
There exists a special temperature $T_r$ at which the branch of fat ($\nu>1/2)$ black rings 
terminates and is smoothly joined by the branch of thin black rings. 
At exactly the same temperature, $G$ of MP black holes has a cusp. 
For temperatures below $T_r$ we observe three branches of black holes: {\em small MP} black holes with positive $C_P$ and the lowest Gibbs free energy, {\em fat black rings} with slightly higher Gibbs free energy and positive specific heat,  and {\em large MP} black holes with the highest $G$ and negative $C_P$, cf. $q=0$ curves in fig.~\ref{fig:5dKerr}. 
Above $T_r$ black holes are no longer possible and only a branch of thin black rings with negative $C_P$ exists.
We remark that very thin black rings are expected to be classically unstable due to the Gregory--Laflamme instability.

Our considerations  imply that spherical black holes are thermodynamically preferable
to fat black rings. This is supported by the study of the horizon areas, e.g., \cite{EmparanReall:2002a, Emparan:2004wy, Emparan:2006mm, ElvangEtal:2007,Emparan:2007wm,Emparan:2010sx}. It is an
open question as to whether there actually might be a phase transition from fat black rings to small MP black holes. Likewise one can ask if  all  thin rings, besides being thermodynamically unstable, are also classically unstable, see \cite{Elvang:2006dd}.  If so then are the small MP solutions the only stable vacuum black holes in five dimensions?

\subsection{Black Saturn}
This exact multi-black hole solution was constructed by Elvang and Figueras \cite{ElvangFigueras:2007} and its `phases' and the first law were studied in \cite{ElvangEtal:2007}.
The solution consists of a central rotating spherical black hole surrounded by a black ring. The solution depends on one dimensionful parameter $L$, and the dimensionless parameters $0\leq \kappa_3\leq \kappa_2<\kappa_1\leq 1$. The thermodynamic quantities associated with  the two horizons are (using an abbreviation $\hat \kappa_i=1-\kappa_i$) 
\ba
\Omega_{BH}&=&\frac{1\!+\!\kappa_2c}{L}\sqrt{\frac{\kappa_2\kappa_3}{2\kappa_1}}
\frac{\kappa_3\hat \kappa_1-\kappa_1\hat \kappa_2\hat \kappa_3c}
{\kappa_3\hat \kappa_1+\kappa_1\kappa_2\hat \kappa_2\hat \kappa_3c^2}\,,\nonumber\\
\Omega_{BR}&=&\frac{1\!+\!\kappa_2c}{L}\sqrt{\frac{\kappa_1\kappa_3}{2\kappa_2}}
\frac{\kappa_3-\kappa_2\hat \kappa_3c}
{\kappa_3-\kappa_3(\kappa_1-\kappa_2)c+\kappa_1\kappa_2\hat \kappa_3c^2}\,,\nonumber\\
S_{BH}&=&\frac{\pi^2L^3}{(1+\kappa_2 c)^2}\sqrt{\frac{2\hat \kappa_1^3}{\hat \kappa_2\hat \kappa_3}}
\Bigl(1+\frac{\kappa_1\kappa_2\hat \kappa_2\hat \kappa_3 c^2}{\kappa_3 \hat \kappa_1}\Bigr)\,,\nonumber\\
S_{BR}&=&\frac{\pi^2L^3}{(1+\kappa_2 c)^2}\sqrt{\frac{2\kappa_2(\kappa_2-\kappa_3)^3}{\kappa_1(\kappa_1-\kappa_3)\hat \kappa_3}}\times\nonumber\\
&&\times \Bigl(1-(\kappa_1-\kappa_2)c+\frac{\kappa_1\kappa_2\hat \kappa_3c^2}{\kappa_3} \Bigr)\,,\\
T_{BH}&=&\frac{1}{2\pi L}\sqrt{\frac{\hat \kappa_2\hat \kappa_3}{2\hat \kappa_1}}
\frac{(1+\kappa_2c)^2}{1+\frac{\kappa_1\kappa_2\hat \kappa_2\hat \kappa_3}{\kappa_3\hat \kappa_1}c^2}\,,\nonumber\\
T_{BR}&=&\frac{1}{2\pi L}\sqrt{\frac{\kappa_1\hat \kappa_3(\kappa_1-\kappa_3)}{2\kappa_2(\kappa_2-\kappa_3)}}
\frac{(1+\kappa_2c)^2}{1-(\kappa_1-\kappa_2)c+\frac{\kappa_1\kappa_2\hat \kappa_3}{\kappa_3}c^2}\,,\nonumber
\ea
whereas the ADM mass $M$ and angular momentum $J$ are
\ba
M&=&\frac{3\pi L^2}{4\kappa_3(1+\kappa_2c)^2}\Bigl(\kappa_3(\hat \kappa_1+\kappa_2)
-2\kappa_2\kappa_3(\kappa_1-\kappa_2)c\nonumber\\
&&+\kappa_2\bigl[\kappa_1-\kappa_2\kappa_3(\hat \kappa_2+\kappa_1)\bigr]c^2\Bigr)\,,\nonumber\\
J_{BH} &=&  -\frac{\pi L^3 c}{\kappa_3(1+\kappa_2c)^2}\sqrt{\frac{\kappa_1\kappa_2}{2\kappa_3}}
\left(\kappa_3\hat \kappa_1+\kappa_1\kappa_2\hat \kappa_2\hat \kappa_3c^2\right)\,,
\nonumber\\
J_{BR} &=& \frac{\pi L^3}{\kappa_3(1\!+\!\kappa_2c)^3}\sqrt{\!\frac{\kappa_2}{2\kappa_1\kappa_3}}
\bigl[\kappa_3\!-\!\kappa_3(\kappa_1\!-\!\kappa_2)c\!+\!\kappa_1\kappa_2\hat \kappa_3c^2\bigr] \nonumber\\
&&\times \bigl[\kappa_3-\kappa_2(\kappa_1-\kappa_3)c+\kappa_1\kappa_2\hat \kappa_2 c^2\bigr]\,,\nonumber\\
J&=&J_{BR}+J_{BH} = \frac{\pi L^3}{\kappa_3(1+\kappa_2c)^2}\sqrt{\frac{\kappa_2}{2\kappa_1\kappa_3}}\nonumber\\
&&\times \Bigl(\kappa_3^2-c\kappa_3\bigl[(\kappa_1-\kappa_2)(\hat \kappa_1+\kappa_3)+\kappa_2\hat \kappa_3\bigl]\nonumber\\
&&+c^2\kappa_2\kappa_3\bigl[(\kappa_1-\kappa_2)(\kappa_1-\kappa_3)+\kappa_1(\hat \kappa_2+\kappa_1-\kappa_3)\bigr]\nonumber\\
&&-c^3\kappa_1\kappa_2\bigl[\kappa_1-\kappa_2\kappa_3(\kappa_1-\hat \kappa_2-\hat \kappa_3)\bigr]\Bigr)\,,
\ea
and a generalization of the Smarr formula \eqref{Smarr}\,, 
\be
\frac{3}{2} M =  T_{BR}S_{BR}+ T_{BH}S_{BH} +\Omega_{BR}J_{BR}+ \Omega_{BH}J_{BH}
\ee
is again satisfied \cite{ElvangFigueras:2007}.
The absence of conical singularities requires
\be
c=\frac{1}{\kappa_2}\Bigl(\varepsilon \frac{\kappa_1-\kappa_2}{\sqrt{\kappa_1\hat \kappa_2\hat \kappa_3(\kappa_1-\kappa_3)}}-1\Bigr)\,,
\ee
with $\varepsilon=+1$ when $c>-1/\kappa_2$ and $\varepsilon=-1$ when $c<-1/\kappa_2$;
$c=-1/\kappa_2$ is nakedly singular.  

\begin{figure}
\begin{center}
\rotatebox{-90}{
\includegraphics[width=0.39\textwidth,height=0.34\textheight]{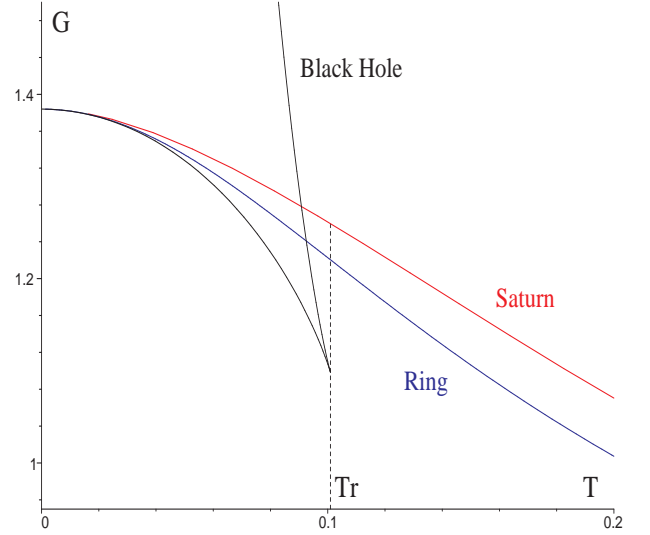}
}
\caption{{\bf Gibbs free energy: $5d$ vacuum black holes.}
The Gibbs free energy of black saturn in thermodynamic and mechanical equilibrium (red curve) is compared to the Gibbs free energy of black ring (blue curve) and the MP spherical black hole (black curve). For temperatures below $T_r$, all three branches are possible, however, the spherical black hole branch globally minimizes the Gibbs free energy.   We have set $J=1$.
}
\label{fig:GSaturn}
\end{center}
\end{figure}
To ensure thermodynamic and mechanical equilibrium, we require
\be
T\equiv T_{BH}=T_{BR}\,,\quad \Omega\equiv \Omega_{BH}=\Omega_{BR}\,.
\ee
This eliminates two of the dimensionless parameters, leaving one; let us say $\kappa_2$. In this case the system can be assigned a ``total'' Gibbs free energy
\be
G(T,J)=M-T(S_{BH}+S_{BR})\,.
\ee
As with the black ring, this can be plotted parametrically. We  display the result in fig.~\ref{fig:GSaturn}, where it is compared to the Gibbs free energy of an MP black hole and black ring. Similar to the discussion in the previous subsection, when treated as different phases of one thermodynamic system, the spherical black hole phase is preferable in the range of temperatures $T\in(0,T_r)$ whereas the stability of the other black objects at higher temperatures is not known.

\section{Thin black rings in AdS}\label{sec:rings}
\subsection{Review of the construction
}
Following \cite{CaldarelliEtal:2008} we briefly recapitulate the perturbative
construction of an asymptotically AdS $d\geq 5$ singly spinning thin black ring with horizon topology $S^1\times S^{d-3}$ using the 
blackfold approach \cite{Emparan:2007wm, Emparan:2009at}.
Such a ring is characterized by the two radii $R$, corresponding to $S^1$, and $r_0$, associated with $S^{d-3}$.
The idea of the construction is to ``bend'' a straight thin boosted black string of width $r_0$ into a circular shape of radius $R$. In the far region the ring is described by a distributional source of energy-momentum of a boosted string in the given (global) AdS$_d$ background
\ba
ds^2&=&-fd\tau^2\!+\!\frac{d\rho^2}{f}
\!+\!\rho^2(d\theta^2\!+\!\sin^2\!\theta d\Omega_{d-4}^2\!+\!\cos^2\!\theta d\psi^2)\,,\nonumber\\
f&=&1+\frac{\rho^2}{l^2}\,,
\ea
placed at $\rho=R$ on the $\theta=0$ plane.
The magnitude of the boost of the string is determined from the equilibrium condition \cite{Carter:2000wv}  and provides the correct centrifugal force
to balance the ring tension and the AdS gravitational potential. Once the correct magnitude of the boost is known,\footnote{Due to the presence of the cosmological constant the boost increases with the radius $R$ of the ring and approaches the speed of light in the limit $R\to\infty$.} one can calculate the asymptotic charges using the energy momentum distribution. This  gives the mass $M$ and the angular momentum $J$ of the ring. On the other hand the properties of the horizon, such as its area, angular velocity, or temperature, are determined from the characteristics of the boosted black string, as we shall point out below.
Once constructed, the solution is valid provided the following approximation holds:
\be\label{approx}
r_0<<\mbox{min}(R,l)\,.
\ee
This condition will play an important role in our discussion of the equation of state for these rings.
Note that there is no restriction on the relation between $R$ and $l$ and so both large and small (with respect to the AdS radius) black rings
are possible.

\begin{figure}
\begin{center}
\rotatebox{-90}{
\includegraphics[width=0.39\textwidth,height=0.34\textheight]{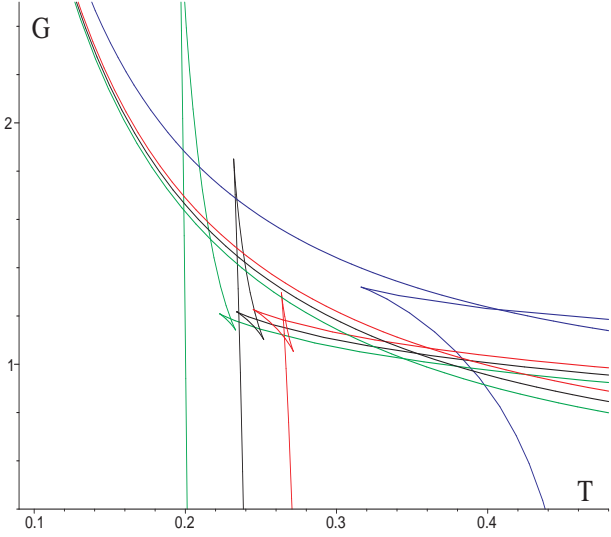}
}
\caption{{\bf Gibbs free energy: thin AdS black rings vs. Kerr-AdS black holes in $d=6$.}
The Gibbs free energy of thin black rings in the blackfold approximation is displayed for $P=\{0.2, 0.073, 0.0564, 0.04\}$ (the upper blue, red, black, and green curves that emerge from the left top corner) and is compared to the corresponding 
exact Gibbs free energy of Kerr-AdS black holes for the same pressures, cf. fig.~\ref{Fig:6DKerr}. For large temperatures the branch of large Kerr-AdS black holes is thermodynamically preferred, whereas in the regime of small temperatures we cannot trust our blackfold approximation. 
}
\label{Fig:6DKerrRingG}
\end{center}
\end{figure}

\subsection{Thermodynamics}
In terms of the dimensionless parameter
\be
\mathsf{R}=\frac{R}{l}\,,
\ee
the thermodynamic quantities associated with the perturbative solution for a thin black ring in AdS are \cite{CaldarelliEtal:2008}
\begin{eqnarray}\label{TdSBRAdS}
M&=&\frac{r_{0}^{d-4}l}{8}\omega _{d-3}(d-2)\mathsf{R}\left(1+\mathsf{R}^2\right)^{\frac{3}{2}}\,, \nonumber\\
J&=&\frac{\omega _{d-3}r_{0}^{d-4}l^2 \mathsf{R}^2}{8}\sqrt{\bigl(1+(d\!-\!2)\mathsf{R}^2\bigr)\bigl(d\!-\!3+(d\!-\!2)\mathsf{R}^2\bigr)}\,, \nonumber\\
S&=&\frac{1}{2}\pi lr_{0}^{d-3}\omega _{d-3}\mathsf{R} \sqrt{\frac{d-3+(d-2)\mathsf{R}^2}{d-4}}=\frac{A}{4}\,, \nonumber \qquad\\
{\Omega}&=&\frac{1}{l}\sqrt{\frac{(1+\mathsf{R}^2)(1+(d-2)\mathsf{R}^2)}{\mathsf{R}^2(d-3+(d-2)\mathsf{R}^2)}}\,,\nonumber \\
T &=& \frac{(d-4)^{\frac{3}{2}}\sqrt{1+\mathsf{R}^2}}{4\pi r_{0}\sqrt{d-3+(d-2)\mathsf{R}^2}}\,.
\end{eqnarray}
The thermodynamic quantities of asymptotically flat thin black rings in higher dimensions \cite{Emparan:2007wm} are obtained from these expressions by setting $l\to \infty$.

To plot the Gibbs free energy 
\be
G=\frac{r_{0}^{d-4}l \mathsf{R}}{8}\omega _{d-3}\sqrt{1+\mathsf{R}^2}
\Bigl[2+(d-2)\mathsf{R}^2\Bigr]\,,
\ee
for fixed $J$ and $P$  we eliminate $r_0$ from the $J$ equation, 
\ba\label{r0}
r_0&=&\Bigl[\frac{128 \pi}{(d-1)(d-2)\omega_{d-3}}\frac{P}{\mathsf{R}^2}\frac{J}{\sqrt{1+(d-2)\mathsf{R}^2}}\nonumber\\
&&\quad\times \frac{1}{\sqrt{d-3 +(d-2)\mathsf{R}^2}}\Bigr]^{\frac{1}{d-4}}\,,
\ea
to obtain $G$ and $T$ as functions of $R$, and plot parametrically.
The behaviour of $G$ for a thin AdS black ring in $d=6$ is illustrated in 
fig.~\ref{Fig:6DKerrRingG} where it is compared to the corresponding Gibbs free energy of a singly spinning Kerr-AdS black hole.
We observe that in the region of high temperature, where the blackfold approximation is valid, large Kerr-AdS black holes are thermodynamically preferred to thin AdS black rings. In the regime of small temperatures we cannot trust the blackfold approximation and the corresponding curve is unphysical.
The Gibbs free energy of asymptotically flat thin rings is discussed in the next section, where it is compared to the Gibbs free energy of an MP black hole, leading to a more interesting behaviour and a possible phase transition between black holes and black rings as shown in fig.~\ref{Fig:5DKerrRingG1}.

The specific heat of AdS black rings can be analyzed in the same way as for black holes. We find that in a given approximation 
$C_P$ is always negative. Hence higher-dimensional AdS thin black rings in this (fast spinning) approximation are thermodynamically unstable. 
We expect that this may not be longer true for the full, not necessarily thin, exact black ring solution, and perhaps `fat 
AdS ring analogues' with positive $C_P$ exist in higher dimensions, cf. discussion of $d=5$ exact asymptotically flat black rings in the previous section.

\subsection{Thermodynamic volume}
Using the thermodynamic quantities \eqref{TdSBRAdS} and the Smarr relation \eqref{Smarr}, the thermodynamic volume of thin AdS black rings reads 
\ba\label{volumeAdSBR}
V_{\mbox{\tiny  AdS}}&=&\frac{\pi \omega_{d-3}l^3}{d-1} \mathsf{R}^3r_0^{d-4}\sqrt{1+\mathsf{R}^2}\nonumber\\
&=&\frac{\pi \omega_{d-3}}{d-1} R^3r_0^{d-4}\sqrt{1+R^2/l^2}\,. 
\ea
One can then readily verify that the first law \eqref{1st} is satisfied.
It is obvious that the last formula has a smooth limit for $l\to \infty$. That is, a thermodynamic volume of an asymptotically flat black ring is
\be\label{volumeflatBR}
V_{\mbox{\tiny  flat}}=\frac{\pi \omega_{d-3}}{d-1} R^3r_0^{d-4}\,,
\ee
while its horizon area reads
\be
A=2\pi r_{0}^{d-3}\omega _{d-3}R \sqrt{\frac{d-3}{d-4}}\,.
\ee

Let us briefly comment on the structure of formula for $V_{\mbox{\tiny  flat}}$, \eqref{volumeflatBR}. We first note that a naive geometric volume of a string ``bent to'' a radius $R$ goes as
\be\label{RingV'}
V'\propto 2\pi R r_0^{d-2}\,.
\ee
This is fundamentally different from the computed thermodynamic volume \eqref{volumeflatBR}.
However, we have seen in Sec.~\ref{sec:KerrAdS} that even for spherical black holes in the presence of rotation  the two volumes, the geometric one and the thermodynamic one, are not the same and differ by a term proportional to $J^2/M$, see \eqref{VKerr}.\footnote{We are grateful to Jennie Traschen and David Kastor for reminding us of this fact.}
For the black ring we find
\be
\frac{J^2}{M}=\frac{d-3}{8(d-2)}\omega_{d-3}r_0^{d-4}R^3\,.
\ee
Obviously, in the validity of our approximation, \eqref{approx}, $R>>r_0$ and this term is much bigger than the contribution of the geometric volume $V'$, 
\eqref{RingV'}. This indicates that we are in fact in the ultraspinning regime. We find
\be
V_{\mbox{\tiny  flat}}=\frac{8\pi}{d-1}\frac{d-2}{d-3}\frac{J^2}{M}\,,
\ee
which replaces Eq.~\eqref{VeqV'} valid for ultraspinning spherical black holes.
To summarize, the structure of the formula \eqref{volumeflatBR} is governed by the ultraspinning contribution rather than the
geometric part of the volume, which explains otherwise puzzling dependence $V\propto R^3r_0^{d-4}$. We expect that more generally, outside of the ultraspinning regime, one would have
\be
V_{\mbox{\tiny  flat}}=V'+\frac{8\pi}{d-1}\frac{d-2}{d-3}\frac{J^2}{M}\,,
\ee
in parallel with formula \eqref{VKerr} for spherical black holes.

\subsection{Isoperimetric inequality}\label{isop}
It was conjectured in \cite{CveticEtal:2010} that a {\em reverse isoperimetric inequality} holds
for a thermodynamic volume of any asymptotically AdS black holes, the statement being verified for a variety of
(charged rotating) black holes with the horizon of spherical topology. Here we show that this conjecture remains true for asymptotically AdS black rings, at least in the thin ring (fast spinning) approximation.

Let us calculate the ratio ${\cal R}$ in  \eqref{ratio}, identifying
${\cal A}=A$ and ${\cal V}=V$ as given by \eqref{TdSBRAdS} and \eqref{volumeAdSBR}\,.
Then we have
\ba
{\cal R}^{(d-1)(d-2)}=\frac{\omega_{d-2}}{2^{d-1}\pi \omega_{d-3}} \left(\frac{R}{r_0}\right)^{2d-5}\!\!\!\!
\frac{\bigl(1+\mathsf{R}^2\bigr)^{\frac{d-2}{2}}}{\bigl(\frac{d-3}{d-4}+\frac{d-2}{d-4}\mathsf{R}^2\bigr)^{\frac{d-1}{2}}}\,\nonumber\\
\geq\frac{\omega_{d-2}}{2^{d-1}\pi \omega_{d-3}} \left(\frac{R}{r_0}\right)^{2d-5}\;
\left(\frac{d-4}{d-2}\right)^{\frac{d-1}{2}} \!\!\!\!\frac{1}{\sqrt{1+\mathsf{R}^2}}\,.\qquad\quad
\ea
Now for $R\leq l$ we have $\sqrt{1+\mathsf{R}^2}\leq \sqrt{2}$ and hence the ratio  \eqref{ratio} is governed by
\be
{\cal R}^{(d-1)(d-2)}\geq \mbox{const}\times \left(\frac{R}{r_0}\right)^{2d-5}\gg 1
\ee
by the assumption of the approximation, \eqref{approx}. For $R>l$, $\sqrt{1+\mathsf{R}^2}\leq \sqrt{2}\mathsf{R}\ll\sqrt{2}R/r_0$. In this case, we have
\be
{\cal R}^{(d-1)(d-2)}\gg \mbox{const}\times \left(\frac{R}{r_0}\right)^{2d-6}\gg 1\,.
\ee
So in both cases the reverse isoperimetric inequality is comfortably satisfied in the given approximation \eqref{approx}.
Note that, similar to the Kerr-AdS case in the ultraspinning regimes, see Sec.~\ref{sec:KerrAdS}, the ratio is much bigger than one; in fact in the full ultraspinning limit in both instances it diverges.

\subsection{Equation of state}
In order to find the equation of state for our black rings, we start from the expressions for $J$, $T$, and $V$ in Eqs.~\eqref{TdSBRAdS}.
Expressing $r_0$ from the $J$ equation as in \eqref{r0}, the $V$ equation becomes a quadratic equation for $\mathsf{R}$, and gives
\ba\label{R2}
\mathsf{R}^2&=&\frac{-1\pm \sqrt{1-4k}}{2}\,,\quad k=\frac{\alpha(d-3)}{\alpha(d-2)^2-1}\,,\nonumber\\
\alpha&=&\frac{1}{4\pi}\frac{d-1}{d-2}\frac{PV^2}{J^2}\geq 0\,.
\ea
In order $\mathsf{R}^2$ be non-negative, we must have $k\leq 0$, i.e., $\alpha\leq 1/(d-2)^2$, or, in terms of the 
physical quantities\footnote{This bound is attained for ultraspinning black rings of $\mathsf{R}$. Such black rings have in fact an infinite temperature, see next subsection.} 
\be\label{PultraBR}
P\leq P_{\mbox{\tiny ultra}}\equiv \frac{4\pi}{(d-1)(d-2)}\frac{J^2}{V^2}\,.
\ee
The $T$ equation \eqref{TdSBRAdS}, 
\be\label{stateRing}
T = \frac{(d-4)^{\frac{3}{2}}\sqrt{1+\mathsf{R}^2}}{4\pi r_{0}\sqrt{d-3+(d-2)\mathsf{R}^2}}\,,
\ee
with $r_0$ and $\mathsf{R}$ expressed by \eqref{r0} and \eqref{R2}, becomes an implicit equation of state, $P=P(V,T,J)$.

Note that, in the region of validity of our approximation we must have
\be\label{ABR}
\frac{R}{r_0}=\frac{\sqrt{(d-1)(d-2)}}{4\sqrt{\pi P}}\frac{\mathsf{R}}{r_0}\gg 1\,,\quad \frac{l}{r_0}\gg 1\,,
\ee
while at the same time the condition \eqref{PultraBR} must be satisfied. For fixed $J$ these conditions pick up a very narrow region in $P-v$ space of possible black ring configurations where we can trust our equation of state.   

\begin{figure}
\begin{center}
\rotatebox{-90}{
\includegraphics[width=0.39\textwidth,height=0.34\textheight]{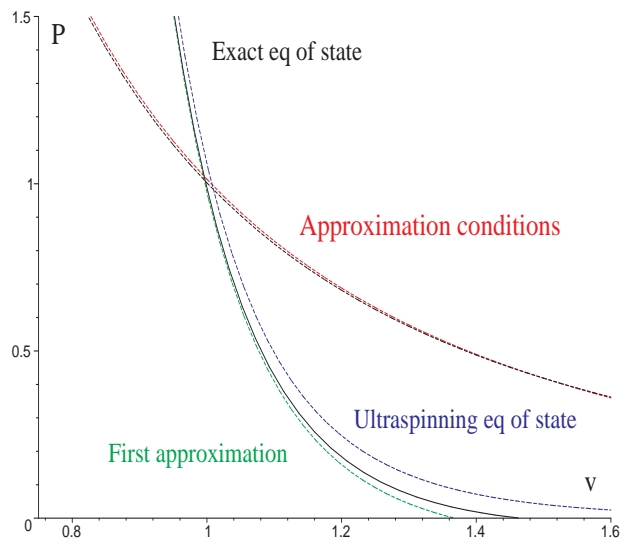}
}
\caption{{\bf Equation of state: thin AdS black ring.}
The ``exact'' equation of state, given implicitly by \eqref{stateRing}, for $T\approx 0.133$ is displayed by the thick solid black curve, and compared to the ultraspinning equation of state \eqref{PultraBR}, displayed by the thick dashed blue curve, and the first correction equation of state \eqref{PexpansionBR}, displayed by the thick dashed green curve. The two conditions \eqref{ABR} are respectively displayed by thin dashed red and black curves.  They almost coincide:  setting $R/r_0=N=l/r_0$, the approximation is valid when $N\gg 1$.  As in the AdS/CFT correspondence, we set $N=3\gg 1$, which gives the curves; the approximation is valid above these curves. This means that we can trust our ``exact'' equation of state in the top left corner of the figure. In this region all three equations of state basically coincide.     
}
\label{Fig:5DPvRing}
\end{center}
\end{figure}

The 
resulting $P-v$ diagram is `trivial', as depicted in fig.~\ref{Fig:5DPvRing} for $d=5$.  In higher dimensions the situation is similar. The critical point, if it exists for AdS black rings, is outside of the domain of validity of our approximation. To see this we will have to wait until the exact AdS black ring solution is known.

Similar to the spherical black hole case we shall now study the ultraspinning regime. (The regime of slow rotation is excluded by the blackfold approximation.)

\subsection{Ultraspinning expansion}
By inspection, the ultraspinning limit, for which $\Omega l\to 1$, corresponds to the limit where we keep the mass of the black ring $M$ fixed and let
\be
\mathsf{R}\to \infty\,,
\ee
that is, the limit of the ultra-large black rings. In this limit, $r_0\sim 1/\mathsf{R}^{\frac{4}{d-4}}\to 0$, $T\sim \mathsf{R}^{\frac{4}{d-4}}\to \infty$, whereas the quantities $V$ and $J$ remain finite.

To write down the equation of state in this limit we proceed similar to the black hole case. Namely, we express $r_0$ from the $M$ equation \eqref{TdSBRAdS}\,,
\be
r_0=\Bigl[\frac{8M}{l\omega_{d-3}(d-2)}\frac{1}{\mathsf{R}(1+\mathsf{R}^2)^{3/2}}\Bigr]^\frac{1}{d-4}\,.
\ee
This is inserted into the equations for $J$, $T$, and $V$ in \eqref{TdSBRAdS}, and the result is expanded in large
$\mathsf{R}$, while we keep the quantities $M$ and $l$ fixed. The first few terms are
\ba
J&=&Ml\Bigl(1-\frac{1}{\mathsf{R}^2}+\frac{(d-1)(2d-5)}{2(d-2)^2\mathsf{R}^4}+\dots\Bigr)\,,\nonumber\\
T&=&\frac{(d-4)^{3/2}}{4\pi \sqrt{d-2}}\Bigl(\frac{l\omega_{d-3}(d-2)}{8M}\Bigr)^{\frac{1}{d-4}}
\mathsf{R}^{\frac{4}{d-4}}\times\nonumber\\
&&\times\Bigl(1+\frac{2d-5}{(d-4)(d-2)}\frac{1}{\mathsf{R}^2}\nonumber\\
&&\qquad-\frac{5d^3-53d^2+172d-178}{4(d-4)^2(d-2)^2}\frac{1}{\mathsf{R}^4}+\dots\Bigr)\,,\nonumber\\
V&=&\frac{8\pi l^2M}{(d-1)(d-2)}\bigl(1-\frac{1}{\mathsf{R}^2}+\frac{1}{\mathsf{R}^4}+\dots\bigr)\,.
\ea
Consequently, we find the following approximate black ring equation of state:
\be\label{PexpansionBR}
P=P_{\mbox{\tiny ultra}}-\frac{2\omega_{d-3}(d\!-\!1)(d\!-\!3)(d\!-\!4)^{\frac{3d-12}{2}}}
{4^d\pi^{d-3}(d-2)^{\frac{d-4}{2}}J T^{d-4}}+O\Bigl(\frac{1}{\mathsf{R}^4}\Bigr)\,, \qquad\ \
\ee
with $P_{\mbox{\tiny ultra}}$ defined in \eqref{PultraBR}.
This is displayed and compared to the ``exact'' equation of state \eqref{stateRing} in fig.~\ref{Fig:5DPvRing}.

Note the interesting coincidence that the first term, $P_{\mbox{\tiny ultra}}$, in the expansion coincides with the 
ultraspinning limit of the Kerr-AdS black hole equation of state. Therefore both the ultraspinning Kerr-AdS black holes and AdS black rings 
have infinite temperature and share the `same equation of state'.

\section{Beyond Thermodynamic Instabilities}\label{sec:Instab}

Thus far we have considered only the thermodynamic behaviour of black holes, using the global minimum of the Gibbs free energy (and the positivity of the specific heat) as a criterion for thermodynamic stability.  However rotating black holes are subject to other kinds of (classical) instabilities.  {\em Ultraspinning instabilities} can lead to bifurcations to new stationary black hole families \cite{EmparanMyers:2003,
Dias:2009iu, Dias:2010a, Dias:2010maa, DiasEtal:2010, Dias:2011jg}.  {\em Superradiant instabilities} \cite{CardosoEtal:2004ho, CardosoDias:2004, CardosoEtal:2006ho} amplify waves that scatter off of the AdS black hole, extracting its rotational energy and driving it to a new state. 

A third  instability is the so called (non-axisymmetric) {\em bar-mode instability}  \cite{EmparanMyers:2003, ShibataYoshino:2010, ShibataYoshino:2010b, HartnettSantos:2013}, which possibly occurs for both MP and  Kerr-AdS black holes. However, nothing is known about the latter case, and even in the former case the mechanism as well as the endpoint of such instability  (apart from the slowly spinning case which results in a slowly spinning MP black hole)
is far from being well understood. It may result in the irradiation of the excessive angular momenta or even a fragmentation into a multiple black hole configuration  \cite{HartnettSantos:2013}. We shall not consider this instability in  further discussion.

\subsection{Ultraspinning instability}
 Ultraspinning (axisymmetric) instabilities occur for both asymptotically flat MP and Kerr-AdS black holes in all  dimensions $d\geq 6$. It is related to
the fact that in $d\geq 6$ the angular momentum of the black hole is not limited from above and in principle can grow all the way to 
infinity.\footnote{In fact, even when the angular momenta are limited (for instance in the equal spinning MP case) the black hole can enter the ultraspining regime.}
 Consequently as the angular momentum grows larger, the black hole horizon flattens, becomes pancake like, and the black hole enters the region of black membrane like behavior (see also Sec.~\ref{sec:KerrAdS}). 
The resulting object is subject to the Gregory--Laflamme instability and the  axisymmetric perturbations can lead to bifurcations to new stationary black hole families. In the absence of a decoupled master equation for gravitational perturbations in higher dimensions, the final endpoint as well as the exact onset of ultraspinning instabilities have to be investigated numerically \cite{Dias:2009iu, Dias:2010a, Dias:2010maa, DiasEtal:2010, Dias:2011jg}.
However, for our purposes   we just need to know when such an instability  approximately ``kicks in'' and may have some effect on the thermodynamic considerations studied in previous sections.

 The method for investigating   ultraspinning instabilities consists of finding {\em zero modes} for stationary perturbations that preserve rotational symmetries. The onset of such instabilities can be assigned a `harmonic number' $k$. 
The $k=0,1$ modes respectively correspond to metric perturbations with $(0,1)$ nodes. These are the so called {\em thermodynamic zero modes} and can be determined from the thermodynamic Hessian, as we shall discuss below. However, these modes do not correspond to bifurcations to new families of stationary black holes; they can only change the black hole into another black hole of the same family, for example an MP black hole into another MP black hole.

 The actual bifurcations occur for $k=2$ and higher; the case of $k=2$ corresponds to a new branch of black holes with a central pinch. For higher $k$ further axisymmetric pinches appear. These cannot be obtained from thermodynamic considerations. However it has been conjectured   \cite{Dias:2009iu} that
the zero modes with $k\geq 2$ can only occur for rotations higher than that of the $k=1$ modes. 
The $k=1$ surface, calculable from thermodynamic considerations, is called the {\em ultra-spinning surface}:  it encloses the region where the thermodynamic Hessian has less than two negative eigenvalues. For higher rotations the black holes are subject to ultraspinning instabilities leading to bifurcations to new black hole families \cite{DiasEtal:2010}.

 Let us first illustrate this for a singly spinning $d=6$ MP black hole.

\subsubsection{Bifurcations of singly spinning MP black holes} 

 A simple criterion  \cite{EmparanMyers:2003} for the onset of ultraspinning instabilities of singly spinning MP black holes involves a consideration of the behaviour of the temperature $T$ for a fixed mass $M$. 
For slowly spinning black holes this decreases, as is familiar for the Kerr solution, and in $d<6$ reaches zero. However, for $d\geq 6$
the temperature instead reaches its minimum and diverges in the ultraspinning limit.
This  temperature growth is typical for black membranes. 

Let us define a dimensionless parameter $x=\frac{a}{r_+}$, in terms of which we have,  cf. $J$ equation \eqref{TDsMP},
\be\label{rplusJ}
r_+=\Bigl[\frac{8\pi}{\omega_{d-2}}\frac{J}{x(x^2+1)}\Bigr]^{\frac{1}{d-2}}\,.
\ee
The temperature of a singly spinning MP black hole is given by 
\be
T=\frac{1}{4\pi}\Bigl(\frac{r_+^{d-4}}{m}+\frac{d-5}{r_+}\Bigr)\,,
\ee
and reaches its minimum at $r_+=r_1$, where
\be
m=\frac{d-4}{d-5}r_1^{d-3}\,\quad \Leftrightarrow\quad x_1=\frac{a}{r_1}=\sqrt{\frac{d-3}{d-5}}
\ee
provided $d\geq 6$. For a fixed angular momentum, using \eqref{rplusJ}, this gives the following black hole radius:
\be\label{r1MP}
r_1=\Bigl[\frac{4\pi}{\omega_{d-2}}\frac{d-5}{d-4}\sqrt{\frac{d-5}{d-3}}J\Bigr]^{\frac{1}{d-2}}\,.
\ee 
The thermodynamic argument below leads to the same value of $r_1$---it gives the $k=1$ zero mode. The $k=0$ zero mode 
for singly spinning MP black holes is present for arbitrary spin.

\begin{figure}
\begin{center}
\rotatebox{-90}{
\includegraphics[width=0.39\textwidth,height=0.34\textheight]{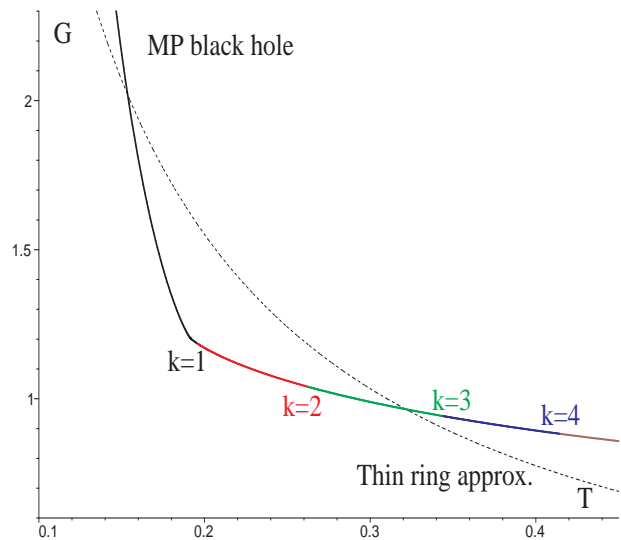}
}
\caption{{\bf Onsets of ultraspinning instabilities: $d=6$ MP black holes.}
The Gibbs free energy of a singly spinning MP black hole with $J=1$ (displayed by solid colored curves) is compared to the Gibbs free energy of a
thin black ring constructed in the blackfold approximation (displayed by a black thin dashed curve); $r_+$ decreases from left to right. The onsets of ultraspinning instabilities \eqref{ys} are displayed by various colors as follows. 
The black solid line corresponds to MP black holes outside of the ultraspinning region (with only the $k=0$ mode present).
The $k=1$ zero mode appears for $r_+=r_1\approx 0.605$ indicated by a point where solid red and black curves join together; the appearance of this zero mode can be predicted from thermodynamic considerations.
The actual onsets of ultraspinning instabilities occur for $k=2,3, 4,\dots$, and are displayed by onsets of green, blue, brown,$\dots,$ curves; these indicate possible bifurcations to
new families of stationary black holes. It is reasonable to conjecture that the actual $G$ of an exact black ring is slightly smaller than that of the displayed approximation and would cross the $G$ of an MP black hole close to $r_+=r_2$, indicating a possible first order phase transition between the two kinds of black holes. 
}
\label{Fig:5DKerrRingG1}
\end{center}
\end{figure}

 The actual onsets of instabilities for perturbations preserving the original Killing symmetries of singly spinning MP black holes in $d=6$ have been studied \cite{Dias:2010maa} and, in terms of a new parameter $y=a/r_m$, where $r_m^{d-3}=2m$,  were found to be
\ba\label{ys}
y_1=y(k=1)&\approx& 1.097\,,\quad y_2=y(k=2)\approx 1.572\,,\nonumber\\
y_3=y(k=3)&\approx& 1.849\,,\quad y_4=y(k=4)\approx 2.036\,.\quad
\ea
Since   $y^{d-3}=x^{d-3}/(x^2+1)$, this gives 
\be\label{xs}
x_1\approx 1.751,\ x_2\approx 4.114,\ x_3\approx 6.472,\ x_4\approx 8.555\,. 
\ee
To each $x_k$ in \eqref{xs}, we use \eqref{rplusJ} to find the black hole radius $r_k$ corresponding to the onset of instabilities with the harmonic number $k$. 
In particular, for $J=1$ this gives $r_1\approx 0.605, r_2\approx 0.337, r_3\approx 0.242, r_4\approx 0.197$. The onset of ultraspinning instabilities leading to bifurcations to new branches of black holes occurs for $r\leq r_2$. The situation is displayed 
in fig.~\ref{Fig:5DKerrRingG1}, where the exact Gibbs free energy of MP black holes, with corresponding onsets, is compared to the Gibbs free energy of approximate highly spinning asymptotically flat black rings constructed in the blackfold approximation of the previous section. 
From the discussion in the previous section we may expect that the approximation is valid for high temperatures, whereas it is completely off for small temperatures. In the middle region, we expect that the free energy $G$ of the actual black ring  may be slightly lower than that of the approximation. 
If so, it may be reasonable to expect that the crossover between $G$ of an MP black hole and that of its exact  (so far analytically unknown, but see \cite{Kleihaus:2012xh} for numerical work) black ring counterpart may occur close to (or perhaps exactly at) $r_2$, where a first order phase transition in between the two kinds of black holes might occur.

\subsubsection{Thermodynamic argument and other examples}
\begin{figure}
\begin{center}
\rotatebox{-90}{
\includegraphics[width=0.39\textwidth,height=0.34\textheight]{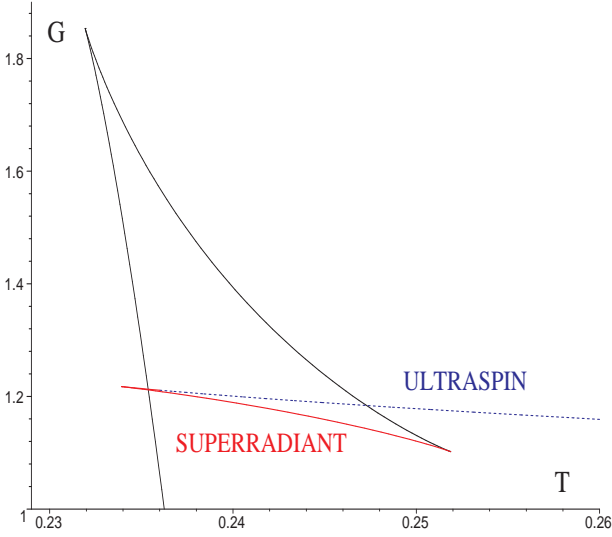}
}
\caption{ {\bf RPT and classical instabilities: $d=6$ singly spinning Kerr-AdS black holes.} The Gibbs free energy for the $d=6$ singly spinning Kerr-AdS black hole is displayed for $J=1$ and $P=0.0564\in (P_t P_z)$ for which the RPT is present, cf. fig.~\ref{Fig:6DKerrZorder} c). Black holes displayed by the solid black curve are classically stable (potentially subject to the bar-mode instability). The red curve indicates the branch of black holes subject to the superradiant instability, $r_+<1.11$. The blue dotted curve displays black holes in the ultraspinning region ($k\geq 1$ zero modes) derived from the thermodynamic argument, $r_+<0.55$. Such black holes are also superradiant unstable. 
We observe that black holes that play a role in the RPT are ultraspinning stable but some of them are subject to the superradiant instability. 
} \label{fig:ultrasupersingle}
\end{center}
\vspace{-0.7cm}
\end{figure}
\begin{figure}
\begin{center}
\rotatebox{-90}{
\includegraphics[width=0.39\textwidth,height=0.34\textheight]{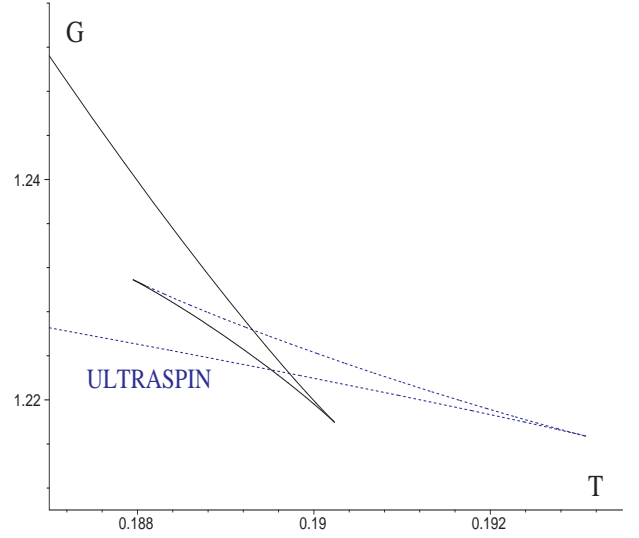}
}
\caption{ {\bf RPT and ultraspinning instability: $d=6$ doubly spinning MP black holes.} 
The Gibbs free energy for the $d=6$ doubly spinning MP black hole is displayed for $q=0.112\in (q_t,q_z)$ for which the RPT is present, 
cf. fig.~\ref{fig:q} d).
The black curve is classically stable (potentially subject to the bar-mode instability). The blue dotted curve indicates the branch of black holes subject to the ultraspinning instability, $r_+<0.613$. Hence, contrary to the Kerr-AdS case the MP black holes that play the role in the RPT are ultraspinning unstable.
} \label{fig:ultrasupersingle2}
\end{center}
\vspace{-0.7cm}
\end{figure}

For more complicated black hole spacetimes, such as  $d=6$ doubly spinning Kerr-AdS black holes, the numeric results are not available.\footnote{ See, however, \cite{Dias:2011jg} for the multiply spinning MP case.} However, 
the ultraspinning surface can be obtained from the following thermodynamic argument, e.g., \cite{MonteiroEtal:2009, Dias:2010a, Dias:2010maa, Dias:2010gk} (see also \cite{Dolan:2013yca}).  

Let $I$ be the (grand-canonical ensemble) Euclidean action, dependent on parameters $y^\alpha$  that completely characterize the black hole in this ensemble: $y^\alpha=y^\alpha(M,J_i)$. Then we consider the Hessian of the action, given by \cite{Dias:2010a} (keeping $\Omega_i$ and $T$ fixed)
\beq\label{U2}
I_{\alpha\beta} \equiv
\frac{\partial ^2 I}{\partial y^\alpha \partial y^\beta}=\left(\frac{1}{T}\frac{\partial ^2 M}{\partial y^\alpha \partial y^\beta}- \frac{\partial ^2 S}{\partial y^\alpha \partial y^\beta}-\frac{\Omega_i}{T} \frac{\partial ^2 J_i}{\partial y^\alpha \partial y^\beta}\right).
\eeq
In particular, as per usual, if we choose the parameters to be $x^\alpha=\{M,J_i\}$, we recover the familiar 
\beq\label{U3}
\left. \frac{\partial ^2 I}{\partial y^\alpha \partial y^\beta}\right\vert_{y=x}=-\frac{\partial ^2 S}{\partial x ^\alpha \partial x ^\beta}=
 -S_{\alpha \beta} (M,J_i)\,.
\eeq
The black hole has a thermodynamic negative mode, corresponding to a thermodynamic instability, if $I_{\alpha \beta}$ given by \eqref{U2} has a negative eigenvalue; thermodynamic zero modes correspond to vanishing eigenvalues.

 For every asymptotically flat vacuum black hole $I_{\alpha\beta}$ possesses at least
one negative eigenvalue and so  all such black holes are locally thermodynamically unstable \cite{Dias:2010a}. For $d\geq 6$ singly spinning MP black holes, as their rotation increases, we obtain an additional thermodynamic instability at precisely $r_+=r_1$, as given by \eqref{r1MP} above. This determines the ultraspinning surface $k=1$.

 For singly spinning Kerr-AdS black holes in $d=4$ and $d=5$, one of the eigenvalues of the $2\times 2$ Hessian matrix $I_{\alpha\beta}$ is always positive and the second one changes sign. We therefore have the $k=0$ thermodynamic zero mode at $r_+=r_0$, given by \eqref{U4} below; the $k=1$ zero mode does not exist and the black holes are ultraspinning stable.
 
 For the $d\geq 6$  singly spinning Kerr-AdS case one can show that two eigenvalues of the Hessian matrix change sign, giving two thermodynamic zero modes, $k=0$ and $k=1$. Solving for the zero eigenvalues of $I_{\alpha\beta}$ in Eq. (\ref{U2}), taking $y^\alpha=\{r_+,a\}$, we obtain \cite{DiasEtal:2010}
\ba\label{U4}
r_0^{2}&=&\frac{d-3}{2(d-1)}\left(a^2+l^2+ 
  \sqrt{a^4-\gamma_d a^2 l^2+l^4}\right)\,, \nonumber\\
  r_1^{2}&=&\frac{d-3}{2(d-1)}\left(a^2+l^2- 
  \sqrt{a^4-\gamma_d a^2 l^2+l^4}\right)\,, \quad
\ea
where $\gamma_d \equiv \frac{2(d^2-6d+1)}{(d-3)^2}$.  
For very large $r_+$, corresponding to slowly spinning Kerr-AdS black holes, there are no negative thermodynamic modes and the black holes are thermodynamically stable.
As we decrease $r_+$, the $k=0$ zero mode appears at $r_0$. Decreasing $r_+$ even further, an additional thermodynamic zero mode $k=1$ occurs 
at $r_1$. Black holes with $r_+<r_1$ are in the ultraspinning regime---they are subject to $k\geq 2$ zero modes and suffer from ulraspinning instabilities. We show in fig.~\ref{fig:ultrasupersingle} that the singly spinning Kerr-AdS black holes that play the role in the RPT studied in Sec.~\ref{sec:KerrAdS} are ultraspinning stable, though some of them may be subject to the superradiant instability, as studied in the next subsection.

We demonstrated in Sec.~\ref{sec:MP} that an RPT also occurs for the $d=6$ doubly spinning MP black holes. Such black holes always possess a $k=0$ negative mode. The thermodynamic argument gives the following formula for the appearance of the $k=1$ zero mode:
\be
r_1^2=\frac{1}{6}\Bigl(a^2+b^2+\sqrt{a^4+14a^2b^2+b^4}\Bigr)\,,
\ee 
where $a$ and $b$ are the two rotation parameters. In this case we find, see fig.~\ref{fig:ultrasupersingle2}, that the doubly spinning MP black holes that play the role in the RPT studied in \ref{sec:MP} are in the ultraspinning regime.

\begin{figure}
\begin{center}
\rotatebox{-90}{
\includegraphics[width=0.39\textwidth,height=0.34\textheight]{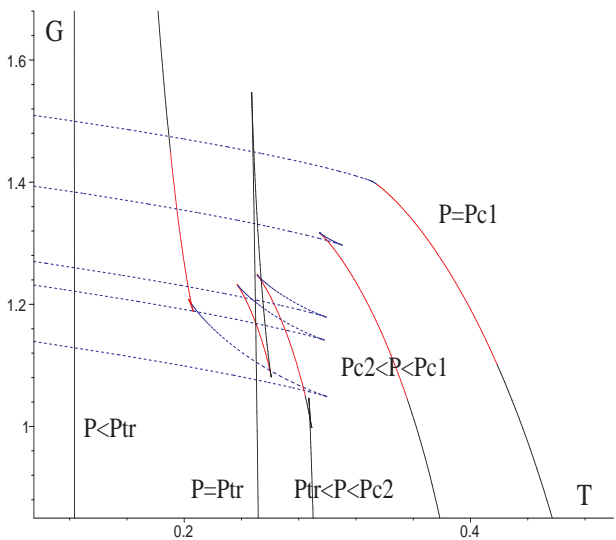}
}
\caption{ {\bf Classical instabilities: doubly spinning $d=6$ Kerr-AdS black hole.} The Gibbs free energy for the $d=6$ doubly spinning Kerr-AdS black holes is displayed for $q=0.05$ and  various pressures (from top to bottom) $P=0.260, 0.170, 0.087, 0.0642, 0.015$, for which we observe the SBH/IBH/LBH phase transition, cf. fig.~\ref{fig:5}. 
The black solid curves denote classically stable black holes (potentially subject to the bar-mode instability). The blue color indicates branches of black holes subject to both the ultraspinning and superradiant instabilities. The red curves correspond to black holes that are superradiant unstable and ultraspinning stable. We conclude that the Kerr-AdS black holes that play the role in the SBH/IBH/LBH phase transitions are potentially subject to both classical instabilities.
} \label{fig:ultrasuperKerrAdSdouble}
\end{center}
\vspace{-0.7cm}
\end{figure}

Turning finally to the doubly spinning Kerr-AdS black holes and the corresponding `solid/liquid/gas' phase transition studied in Sec.~\ref{sec:KerrAdS}, no simple analytic formula for the appearance of thermodynamic zero modes exists, and we proceed numerically as displayed 
in fig.~\ref{fig:ultrasuperKerrAdSdouble}. We find that the part of the Gibbs diagram important for the existence of
multiple first order phase transitions and possibly a triple point occurs in the ultraspinning region and is subject to the superradiant instability. Hence the observed thermodynamic phenomena have to `compete' with both the ultraspinning and 
the superradiant classical instabilities.

\subsection{Superradiant instabilities}

 Superradiant instabilities are potentially present for rapidly spinning Kerr-AdS black holes in all $d\geq 4$ dimensions.\footnote{Interestingly, they also occur for the massive scalar field in the asymptotically flat $d=4$ Kerr black hole spacetime \cite{Detweiler:1980,Dolan:2007mj}.}
 They also A superradiant instability occurs for \cite{Hawking:1999dp}
\be\label{super}
\Omega_il>1 
\ee  
and is associated with perturbations that break the spacetime axisymmetry. For example  a gravitational wave of frequency $\omega$ 
co-rotating with the background black hole can extract rotational energy from the black hole if   $\omega<m\Omega_i$, where $m$ is the wave angular momentum. The effect is called superradiance and represents a `field analogue' of the Penrose process. Superradiance occurs also in the asymptotically flat case. However, in the presence of the `AdS confining box', the amplified field reflects back on black hole and is amplified again and again, driving the system unstable whenever the condition \eqref{super} is satisfied. 

\begin{figure}
\begin{center}
\rotatebox{-90}{
\includegraphics[width=0.39\textwidth,height=0.34\textheight]{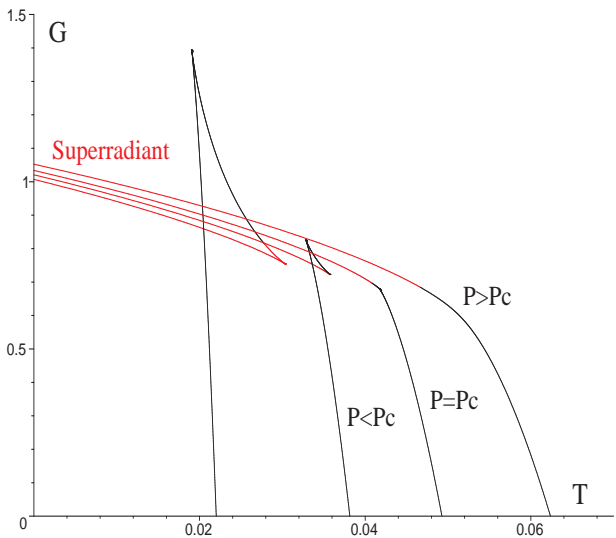}
}
\caption{{\bf Superradiant instability: $d=4$ Kerr-AdS black hole.}
The Gibbs free energy for the Kerr-AdS black hole in $d=4$ is displayed for various pressures, $P/P_c=\{1.6, 1, 0.6, 0.2\}$, $P_c\approx 0.002856$, and fixed $J=1$. The red curves display black holes that are subject to the superradiant instability, the black ones are superradiant stable. 
The figure clearly demonstrates that even the Van der Waals like phase transition occurring for $d=4$ Kerr-AdS black holes, discussed in Sec.~\ref{sec:4d},  is subject to the superradiant instability.
} \label{fig34}
\end{center}
\vspace{-0.7cm}
\end{figure}

 Superradiant regions play an important role for the  thermodynamic phase transitions we have studied.
These instabilities are indicated in the $G-T$ diagrams for $d=6$ singly and doubly spinning Kerr-AdS black hole in figs.~\ref{fig:ultrasupersingle} and \ref{fig:ultrasuperKerrAdSdouble}. We find that the corresponding thermodynamic effects we have discovered  so far must compete with the superradiant instability. Interestingly, this is even true for the Van der Waals like phase transition that occur in $d=4$ Kerr-AdS black hole spacetimes, see fig.~\ref{fig34}.
However the time scale of these latter instabilities can vary widely, and so it may happen that the thermodynamic effects we have discovered can actually take place.

\section{Discussion}\label{sec:Discussion}
\begin{figure}
\begin{center}
\rotatebox{-90}{
\includegraphics[width=0.39\textwidth,height=0.34\textheight]{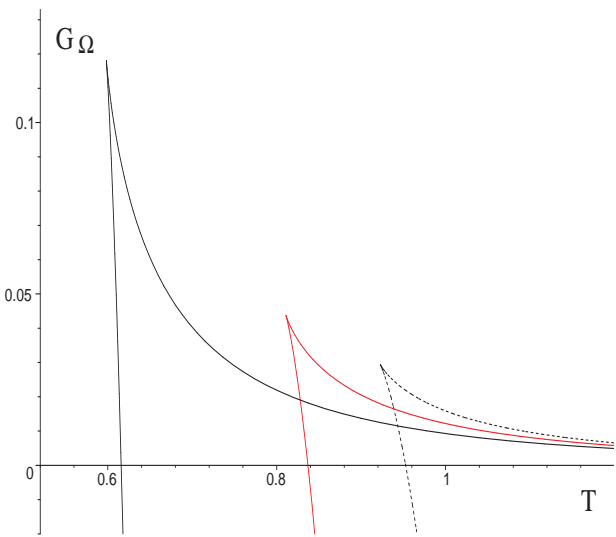}
}
\caption{{\bf Grand-canonical thermodynamic potential.} The grand-canonical thermodynamic potential $G_{\Omega}$ is displayed for $d=6$ singly spinning Kerr-AdS black hole for fixed $\Omega=1$ and various pressures (from left to right) $P=0.5, 0.8, 1$.  Its behaviour is qualitatively similar to the behavior of $G$ for the Schwarzschild-AdS black hole. Namely, although we may observe an analogue of the Hawking--Page transition, there is no swallow tail present and hence no first order phase transition of the Van der Waals type. 
} \label{fig:GrandCan}
\end{center}
\vspace{-0.7cm}
\end{figure}

We have carried out a thorough study of the thermodynamics of higher-dimensional asymptotically flat and AdS rotating black holes with various horizon topologies in the {\em canonical} (fixed $J$) ensemble, interpreting the cosmological constant as thermodynamic pressure. 
By studying the corresponding thermodynamic potential---the Gibbs free energy---we uncovered a number of novel thermodynamic phase transitions in these black hole spacetimes.

We have shown that multiply-rotating Kerr-AdS black holes exhibit a rich set of interesting thermodynamic phenomena known from the every day thermodynamics of simple substances. Specifying to six dimensions, we have demonstrated that reentrant phase transitions, triple points (tricritical points), multiple first-order transitions, solid/liquid/gas phase transitions, and Van der Walls liquid/gas phase transitions can all occur depending on the ratio of the two angular momenta. We have evidence that such phenomena will take place in dimensions $d> 6$, though the details remain to be worked out.
Intriguingly, we have shown that reentrant phase transitions occur even for   asymptotically flat Myers--Perry black holes. The cosmological constant is not necessary to observe this phenomenon.

We have also studied the thermodynamic volume, the corresponding equation of state, $P=P(V,T)$, as well as the associated reverse isoperimetric inequality. For the first time, we verified the validity of this inequality for the black hole spacetimes with non-spherical horizon topology, providing one more piece of evidence for the conjecture  \cite{CveticEtal:2010} that a 
reverse isoperimetric inequality  holds for any asymptotically AdS black hole.

In this paper we concentrated on a canonical ensemble where the angular momentum is fixed and the horizon angular velocity is allowed to vary. We showed that in this ensemble all possible kinds of interesting phase transitions happen. Alternatively, one could  
consider the thermodynamic behaviour of above black holes in the {\em grand-canonical ensemble}, where the angular velocity of the horizon is fixed
and the angular momentum is free to vary, e.g., \cite{ChamblinEtal:1999a, HawkingEtal:1999, CaldarelliEtal:2000, PecaLemos:1999}. The corresponding thermodynamic potential $G_{\Omega}$ is related to the Gibbs free energy \eqref{G} by the Legendre transformation 
\be
G_{\Omega}\equiv G-\sum_i \Omega_i J_i-\Phi Q=G_{\Omega}(T,P, \Omega_i, \Phi)\,.
\ee 
However, for charged-AdS black holes it has previously been demonstrated \cite{ChamblinEtal:1999a} that the grand-canonical ensemble does not admit first-order phase transitions ala Van der Waals. Indeed, the behaviour of $G_{\Omega}$ is rather `boring', reminiscent of the 
Schwarzschild-AdS behaviour. We have checked for singly-spinning Kerr-AdS black holes in $d=6$ that a similar conclusion remains valid in the (higher-dimensional) rotating case (see also \cite{HawkingEtal:1999}). Specifically, the Gibbs free energy $G_{\Omega}$ exhibits Schwarzschild-like behaviour without any Van der Waals like first-order phase transitions, as shown in fig.~\ref{fig:GrandCan}. A similar conclusion remains valid for  multiply-spinning Kerr-AdS black holes in $d=6$. For these reasons we expect that the phase structure of
the grand-canonical ensemble is not ``as interesting'' as that of the canonical ensemble. (See, however, 
recent work \cite{Dutta:2013dca}.)

Another interesting open possibility left for future is to use an ensemble where the thermodynamic volume $V$ rather than the cosmological constant $\Lambda$ is fixed, with a corresponding thermodynamic potential $G_V$ related to $G$ by 
$G_V\equiv G- PV=G_V(T,V,J_i, Q)\,;$ 
and similarly for the grand-canonical ensemble. The question whether such an ensemble can be `prepared' for black hole spacetimes remains open.

Finally, we stress that all the results of this paper are entirely based on thermodynamic considerations.  The phase transitions uncovered are thermodynamic phase transitions---determined entirely from the behaviour of the black hole Gibbs free energy. 
In  ``non-ideal'' black hole geometries  perturbations will be present that  may lead to classical instabilities and additional/alternative 
phase transitions. One may, for example, anticipate for small temperatures the formation of hairy black holes, or appearance of 
ultraspinning unstable modes for rapidly spinning black holes.   Similarly, in the presence  of radiation/matter, one might expect radiation/black hole phase transitions of the Hawking--Page type. The question as to which transition would actually `win' and take place depends on the timescales of the corresponding phenomena.
This, however, goes beyond the scope of the present paper.

\section*{Acknowledgments}
We would like to thank Brian~P.~Dolan, Pau~Figueras, David~Kastor, Don~N.~Page, and Jennie~H.~Traschen for discussions and helpful comments on this work and Jorge~E.~Santos for useful comments and reading the manuscript.
This research was supported in part by Perimeter Institute for Theoretical Physics and by the Natural Sciences and Engineering Research Council of Canada. Research at Perimeter Institute is supported by the Government of Canada through Industry Canada and by the Province of Ontario through the Ministry of Research and Innovation.



\providecommand{\href}[2]{#2}\begingroup\raggedright\endgroup

\end{document}